\documentclass[twocolumn,superscriptaddress,longbibliography,prx, nofootinbib]{revtex4-2}
\usepackage[pdftitle={NonAbelianswSSB}]{hyperref}
\usepackage{amsmath,amssymb,amsfonts,amsthm,mathtools}

\usepackage{graphicx}
\usepackage{bm}
\usepackage{mathtools}
\usepackage{physics}
\usepackage{xcolor}

\usepackage{bbm}
\usepackage{ulem}
\usepackage{hyperref}
\usepackage{enumitem}
\usepackage{tikz}
\usetikzlibrary{decorations.pathreplacing,calc, decorations.markings,matrix,arrows.meta,fadings}
\definecolor{cred}{RGB}{178,24,24}     
\definecolor{cgray}{RGB}{130,130,130} 
\definecolor{clatt}{RGB}{215,215,215} 
\definecolor{curveblue}{RGB}{0,0,205}
\definecolor{curvegreen}{RGB}{34,139,34}
\definecolor{guidegray}{RGB}{170,170,170}
\usepackage[ruled,vlined]{algorithm2e}
\usepackage{dsfont}

\usepackage{framed}
\usepackage{comment}

\definecolor{lvlone}{RGB}{204,204,255}   
\definecolor{lvltwo}{RGB}{198,245,198}   
\definecolor{lvlthree}{RGB}{255,250,185} 
\definecolor{lvlfour}{RGB}{255,220,170}  
\definecolor{lvlfive}{RGB}{255,190,190}  

\tikzset{
  node base/.style={circle, draw=black, line width=0.6pt, inner sep=0pt},
  dotV/.style   ={node base, fill=lvlone,   minimum size=5.5pt},
  pentR/.style  ={node base, fill=lvltwo,   minimum size=6.5pt},
  squareR/.style={node base, fill=lvlthree, minimum size=7.5pt},
  triR/.style   ={node base, fill=lvlfour,  minimum size=8.5pt},
  boxR/.style   ={node base, fill=lvlfive,  minimum size=11pt, font=\scriptsize},
  lvl/.style    ={line width=0.8pt, black},
  flow/.style   ={->, >=Stealth, line width=0.7pt},
}
 
\newcommand{\panel}[3]{%
  \begin{scope}[shift={(#2,#3)}]
    \ifnum#1>1
      \foreach \j in {0,1,2,3} \foreach \i in {0,2} { \draw[lvl] (\i,\j) -- (\i+1,\j); }
    \fi
    \ifnum#1>2
      \foreach \j in {0,2} \foreach \i in {0.5,2.5} { \draw[lvl] (\i,\j) -- (\i,\j+1); }
    \fi
    \ifnum#1>3
      \foreach \j in {0.5,2.5} { \draw[lvl] (0.5,\j) -- (2.5,\j); }
    \fi
    \ifnum#1>4
      \draw[lvl] (1.5,0.5) -- (1.5,2.5);
    \fi
    \foreach \j in {0,1,2,3} \foreach \i in {0,1,2,3} { \node[dotV] at (\i,\j) {}; }
    \ifnum#1>1
      \foreach \j in {0,1,2,3} \foreach \i in {0.5,2.5} { \node[pentR] at (\i,\j) {}; }
    \fi
    \ifnum#1>2
      \foreach \j in {0.5,2.5} \foreach \i in {0.5,2.5} { \node[squareR] at (\i,\j) {}; }
    \fi
    \ifnum#1>3
      \foreach \j in {0.5,2.5} { \node[triR] at (1.5,\j) {}; }
    \fi
    \ifnum#1>4
      \node[boxR] at (1.5,1.5) {0};
    \fi
  \end{scope}
}

\newcommand{\DG}{\mathcal{D}(G)}

\newcommand{\LG}{L_2(G)}
\newcommand{\tj}[6]{\begin{pmatrix} #1 & #2 & #3 \\ #4 & #5 & #6 \end{pmatrix}}

\theoremstyle{remark}

\newcommand{\E}{\mathcal{E}}
\newcommand{\R}{\mathcal{R}}
\newcommand{\1}{\mathbbm{1}}

\newcommand{\beq}{\begin{equation}}
\newcommand{\eeq}{\end{equation}}
\newcommand{\beqn}{\begin{eqnarray}}
\newcommand{\eeqn}{\end{eqnarray}}

\begin{document}

\title{Climbing the Fusion Tree: Center Ordering and Decoding Hierarchy \\in Non-Abelian Mixed States}

\author{Pablo~Sala}
\thanks{P.S. and V.T. contributed equally to this work. \\
\href{mailto:pablo.sala@ift.csic.es}{pablo.sala@ift.csic.es}\\
\href{mailto:vlad.temkin@berkeley.edu}{vlad.temkin@berkeley.edu}
}
\affiliation{Department of Physics, University of California, Berkeley, CA 94720, USA}
\affiliation{Simons Institute for the Theory of Computing, University of California at Berkeley}
\affiliation{Instituto de Física Teórica UAM-CSIC, Universidad Autónoma de Madrid, Cantoblanco, 28049, Madrid, Spain}

\author{Vlad~Temkin}
\thanks{P.S. and V.T. contributed equally to this work. \\
\href{mailto:pablo.sala@ift.csic.es}{pablo.sala@ift.csic.es}\\
\href{mailto:vlad.temkin@berkeley.edu}{vlad.temkin@berkeley.edu}
}
\affiliation{Department of Physics, University of California, Berkeley, CA 94720, USA}

\author{Cenke~Xu}
\affiliation{Department of Physics, University of California, Santa Barbara, California 93106, USA}

\author{Daniel~Podolsky}
\affiliation{Physics Department, Technion, Haifa 32000, Israel}

\author{Ehud~Altman}
\affiliation{Department of Physics, University of California, Berkeley, CA 94720, USA}
\affiliation{Materials Sciences Division, Lawrence Berkeley National Laboratory, Berkeley, California 94720, USA}

\date{\today}

\begin{abstract}

We develop an operational approach to strong-to-weak spontaneous symmetry breaking (sw-SSB) of a non-Abelian symmetry group $G$. We formulate it as the decoding task of distinguishing two strongly symmetric states that differ by a pair of inserted charges, after both are corrupted by strongly symmetric decoherence. For finite $G$, this task is dual to distinguishing logical states of the quantum double $D(G)$ under pure-charge noise. We derive a hierarchy of statistical mechanics models describing optimal decoding from data that include the fusion outcomes measured on successively larger nested regions. As for maximum-likelihood decoders of Abelian codes, the degrees of freedom are local group elements, and the measured syndromes act as quenched disorder obeying a Nishimori-like condition. Strikingly, because local syndromes do not determine how they fuse globally, the strong symmetry is realized only nonlocally, and irreps with trivial central character can hide inside noise-generated irreps. As a result, for local syndromes, the $G\times G$ symmetry spontaneously breaks to $Z(G)\times G_{\rm diag}$ at \emph{any} nonzero error rate. Here $Z(G)$ is the center of $G$, and $G_{\rm diag}$ is the weak symmetry, which cannot be broken. The only transition at nonzero error rate is therefore an ordering of the center, marking the failure to distinguish states carrying irreps with nontrivial central character, and its threshold cannot decrease as the hierarchy is ascended. For groups like $S_3$ and ${\rm SU}(2)$ this conclusion holds at any level of the hierarchy that does not scale with system size. We confirm these predictions with tensor-network calculations for $S_3$ and ${\rm SU}(2)$. At the first level, the $S_3$ model, whose center is trivial, shows no transition, and the fidelity correlations of all irreps decay to positive constants. The ${\rm SU}(2)$ model exhibits a $\mathbb{Z}_2$ transition in the Nishimori-Ising universality class, where the fidelity correlations of spin-$1/2$ irreps acquire long-range order. The threshold rises from $p_c \approx 0.174$ at the first level to $p_c \approx 0.202$ at the second level of the hierarchy. Our results provide a systematic framework for studying non-Abelian sw-SSB and for constructing decoders for non-Abelian mixed states.

\end{abstract}

\maketitle

\section{Introduction}\label{sec:intro}

Advances in quantum computing platforms have brought into focus collective phenomena associated with encoding and processing of quantum information. Measurement-induced phase transitions~\cite{li2018zeno, skinner2019mipt, li2019mipt, bao2020MIPT}, failure of quantum error correction~\cite{dennisTQM2002, fanDiagnostics2024, terhal2015QECreview}, and strong-to-weak spontaneous symmetry breaking (sw-SSB)~\cite{lee2023swssb, Lessa_swssb, sala2024swssb} can all be understood as phase transitions related to the ability to learn or recover information from the system. 

A case in point is sw-SSB, signaled by loss of distinguishability between two mixed states~\cite{Lessa_swssb} subject to decoherence. The states to be distinguished, $\rho$ and $\sigma$, both transform trivially under a global symmetry $G$, but differ by insertion of a pair of distant charges (irreps of $G$) into $\sigma$. How well they can be distinguished is measured by the Uhlmann fidelity $F(\rho,\sigma)=\big\|\sqrt{\rho}\sqrt{\sigma}\big\|_1$, which can be formulated as a special kind of two point correlation function associated with the charge insertions. 
Onset of non decaying fidelity correlations above a critical decoherence strength indicates that the two states can no longer be sharply distinguished, signaling SW-SSB.

While the Uhlmann fidelity is a complex quantum quantity, previous work exemplified sw-SSB in special cases that allow a significant simplification~\cite{Lessa_swssb, sala2024swssb, temkin2025charge, vijay2025info, hauser2026swssb, weinstein2025efficient, Whightman_corr}. Specifically, these works considered Abelian symmetries, where the strongly symmetric decoherence injects neighboring charge pairs (dipoles) that corrupt the original state, while leaving the density matrix diagonal in the local charge basis. Hence the decoherence process is described by classical stochastic dynamics and the Uhlmann fidelity collapses to a much simpler classical fidelity between probability distributions. 
Operationally this implies an optimal decoding scheme for distinguishing the two states: measure the symmetry charge on all sites to fix an eigenstate of $\rho$, then compute the probabilities of each initial state conditioned on the measurement outcome. This Bayesian inference problem can be formulated as a statistical mechanics model with a phase transition indicating failure of the decoder. 
For example, when $G=\mathbb{Z}_2$ this model is a random-bond Ising model (RBIM)~\cite{lee2023swssb, Lessa_swssb, dennisTQM2002, weinstein2025efficient, Whightman_corr} and for ${\rm U}(1)$ symmetry it is a special kind of a random XY model~\cite{temkin2025charge}. It is worth noting that these decoding schemes are equivalent through a Krammers-Wannier duality to decoding of Abelian quantum error correction (QEC) codes~\cite{lee2023swssb, fanDiagnostics2024, temkin2025charge, vijay2025info, chen2024separability, Tantivasadakarn_2023}. 
Crucially, the classical nature of the problem, encapsulated in the fact that a measurement projects onto eigenstates of the density matrix, ensures that the decoding transition reflects the intrinsic sw-SSB with Abelian symmetries and the intrinsic error threshold in Abelian QEC. 

In this paper we derive effective models that elucidate the nature of non-Abelian sw-SSB, where the aforementioned classical description breaks down. At the heart of the problem is the fact that quantum channels with strong non-Abelian symmetry necessarily produce entanglement. Specifically, they generically create pairs of multi-dimensional irreps from the vacuum and such pairs must be entangled in order to fuse to the trivial representation. For example, a channel with strong ${\rm SU}(2)$ symmetry injecting pairs of neighboring spin-$\tfrac{1}{2}$ must create them in singlet Bell states. 
The density matrix produced by such a channel is not diagonal in any local basis, thus performing local measurements does not project onto its eigenstates.  

The presence of multi-dimensional irreps also complicates the operational definition of sw-SSB in terms of the ability to recover the uncorrupted state from measurement data. In the Abelian recovery scheme, we can measure the charge on each site and determine with certainty which pairs of charges add to zero and can therefore be matched and annihilated into the vacuum. In the non-Abelian case, by contrast, measuring the local irreps does not determine the fusion of pairs. For example, two spin-$\tfrac{1}{2}$ defects could fuse into spin $0$ or $1$. 
The same problem arises in decoding of non-Abelian quantum error correction codes when attempting to match non-Abelian anyons created by the errors~\cite{Wootton_14, Dauphinais_2017, Schotte_22}. 
In either case it is clear that measuring only the local irrep or anyon type is not optimal.
One can add information to improve the decoder by also measuring the irrep associated with all disjoint pairs of sites to determine their fusion channel, then measure all pairs of pairs and so on, with increasingly non-local measurements fixing higher levels of the fusion tree. Thus, the optimal measurement protocol that would reproduce the intrinsic sw-SSB transition, defined through the Uhlmann fidelity $F(\rho,\sigma)$, is necessarily non-local. The sub-optimality of measuring local syndromes in a non-Abelian code was recently pointed out for the $D_4$ quantum double model~\cite{pabloD4,Jing_2026}. While the intrinsic threshold under a specially chosen error model appears to be $p_c=1/2$~\cite{Hong26}, the optimal scheme based on strictly local measurements has a threshold of $p_c\approx0.21$~\cite{Jing_2026}. 

To elucidate how the novel features of the non-Abelian problem affect decoding of non-Abelian codes and sw-SSB, we derive statistical mechanics models for the optimal decoders based on a family of natural measurement protocols.
We then focus on the special problem of sw-SSB of strong $S_3$ and ${\rm SU}(2)$ symmetries. The former corresponds to the smallest non-Abelian group, which will already show the main differences with respect to Abelian groups. The latter presents a particularly intriguing puzzle. Recall that in the case of Abelian symmetries sw-SSB can be described by disordered statistical mechanics models with the same symmetry and in the same number of spatial dimensions. 
If this pattern generalizes to non-Abelian symmetries, then we might naively expect that the Hohenberg--Mermin--Wagner theorem~\cite{hohenberg1967LRO, mermin1966absence} prevents sw-SSB of a strong ${\rm SU}(2)$ symmetry in two spatial dimensions even for sub-optimal measurement protocols, thereby extending the decodable phase to arbitrary decoherence strength. Our approach suggests an interesting resolution to this puzzle.

Before proceeding we summarize previous work that also touches on the topic of non-Abelian sw-SSB. To the best of our knowledge previous related works have either focused on Rényi quantities~\cite{sala2024swssb,ziereis2025}, with unclar operational meaning, or characterized the stationary states at maximal decoherence ~\cite{pabloD4,li2025entanglement,Sahu_24}. More recently, Ref.~\cite{ma2026} studied the monitored dynamics with SU$(2)$ symmetry in one spatial dimension focusing on learning or ``sharpening" of the total spin. Their analysis applies to the time evolution of pure states without decoherence. 
Because the spin learning transition does not occur inside the volume law phase and exhibits a dynamical exponent different from one, it does not seem to be directly related to mixed state transitions like sw-SSB in two dimensions, unlike $U(1)$ charge sharpening \cite{agrawal2022entanglement,barratt2022field}. 

\begin{figure*}[t]
    \centering
    \begin{tikzpicture}[>=stealth,scale=0.9,
    every node/.style={transform shape},
    level1/.style={circle,draw,fill=green!20,inner sep=1pt,font=\scriptsize},
    level2/.style={circle,draw,fill=yellow!35,inner sep=1pt,font=\scriptsize},
    leaf/.style={circle,draw,fill=blue!15,inner sep=1pt,font=\scriptsize},
    root/.style={circle,draw,fill=red!20,inner sep=2pt,font=\small\bfseries}
]

\def\panelSep{2.0}    
\def\panelShift{0.0}  
\def\panelWidth{8.5}  
    \begin{scope}[shift={(\panelShift,0)}]
    \node[anchor=south,font=\bfseries] at (0,4.3) {(a) Abelian ($\mathbb{Z}_2$, U(1))};
    \foreach \i/\lab in {1/$\Gamma_1$,2/$\Gamma_2$,3/$\Gamma_3$,4/$\Gamma_4$,5/$\Gamma_5$,6/$\Gamma_6$,7/$\Gamma_7$,8/$\Gamma_8$}{
        \pgfmathsetmacro{\xpos}{(\i-4.5)*0.95}
        \node[leaf] (AL\i) at (\xpos,0) {\lab};
    }
    \node[rectangle,draw,fill=orange!20,rounded corners,inner sep=4pt,font=\small] (AS) at (0,2.6) {$\sum_v \Gamma_v = 0$};
    \foreach \i in {1,...,8}{
        \draw[gray!70] (AL\i) -- (AS);
    }
    
    \end{scope}
    \begin{scope}[shift={(\panelShift+\panelWidth+\panelSep,0)}]
    \node[anchor=south,font=\bfseries] at (0,4.3) {(b) Non-Abelian ($S_3$, SU(2))};
    \foreach \i/\lab in {1/$\Gamma_1$,2/$\Gamma_2$,3/$\Gamma_3$,4/$\Gamma_4$,5/$\Gamma_5$,6/$\Gamma_6$,7/$\Gamma_7$,8/$\Gamma_8$}{
        \pgfmathsetmacro{\xpos}{(\i-4.5)*1.0}
        \node[leaf] (L\i) at (\xpos,0) {\lab};
    }
    \foreach \i/\xp in {1/-3,2/-1,3/1,4/3}{
        \node[level1] (M\i) at (\xp,1.35) {$\Gamma^{(2)}_{\i}$};
    }
    \draw[thick] (L1) -- (M1);
    \draw[thick] (L2) -- (M1);
    \draw[thick] (L3) -- (M2);
    \draw[thick] (L4) -- (M2);
    \draw[thick] (L5) -- (M3);
    \draw[thick] (L6) -- (M3);
    \draw[thick] (L7) -- (M4);
    \draw[thick] (L8) -- (M4);
    \node[level2] (N1) at (-2,2.7) {$\Gamma^{(3)}_{1}$};
    \node[level2] (N2) at ( 2,2.7) {$\Gamma^{(3)}_{2}$};
    \draw[thick] (M1) -- (N1);
    \draw[thick] (M2) -- (N1);
    \draw[thick] (M3) -- (N2);
    \draw[thick] (M4) -- (N2);
    \node[root] (R) at (0,4.0) {$0$};
    \draw[thick] (N1) -- (R);
    \draw[thick] (N2) -- (R);
    \draw[dashed,rounded corners=6pt,gray,thick]
        (-4.0,-0.35) rectangle (4.0,0.35);
    \node[font=\scriptsize,text=gray,anchor=west] at (4.1,0.0) {level 1};

    \draw[dashed,rounded corners=6pt,gray,thick]
        (-3.6,0.9) rectangle (3.6,1.8);
    \node[font=\scriptsize,text=gray,anchor=west] at (3.7,1.35) {level 2};

    \draw[dashed,rounded corners=6pt,gray,thick]
        (-2.6,2.2) rectangle (2.6,3.2);
    \node[font=\scriptsize,text=gray,anchor=west] at (2.7,2.7) {level 3};

    \draw[dashed,rounded corners=6pt,gray,thick]
        (-0.55,3.6) rectangle (0.55,4.4);
    \node[font=\scriptsize,text=gray,anchor=west] at (0.65,4.0) {global vacuum};
    \draw[->,thick]
    (-4.25,0) -- (-4.25,4.2)
    node[midway,left] {$\log L$};
    \end{scope}
    \end{tikzpicture}

\caption{
    \textbf{Fusion tree structure.}
    \textbf{(a)}~For Abelian groups, every irrep is one-dimensional and fusion is deterministic: the constraint that all irreps fuse to the vacuum reduces to a single additive condition (parity for $\mathbb{Z}_2$, charge conservation $\sum_v k_v = 0$ for $\mathrm{U}(1)$). All on-site irreps connect directly to a single global constraint---there are no intermediate degrees of freedom.
    \textbf{(b)}~For non-Abelian groups such as $S_3$ or SU(2), the singlet projection~\eqref{eq:Haar_proj} has a non-trivial matrix structure.
    The syndrome probability~\eqref{eq:fusion_tree_expansion} involves a sum over all consistent binary fusion trees: on-site irreps $\Gamma_v$ (blue, leaves) are sequentially fused into intermediate irreps $\Gamma^{(2)}$ (green, level~1), then $\Gamma^{(3)}$ (yellow, level~2), terminating at the trivial irrep $0$ (red) at the root.
    Each internal edge carries an irrep that is \emph{not} uniquely determined by the on-site data---this ambiguity is the hallmark of non-Abelian fusion and forms the basis of the measurement hierarchy (Sec.~\ref{sec:hierarchy}). 
    }
    \label{fig:fusion_tree}
\end{figure*}
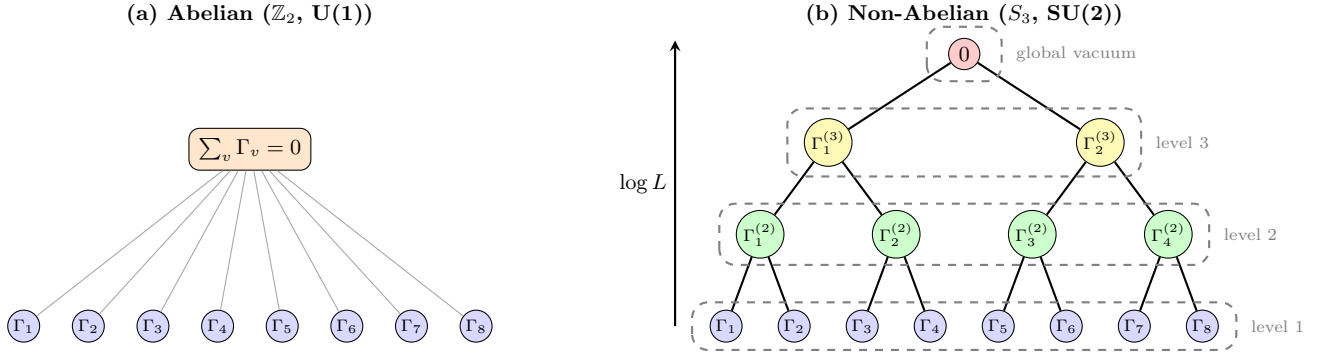

\vspace{15pt}
\subsection{Summary of results}

We consider a model of strongly symmetric decoherence consisting of commuting local quantum channels that with probability $p_\Gamma$ insert pairs of irreps of a group $G$ that fuse to the identity $\Gamma\otimes\bar{\Gamma}$. Through a Krammers-Wannier duality~\cite{Tantivasadakarn_2023, yoshida2017gapped, lootens2023dualities}, the model also applies to decoherence creating pairs of pure charge anyons in a $D(G)$ topological code.
The state corrupted by the noise is  diagnosed by a set of commuting measurements that respect the strong symmetry, which may include measurements of the on site irreps $\Gamma$ as well as of the fusion of subsets of those irreps. 
We construct the optimal decoders for recovering the uncorrupted state from the measured data. The probability $p_\alpha(\{M_j\})$ for a particular recovery $\alpha$ conditioned on the measurement outcomes $\{M_j\}$ (irreps obtained on site or as fusion outcomes at higher levels) is formulated as partition-functions of statistical mechanics models, where the local site variables are group elements $g$. This formulation leads to several new insights.

The first insight concerns decoding from data of local syndrome measurements. 
The partition function in this case takes the form
\begin{equation}
    Z[\{\Gamma_v\}]= \int\!\mathcal{D}g\, W({\bf{g}})\prod_v d_{\Gamma_v}\chi_{\Gamma_v}(g_v),
    \label{eq:Z}
\end{equation}
where $\Gamma_v$ are the irreps detected on lattice sites $v$. The weight $W(\bf{g})$ originates from the noise channels. It is a function only of characters $\chi_{\Gamma_{ij}}(g_i g_j^{-1})$ of group elements on near neighbor sites, where $\Gamma_{ij}$ labels the irrep inserted by the noise. Crucially each term in the weight $W(g)$ is manifestly invariant under multiplication by a global group element $h$, thus encoding the strong symmetry of the underlying quantum problem.  
On the other hand, the syndrome weight in the partition function $S({\bf g}|\{\Gamma_v\})=\prod_v \chi_{\Gamma_v}(g_v)$ is in general not invariant under multiplication by a group element. This leads to 
a crucial difference between the Abelian and non-Abelian cases. For Abelian symmetries, any observed syndrome configuration has a definite total charge (and due to the symmetry of the quantum channel this is trivial). Therefore the syndrome weight $S({\bf g}|\{\Gamma_v\})$ for an observed configuration is manifestly symmetric. For non-Abelian symmetries, an observed set of syndromes $\{\Gamma_v\}$ can fuse into several different global irreps, implying that the weight $S({\bf g}|\{\Gamma_v\})$ is not symmetric. 
More precisely the syndrome weight breaks the strong symmetry $G\times G$ down to $G_{\text{diag}}\times Z(G)$, where $Z(G)$ is the center of $G$ and $G_{\text{diag}}$ is the weak symmetry that cannot be broken by the channel. 
Although the symmetry is restored globally due to integration over the symmetric weight $W({\bf g})$, the non-Abelian syndrome weights act as local symmetry breaking fields that induce long range fidelity correlations in irreps $\Upsilon$ of $G$ on which $Z(G)$ acts trivially. Hence a decoder based on local syndrome measurements has a vanishing threshold for non-Abelian sw-SSB. However such a decoder can still see a transition involving SSB of the center $Z(G)$. The underlying physical mechanism is that, for any non-Abelian group, any irrep of dimension greater than one can hide another one. Hence, for any finite error rate, errors can hide the $\Upsilon$ irreps, which necessarily have trivial central-characters. 

The general principles are nicely illustrated by the two examples of $S_3$ and ${\rm SU}(2)$ strongly symmetric quantum channels decoded from local syndromes. Because $S_3$ has a trivial center, we expect that the fidelity correlations for insertions of all irreps approach a positive constant at long distances. Moreover, for simple decoherence channels we obtain lower bounds for the long distance saturation values of these correlations, based on the probabilities that insertion of each irrep is locally masked by the noise. ${\rm SU}(2)$ symmetry provides a richer example because its center is $\mathbb{Z}_2$. Thus, while the fidelity correlations of integer spin irreps that have a trivial central-character can exhibit long range order for arbitrarily weak noise, the spin-$1/2$ fidelity correlations are expected to become long ranged at a $\mathbb{Z}_2$ ordering transition with non-vanishing threshold $p_c$. The question of the Hohenberg-Mermin-Wagner theorem preventing a transition does not arise here because the local syndromes effectively break the continuous strong symmetry down to the discrete center.

The analysis of decoding from local syndromes raises the question whether we can increase the thresholds or even approach the true sw-SSB transitions by improving the measurement protocol. A natural way to do this is by supplementing the local syndromes with measurements of the fusion of pairs of syndromes. More generally, this suggests a hierarchical set of (commuting) measurements that give the fusion outcomes on ascending levels of a fusion tree, as illustrated in Fig. ~\ref{fig:fusion_tree}. In section \ref{sec:hierarchy} we generalize the partition function \eqref{eq:Z} to ones representing the optimal decoders based on measurements that reach up to level $N$ of the fusion tree [see Eq. \eqref{eq:level_general}]. These partition functions are still of the form of disordered statistical mechanics models, with disorder distributions that become increasingly non-local with the level $N$. At a formal level the data-processing inequality ensures that using measurements from higher levels can only decrease the fidelity correlations. Consequently, the observed decoding thresholds can only increase with $N$. 

We demonstrate and verify the general framework numerically by 
formulating the partition functions describing the decoding using measurements of $N$ levels of the fusion tree as two dimensional tensor networks. Contracting the tensor networks using the boundary matrix product state (MPS) method provides a scalable approach to sample syndrome configurations and compute all fidelity correlations. We apply this scheme to the examples of $S_3$ and ${\rm SU}(2)$ symmetries and the results confirm the predictions discussed above. In particular we observe a phase transition marking the establishment of long-range fidelity correlations in spin-1/2 irreps with critical properties that match those exhibited by the RBIM on the Nishimori line~\cite{Nishimori_81}. Furthermore we demonstrate that the threshold for this transition increases from $p_c^{(1)}\approx 0.174$ from level 1 (local syndromes) to $p_c \approx 0.202$ upon ascending  to level 2 of the fusion tree.

While ascending to a finite level of the fusion tree can improve thresholds, the generalized syndrome weight still breaks the strong symmetry down to $G\times Z(G)$ as in the case of local syndromes. To observe a sw-SSB of the full symmetry (even if not the intrinsic transition) we would need a measurement protocol that fully restores the $G\times G$ symmetry. That is, the syndrome weight should be manifestly invariant under multiplication by a global group element. This would be achieved if we could group the syndromes into clusters that each fuse to the identity irrep (or a one dimensional one). Previous works have shown that heuristic clustering schemes can succeed below a small $p_{\text{cl}}>0$ \cite{ducloscianci2010rgdecoder,Wootton_14}. 
We conjecture that the threshold may be increased by taking a maximum likelihood approach, where the optimal clustering at each level is inferred using the partition function of the lower levels. 

The remainder of the paper is organized as follows. Section~\ref{sec:review} reviews the sw-SSB transition, emphasizing an equivalent operational formulation in terms of distinguishability protocols. Section~\ref{sec:model} then sets up the group-theoretical framework, introducing the decoherence model and providing exact expressions for the resulting decohered density matrices. Section~\ref{sec:syndrome} then derives the probability of syndrome configurations $P(\{\Gamma_v\})$, and its formulation in terms of sum over consistent fusion trees. It then introduces the hierarchical measurement construction, and provides the explicit connection with the study of the stability of non-Abelian topological quantum memories. Section~\ref{sec:Examples} then discusses two examples, $S_3$ and SU$(2)$ sw-SSB. Section~\ref{sec:decoders} provides the necessary numerical methods to study the resulting statistical mechanics models using tensor network techniques, and provides numerical results for the two cases mentioned above. We conclude in Sec.~\ref{sec:discussion} with a summary of our conclusions and a list of open questions. Additional technical derivations and details are deffered to the appendices.
\tableofcontents

\section{Review of strong-to-weak SSB} \label{sec:review}

In this section, we review the notion of strong-to-weak spontaneous symmetry breaking (sw-SSB) from an operational perspective, formulating it as the loss of ability to distinguish two symmetric states in the presence of decoherence.

Let us consider a symmetry group $G$ whose action is generated by a unitary operator $U_g$. Unlike pure states, mixed states can possess two different types of symmetries~\cite{buca2012lindblad,albert2014lindblad}. A mixed state $\rho$ is \emph{strongly symmetric} if $U_g \rho = \rho U_g= e^{i \varphi_g}\rho$ for all group elements $g\in G$. In this case, the state is a mixture of different states within the same charge sector (or a symmetric pure state). On the other hand, a mixed state $\rho$ is \emph{weakly symmetric} if $U_g\rho U_g^\dagger=\rho$, while $U_g\rho $ is not proportional to $\rho$. Clearly, a strong symmetry implies the corresponding weak symmetry but the converse does not hold. An intuitive example is provided by thermodynamic ensembles. For a system conserving the total particle number $N$, the canonical ensemble $\rho_\beta = e^{-\beta H}/Z$ is strongly symmetric, while the grand canonical ensemble $\rho_{\beta,\mu} = e^{-\beta (H-\mu N)}/Z$, in which the particle number fluctuates, is only weakly symmetric. 

The standard notion of spontaneous symmetry breaking (SSB) corresponds to breaking the weak symmetry. It can be diagnosed by the emergence of long-range order in correlation functions of an order parameter $O_x$, e.g., $\lim_{|x-y|\to \infty}{\rm tr}(\rho O_x O_y)\ne 0$. Strongly symmetric mixed states admit another kind of phase transition when subject to strongly-symmetric open-system evolution. Specifically we will consider completely positive trace preserving (CPTP) maps
\begin{equation}
    \mathcal{E}(\rho)=\sum_\alpha K_\alpha \rho K_\alpha^\dagger,
\end{equation}
where $K_\alpha$ are Kraus operators satisfying the trace-preserving condition $\sum_\alpha K^\dagger_\alpha K_\alpha=\mathbf{1}$.
This evolution is strongly symmetric if for each Kraus operator $U_gK_\alpha U_g^\dagger=K_\alpha$ for all elements $g\in G$. In many physical problems the many-body open system dynamics is obtained from composing quantum channels that act locally. For example take channels $\mathcal{E}_{ij}$ acting on bonds $\langle i,j\rangle$ of the lattice and construct $\E(\rho)= \circ_{\langle i,j\rangle }\mathcal{E}_{ij}(\rho)$. 
Under this time evolution the strong symmetry can be spontaneously broken down to the weak symmetry. The transition manifests in information theoretic quantities that measure the ability to extract symmetry related information; it 
cannot be captured by conventional two-point correlation functions linear in the density matrix $\rho$~\cite{Lessa_swssb, lee2023swssb, Makinde_23,  Grover_24, sala2024swssb, weinstein2025efficient, Whightman_corr}.

A concrete signature of sw-SSB, proposed in Ref. \cite{Lessa_swssb}, measures the distinguishability between two states using the fidelity correlator. Consider the two  symmetric initial pure states: a reference state $\rho_0$ and another one, $\rho_{0,xy}$ that differs by the insertion of symmetry charges at locations $x$ and $y$. Now subject these two states to the same strongly symmetric local quantum channel $\rho_\alpha=\mathcal{E}(\rho_{0,\alpha})$.
The ability to distinguish between the two mixed states is quantified by the 
 Uhlmann fidelity~\cite{jozsa1994fidelity, uhlmann1975transition}
\begin{equation}\label{eq:fidelity}
    F^Q(\rho,\rho_{xy}) =\|\sqrt{\rho}\sqrt{\rho_{xy}}\|_1=\mathrm{tr}\sqrt{\sqrt{\rho}\;\rho_{xy}\sqrt{\rho}},
\end{equation}
which in this context defines the fidelity correlation as a function of $|x-y|$. Because the local decoherence can only inject near-neighbor pairs, for weak enough decoherence one expects the fidelity correlation to
vanish at infinite separation, reflecting the fact that a state with isolated charges remains perfectly distinguishable from one without. However upon increasing the decoherence strength the state undergoes a phase transition beyond which the fidelity correlations are non-decaying at long distances. This marks the sw-SSB. Intuitively, the transition occurs when the decoherence is strong enough and the additional pair of charges becomes indistinguishable from the charges created by the noise itself. 

Observing the transition in practice requires a decoder that attempts to discern the states using measured data. To determine which state is more likely, the decoder needs to compute the conditional probabilities $p_{\alpha}(r)\equiv p(r|\alpha)=\text{tr}(\rho_\alpha M_r)$, where $\{M_r\}$ are positive operator-valued measurements (POVM) with possible measurement outcomes labeled by $r$ and $\alpha$ labels the states to be discerned.
The ability to distinguish the two states from the measurement data is then quantified by the classical fidelity~\cite{FuchsCaves_95}
\begin{align} \label{eq:Fidelity_POVM}
     F^C_M(\rho,\rho_{xy}) &=\sum_r \sqrt{\tr(\rho M_r)\tr(\rho_{xy} M_r)}.
\end{align}
Because the classical fidelity associated with a specific measurement protocol is an upper bound on the quantum fidelity a transition observed using such a decoder gives only a lower bound on the critical decoherence strength $p_c$ for the intrinsic sw-SSB~\footnote{The Uhlmann fidelity is attained by measuring in the eigenbasis of the Fuchs-Caves operator $O_{\rm FC}=\rho^{-1/2}\sqrt{\sqrt{\rho}\,\rho_{xy}\sqrt{\rho}}\,\rho^{-1/2}$~\cite{FuchsCaves_95}, which, however, is even harder to find than the quantum fidelity.}. 

Most previous work on sw-SSB considered Abelian groups, where the problem simplifies considerably. In this case, the evolution of the density matrix under decoherence is seen to be essentially classical in the charge basis thus a decoder performing Bayesian inference from charge measurements captures the intrinsic transition exactly. 

As an illustrative example consider a $\mathbb{Z}_2$ sw-SSB. A simple model of this problem has qubits on vertices of a two dimensional lattice. 
The global $\mathbb{Z}_2$ symmetry is generated by
$U=\prod_i \hat{X}_i$. The symmetric initial states we want to discern are: (i) $\rho_0 =\bigotimes_i |+\rangle_i\langle+|$, a product of even parity states on every site. (ii) $\rho_{0,xy}=Z_x Z_y\rho_0 Z_x Z_y$, where a pair of $\mathbb{Z}_2$ charges (odd parity states) have been inserted at locations $x$ and $y$.
Now these initial states are subject to a strongly symmetric channel creating nearest-neighbor pairs of $\mathbb{Z}_2$ charges with a probability $p$ per bond:
\begin{equation}
\label{eq:Z2_example_channel}
    \mathcal{E}_{ij}(\rho)
    =
    (1-p)\rho
    +
    p\,Z_iZ_j\rho Z_iZ_j,
\end{equation}
Both Kraus operators, $\sqrt{1-p}\,\mathbf{1}$ and $\sqrt{p}\,Z_iZ_j$, commute with the symmetry generator $U$ and hence the dynamics is strongly symmetric. Crucially, the resulting
decohered density matrices are  diagonal in the local charge basis ($X$-basis)
\begin{equation}
\rho_\alpha=\sum_{\bm{X}}p_{\alpha}(\bm{X})|\bm{X}\rangle \langle \bm{X}|,
\end{equation}
where $\alpha$ labels the states to be distinguished, $\bm{X}=(X_1,\ldots,X_N)$ and $X_i=\pm 1$ the eigenvalue of $\hat{X}_i$ ($\mathbb{Z}_2$ charge) on a site. This implies that the problem is essentially classical. In particular,
the quantum fidelity between the two states is equal to the classical fidelity based on measurements of the local $\mathbb{Z}_2$ charge:
\begin{equation}    F^Q(\rho,\rho_{xy})=\sum_{\bm{X}}p(\bm{X})\sqrt{\frac{p_{xy}(\bm{X})}{p(\bm{X})}}
\label{eq:NishimoriRBIM}
\end{equation}
In this case the probabilities $p_\alpha(\bm{X})$ are given by the partition function of the RBIM on the dual lattice, with the Ising fluxes fixed to the locations of the negative values of $X_i$~\cite{Lessa_swssb}. 
Hence, the quantum fidelity is captured exactly by a disorder averaged correlation function \eqref{eq:NishimoriRBIM} in the RBIM on the Nishimori line~\cite{Nishimori_81}.

Another example of Abelian sw-SSB studied previously considers a quantum channel with strong ${\rm U}(1)$ symmetry \cite{ temkin2025charge, hauser2026swssb}. To construct commuting U$(1)$ strongly symmetric Kraus operators, one can place a quantum rotor on every vertex, with $\hat{\theta}_i$ the phase operator conjugate to the charge $\hat{n}_i$ on site $i$, i.e., $[\hat{\theta}_i,\hat{n}_j]=i\delta_{ij}$. Here, the global ${\rm U}(1)$ symmetry is generated by $\hat N=\sum_i \hat{n}_i$. The on-site Hilbert space can be spanned in the eigenbasis of $\hat{\theta}$, $\{\ket{\theta}\,|\, \theta \in \textrm{U}(1)\}$; or in the eigenbasis of $\hat{n}$, i.e., $\{\ket{n}\,|\, n\in \mathbb{Z}\}$. The two initial states to be discerned in this case may be taken to be (i) $\rho_0 =\bigotimes_i |0\rangle_i\langle0|$, a product of the zero charge state on every site. (ii) $\rho_{0,xy}=e^{i(\hat{\theta}_x-\hat{\theta}_y)}\rho_0 e^{-i(\hat{\theta}_x-\hat{\theta}_y)}$ corresponding to inserting a particle and a hole at distant sites $x$ and $y$ respectively. A minimal decoherence model involves a local quantum channel inserting particle hole pairs with probability $p$ independently on all nearest neighbors
\begin{equation} \label{eq:U1_example_channel}
    \mathcal{E}_{ij}(\rho)=(1-p)\rho+\frac{p}{2} \left(e^{i(\hat{\theta}_i-\hat{\theta}_j)}\rho e^{-i(\hat{\theta}_i-\hat{\theta}_j)}+\text{h.c.}\right)
\end{equation}
Again we see that the density matrix remains diagonal in the local charge basis. 
In this case the probabilities 
$p_\alpha(\bm{n})$ for 
measuring a charge configuration $\bm{n}$ on the state $\alpha$ are given by the partition functions of a random XY model with a special kind of disorder \cite{temkin2025charge}. Correspondingly, the quantum fidelity is a disorder averaged correlation function in this model, obeying a Nishimori condition.
Interestingly, there is no true sw-SSB in this case due to the continuous ${\rm U}(1)$ symmetry. Instead, the fidelity correlations in the non-decodable phase decay as a power-law and the decoding transition is akin to a Kosterlitz-Thouless transition (with some subtle unique features). We remark that if  instead of a rotor model we used a model with finite dimensional (``qudit") local Hilbert space, then the simplest decoherence channel would have Kraus operators of the form $S^+_iS^-_j$ that do not commute on adjacent bonds. This would complicate the decoherence dynamics, but we do not expect it to change the qualitative features or the universality of the transition. 

The crucial point that generalizes to  sw-SSB with any Abelian symmetry 
is that it is always possible to formulate a minimal model, in which the density matrix remains diagonal and the quantum fidelity maps exactly to the classical fidelity between probability distributions of charge configurations. In this sense the problem of sw-SSB with Abelian symmetries is classical.

The reduction to a classical problem does not hold in the case of non-Abelian symmetries. One novel feature of a quantum channel with strong non-Abelian symmetry is that it can create pairs of multi-dimensional irreps. Such a process necessarily generates entanglement because, to respect the symmetry, the two irreps must be created as a singlet Bell-pair. Hence the density matrix of the decohered state is non-separable~\cite{li2025entanglement, sahu2026entanglement}; it is not equal to a classical probability distribution in any local basis. 
Furthermore, measuring the local symmetry charges (i.e. the irreps occupying lattice sites) does not provide the optimal information to correct errors, and it does not remove coherences because these measurements cannot resolve the internal state of multi-dimensional irreps.  
We will derive the statistical mechanics model describing the decoding based on local syndrome measurements. We'll then generalize to include a hierarchy of increasingly non-local measurement protocols such that the associated  classical fidelity is reduced toward the intrinsic quantum fidelity.   
Before proceeding it is worth noting that a previous work by one of us has provided an exact result for the quantum fidelity in a model with $D_4$ symmetry, but for a very specific choice of decoherence~\cite{pabloD4}.

\section{Strongly-symmetric decoherence model}\label{sec:model}

The goal of this section is to provide a common formalism for analysing sw-SSB under a general group $G$. This allows us to compare how the properties of $G$ affect sw-SSB within a common family of strongly symmetric channels. We consider finite and compact Lie groups, including both Abelian and non-Abelian cases. We treat these groups on the same footing using the regular representation as the basic building block. We then construct a simple family of strongly symmetric quantum channels composed of commuting local channels. In particular, the resulting channels reduce to those in Eqs. \eqref{eq:Z2_example_channel} and \eqref{eq:U1_example_channel} for \(G=\mathbb{Z}_2\) and \(\mathrm{U}(1)\), respectively.

\subsection{Group-element Hilbert space}

Let $G$ be the symmetry group of the system of interest; here we focus on either a finite group of order $|G|$ or a compact Lie group. For finite groups, the local Hilbert space is spanned by group elements $\mathbb{C}[G]= \mathrm{span}\{ \ket{g} : g \in G\}$ with inner product $\langle g,h\rangle = \delta_{g,h}$. For compact Lie groups, the local Hilbert space is the space square-integrable functions on $G$ equipped with the Haar measure: $\LG=\{f:G\to \mathbb{C}\,:\, \int_G d\mu(g) \,|f(g)|^2<\infty\}$. The Haar measure is normalized such that $|G|=\int_G d\mu(g) = 1$, and satisfies $d\mu(hg) = d\mu(gh) = d\mu(g)$, with the inner product $\langle g,h\rangle=\delta(g-h)$.

The group $G$ naturally acts on this Hilbert space by permuting group elements. An element $h$ of a non-Abelian group $G$ can multiply the state from the \emph{left} $L_{h}$ or \emph{right} $R_{h}$:
\begin{equation}\label{eq:left_reg}
    L_{h} \ket{g} = \ket{hg}\,,\qquad R_{h} \ket{g} = \ket{gh^{-1}}\,,
\end{equation}
defining the left and right regular representations respectively~\footnote{Both representations are unitary: $L_{h}^{-1}=L_{h}^\dagger = L_{h^{-1}}$ and $R_{h}^{-1}=R_{h}^\dagger = R_{h^{-1}}$, and $L_{h_1}L_{h_2} = L_{h_1 h_2}$ and $R_{h_1}R_{h_2} = R_{h_1 h_2}$.}. Crucially, left and right actions commute, $[L_{h}, R_{g}] = 0$ for all $h,g\in G$, reflecting the fact that they correspond to independent ``lab-frame'' and ``body-frame'' rotations of a rigid body~\cite{landau2013quantum}.

Consider a $d_\Gamma$-dimensional unitary irreducible representation (irrep) $D^\Gamma$ of the group $G$~\footnote{Notice that when $G$ is a finite group, all representations are equivalent to unitary representations.}. The value of the on-site irrep is provided by measuring the corresponding central idempotents, which are defined as
\begin{equation}\label{eq:projectors}
    P_\Gamma = \frac{d_\Gamma}{|G|}\sum_{g\in G} \chi_\Gamma(g)\, L_{g}\,,\quad P_\Gamma = d_\Gamma \int_G d\mu(g)\, \chi_\Gamma(g)\, L_{g}\,,
\end{equation}
for finite and continuous groups respectively. $\chi_\Gamma(g) = \mathrm{tr}\, D^\Gamma(g)$ are the characters of the group.
These are orthogonal projectors, $P_\Gamma P_{\Gamma'} = \delta_{\Gamma\Gamma'} P_\Gamma$, that resolve to the identity $\sum_\Gamma P_\Gamma = \mathbbm{1}$ (Appendix~\ref{app:transform}).
Their commutativity follows from the rearrangement theorem for finite groups and the fact that characters are class functions, $\chi_\Gamma(g) = \chi_\Gamma(hgh^{-1})$.
Importantly, $P_\Gamma$ commutes with all left symmetry transformations, $[P_\Gamma, L_{g}] = 0$, (and trivially with the right ones $R_g$) making it a legitimate observable in a $G$-symmetric theory.
Notice that if $P_{\Gamma,v}\ket{\psi} = \ket{\psi}$, then site $v$ carries a unique irrep $\Gamma$---this is the non-Abelian generalization of a state with definite local charge.
Throughout this work, when no ambiguity arises, we use the shorthand $\int dg (\cdot)$ to denote the group average, including both the finite and the continuous cases.

An alternative basis can be constructed using the Peter--Weyl theorem~\cite{peter1927PWtheorem}. For each irrep $\Gamma\in\hat G$, the matrix elements $D^\Gamma_{ab}(g)$, with $a,b=1,\dots,d_\Gamma$, form an orthogonal basis of $L_2(G)$ with respect to the Haar measure, and similarly for $\mathbb{C}[G]$. Namely, 
\begin{equation} \label{eq:ONB}
    \int d\mu(g)\, D^\Gamma_{mn}(g)D^{\Gamma'}_{m'n'}(g)=\frac{1}{d_\Gamma}\delta_{\Gamma',\bar{\Gamma}}\delta_{m',m}\delta_{n',n},
\end{equation}
respectively, where $\bar{\Gamma}$ is the dual irrep of $\Gamma$, defined as $D^{\bar{\Gamma}}(g)=D^{{\Gamma}}(g^{-1})^T$. $\bar{\Gamma}$ is the only irrep for which $\Gamma \otimes \bar{\Gamma}$ contains the trivial one. For unitary representations that we consider in this work, the dual coincides with the conjugate representation, i.e., $D_{ab}^{\bar{\Gamma}}(g)=D_{ab}^{{\Gamma,*}}(g)=D^\Gamma_{ba}(g^{-1})$. We therefore define the orthonormal basis
\begin{equation} \label{eq:G_ab}
 \ket{\Gamma,a,b}=\sqrt{d_\Gamma} \int dg\,\,D^\Gamma_{ab}(g)\ket{g},
\end{equation}
with inverse relation
$\ket{g}
=
\sum_{\Gamma,a,b}
\sqrt{\frac{d_\Gamma}{|G|}}\,
D^{\Gamma,*}_{ab}(g)\ket{\Gamma,a,b}$. For Abelian groups all irreps are one-dimensional, so this basis transformation reduces to the standard Fourier transform
$\ket{\Gamma}
=
\frac{1}{\sqrt{|G|}}\sum_{g\in G}\chi_\Gamma(g)\ket{g}$,
and its inverse
$\ket{g}
=
\frac{1}{\sqrt{|G|}}\sum_{\Gamma\in\hat G}
\chi_\Gamma^*(g)\ket{\Gamma}$.

The expression for the projector $P_\Gamma$ in terms of the irrep basis is simply 
\begin{equation} \label{eq:PG_diag}
    P_\Gamma = \sum_{a,b=1}^{d_\Gamma}\ket{\Gamma, a,b}\bra{\Gamma, a,b}.
\end{equation}
Namely, $P_\Gamma |\Gamma',a,b\rangle = \delta_{\Gamma, \Gamma'}|\Gamma',a,b\rangle$, and $L_g=\sum_{\Gamma, a,a',b}D^{\Gamma*}_{aa'}(g)\ket{\Gamma, a, b}\bra{\Gamma, a',b}$. We notice that with the convention used in Eq.~\eqref{eq:G_ab} to define
$\ket{\Gamma,a,b}$, the label $\Gamma$ denotes the matrix coefficient
appearing in the definition of the state. This convention gives the
projector in Eq.~\eqref{eq:PG_diag}, and hence the definition of
$P_\Gamma$ in Eq.~\eqref{eq:projectors}, with $\chi_\Gamma(g)$ rather
than $\chi_\Gamma^*(g)$, as usually defined in the literature (see, e.g., Ref.~\cite{serre1977linear}). The corresponding state transforms under the
left action as $L_g\ket{\Gamma,a,b}=\sum_{c=1}^{d_\Gamma}D^{\bar{\Gamma}}_{ca}(g)\ket{\Gamma,c,b}$, so the
irrep carried by the left action is the dual $\bar{\Gamma}$ (rather than $\Gamma$).

The familiar case when $G={\rm SU}(2)$ provides some physical intuition on the basis defined in Eq.~\eqref{eq:G_ab}. The Wigner $D$-matrices in this case are given by the matrix elements of a rotation operator $\hat{R}(g)$, representing the group element $g$, in the standard angular momentum basis $\ket{J,m}$~\cite{edmondsbook}:
\begin{equation}
        D^J_{mn}(g)=\langle J,m|\hat{R}(g)|J,n\rangle.
\end{equation}
Hence, to fully characterize this rotation one requires two different magnetic indices $m$ and $n$, with $m,n=-J,\dots, +J$. This can be understood as the lab and body frames of a quantum spinning top~\cite{landau2013quantum}, with $g$ the transformation connecting these two frames. 
    
\subsection{Reference vacuum state}

The local Hilbert space introduced in the previous section allows us to define a simple product state carrying the trivial representation $\Gamma=0$ on every site $v$:
$\ket{\Psi_0}=\prod_v\ket{0}_v$, where 
\begin{equation}\label{eq:init_state}
    \ket{0} \equiv \frac{1}{\sqrt{|G|}}\sum_{g\in G}\ket{g}, \quad \ket{0} \equiv \int_G d\mu(g)\, \ket{g}\,
\end{equation}
for finite and compact groups respectively. The reference state translates to a density matrix
\begin{equation}
\rho_0=\ket{\Psi_0}\bra{\Psi_0}=\int\mathcal{D}l\mathcal{D}r\ket{{\bf l}}\bra{{\bf r}},
\end{equation}
where $\mathcal{D}g=\prod_{v} dg_v$ refers to the product of group-averages or Haar measures on each vertex, and ${\bf{g}}=\{g_v\}$ is a configuration of group elements. One can think of this state as the ``vacuum'' from which excitations (non-trivial irreps) are subsequently created by the noise.

We will take this vacuum as one of the two initial states to be distinguished after they are subjected to the decoherence channel. The other initial state has a pair non-trivial irreps that fuse to the identity, created on two distant sites $x$ and $y$. The precise expression for the alternative initial state will be discussed later.

\begin{table*}[t]
    \centering
\begin{tikzpicture}[
  corner/.style ={fill=black!10, text width=2.5cm, align=left},
  hdrA/.style   ={fill=blue!15,  text width=5.2cm, align=center, font=\bfseries},
  hdrN/.style   ={fill=orange!22,text width=7.9cm, align=center, font=\bfseries},
  rlabel/.style ={fill=black!6,  text width=2.5cm, align=left, font=\bfseries},
  cellA/.style  ={fill=blue!4,   text width=5.2cm, align=left},
  cellN/.style  ={fill=orange!5, text width=7.9cm, align=left},
]
\matrix (T) [
  matrix of nodes,
  nodes={anchor=center, outer sep=0, inner xsep=6pt, inner ysep=4pt,
         execute at begin node={\small \thinmuskip=3mu \medmuskip=4mu \thickmuskip=5mu}},
  row sep=2pt, column sep=2pt,
  row 1/.style={nodes={minimum height=0.8cm}},
  row 2/.style={nodes={minimum height=1.55cm}},
  row 3/.style={nodes={minimum height=1.0cm}},
  row 4/.style={nodes={minimum height=1.1cm}},
  row 5/.style={nodes={minimum height=1.6cm}},
  row 6/.style={nodes={minimum height=1.3cm}},
  row 7/.style={nodes={minimum height=1.4cm}},
]{
  |[corner]| {}
    & |[hdrA]| Abelian groups (e.g., U$(1)$)
    & |[hdrN]| Non-Abelian groups (e.g., SU$(2)$) \\
  |[rlabel]| Degrees of freedom
    & |[cellA]| Rotor variable with $[\hat\theta,\hat n]=i$: \newline
                $n\in\mathbb{Z}$ and $\theta\in[0,2\pi)$
    & |[cellN]| Rotation matrix $\ket{g}$, or irreps and magnetic components
                $\ket{J,m,n}$, \newline
                with $2J\in \mathbb{Z}^+$, and
                $m,n=1,\dots,d_J=2J+1$ \\
  |[rlabel]| Initial state on site
    & |[cellA]| Zero density: $\ket{n=0}=\int_0^{2\pi}\tfrac{d\theta}{2\pi}\,\ket{\theta}$
    & |[cellN]| Trivial irrep: $\ket{J=0}=\int_{\mathrm{SU}(2)} dg\,\ket{g}$ \\
  |[rlabel]| Noise channel
    & |[cellA]| $K^{n_{xy}}_{\langle xy\rangle}=e^{in_{xy}(\theta_x-\theta_y)}$
    & |[cellN]| $K^{J_{ij}}_{a,b,\langle xy\rangle}=\frac{1}{\sqrt{d_J}}\int dg_i dg_j D^\Gamma_{ab}(g_i^{-1}g_j)\ket{g_i, g_j}\bra{g_i,g_j}$ \\
  |[rlabel]| Fusion of irreps
    & |[cellA]| Deterministic and additive: \newline
                $n^{(1)}_v\otimes n_v^{(2)}=n^{(1)}_v+n_v^{(2)}$
    & |[cellN]| Probabilistic: $\displaystyle J^{(1)}_v\otimes J_v^{(2)}=\sum_{J_v=|J^{(1)}_v-J_v^{(2)}|}^{J^{(1)}_v+J_v^{(2)}}J_v$ \\
  |[rlabel]| Syndrome measurement
    & |[cellA]| Charge measurement: \newline
                $P_{n_v}=\int\tfrac{d\theta}{2\pi}\,e^{in_v\theta}\,e^{i\hat n_v\theta}$
    & |[cellN]| Irrep projector: \newline
                $P_{J_v}=d_{J_v}\!\int_{\mathrm{SU}(2)}\!dg\,\chi_{J_v}(g)\,L_{g}$ \\
  |[rlabel]| Global \newline constraint
    & |[cellA]| Charge neutrality: \newline
                $\mathbb{P}_{\mathrm{inv}}=\int\tfrac{d\theta}{2\pi}\,e^{i\theta\sum_v n_v}=\delta(\sum_v n_v)$
    & |[cellN]| Sum over fusion trees: \newline
                $\mathbb{P}_{\mathrm{inv}}= \sum_{\{{\boldsymbol{J}}\}} \ket{\mathcal T_{\boldsymbol{J}}}\bra{\mathcal T_{\boldsymbol{J}}}$ with $\boldsymbol{J}_{\textrm{leaves}}=\{J_v\}$\\
};
\end{tikzpicture}

    \caption{
    \textbf{Comparison of the group-based construction between Abelian and non-Abelian groups.} We choose the continuous groups U$(1)$ and SU$(2)$ to make the comparison explicit. }
    \label{fig:A_vs_NA}
\end{table*}
\subsection{Strongly-symmetric quantum channels}\label{sec:channel}

We can characterize quantum channels by how they operate on a general density matrix, written in the group element basis
\begin{equation} \label{eq:rho_0}
    \varrho =\int \mathcal{D}l \mathcal{D}r\,\varrho({\bf{l}}, {\bf{r}})\,\ket{{\bf{l}}}\bra{{\bf{r}}},
\end{equation}

The elementary building block of the quantum channels we will study is a two-site quantum channel $\mathcal{E}^\Gamma_{i,j }$, acting on nearest neighbor sites, that is both strongly $G$-left symmetric and diagonal in the group basis. 
The second property implies the general form 
\begin{align} \label{eq:Exy_general}
    \mathcal{E}_{i,j}(\varrho)=\int \mathcal{D}l \mathcal{D}r\,w(l_i, l_j,r_i,r_j)\varrho({\bf{l}}, {\bf{r}})\,\ket{{\bf{l}}}\bra{{\bf{r}}},
\end{align}
The first property means that $\mathcal{E}^\Gamma_{i,j }$ is invariant under multiplication by ${\bf{L_g}}=\otimes_v L_g$ for any $g\in G$ operating on either bra or ket degrees of freedom. This restricts the kernel $w$ to be a function of only two arguments $ w(l_i^{-1} l_j,r_i^{-1}r_j)$. In addition $\mathcal{E}^\Gamma_{i,j }$ must be a CPTP map. The trace preserving condition fixes $w(g,g)=1$ for all $g\in G$. The same condition also guarantees that the channel is \emph{unital}, i.e., $\mathcal{E}_{\langle ij \rangle}(\mathds{1})=\mathds{1}$. The requirement that $\mathcal{E}_{i,j }(\rho_0)$ is a \emph{self-adjoint} operator implies $w(l,r)^*=w(r,l)$. Finally, \emph{complete positiveness} of the channel requires $w(l,r)=\sum_\alpha k_\alpha(l)k^*_\alpha(r)$, where the eigenvectors satisfy $\sum_\alpha |k_\alpha(g)|^2=1$ for all $g\in G$. $\alpha$ takes infinitely many values for continuous groups, while $\alpha=1,\dots, |G|$  for finite groups (see additional details in App.~\ref{app:weightW}). These constraints determine the general form of the Kraus operators that make up the channel
\begin{equation} \label{eq:gen_Kraus}
    \hat{K}_{\alpha;i,j}=\int dg_i dg_j \,\,k_\alpha(g_i^{-1} g_j)\ket{g_i,g_j}\bra{g_i,g_j}.
\end{equation}

Recall now that the set of representation matrices $D_{ab}^\Gamma$ form an orthonomal basis (Eq.~\eqref{eq:ONB}). Therefore, we can expand $k_\alpha(g_i^{-1}g_j)$ as
\begin{equation} \label{eq:k_in_D}
    k_\alpha(g_i^{-1}g_j)=\sum_{\Gamma\in \hat{G}} \sum_{a,b=1}^{d_\Gamma} \langle \Gamma, a, b|k_\alpha\rangle D^\Gamma_{ab}(g_i^{-1}g_j), 
\end{equation}
with $\langle \Gamma, a, b|k_\alpha\rangle=\int dg \sqrt{d_\Gamma}D^{\Gamma,*}_{ab}(g)k_\alpha(g) $. Hence, general Kraus operators include a sum over different irreps $\Gamma$. 
The physical interpretation is that a symmetric Kraus operator can in general create a coherent superposition of singlet pairs composed of different irreps. For example, a Kraus operator symmetric under ${\rm SU}(2)$ can create a super-position of a spin-$1/2$ singlet pair and a spin-1 singlet pair.

In this work, however, we focus on incoherent channels, where the Kraus operators insert a singlet pair composed of a unique irrep $\Gamma$ and its dual $\bar{\Gamma}$ on nearest neighbor sites.
This corresponds to the choice of Kraus operators~\footnote{An alternative  choice of Kraus operators corresponds to $k_\alpha(g_i^{-1}g_j)=D^\Gamma_{ab}(g_j^{-1}g_i)$, which is the adjoint of Eq.~\eqref{eq:Kraus} up to a swap of representation indices, i.e., $K^\Gamma_{ab;i,j}\to K^{\Gamma,\dagger}_{ba;i,j}$. This choice leads to the same expression of $\rho$ in Eq.~\eqref{eq:rho_explicit} as long as $p_\Gamma=p_{\bar{\Gamma}}$.}
\begin{align}\label{eq:Kraus}
    \hat{K}^\Gamma_{ab;i,j}=\frac{1}{\sqrt{d_\Gamma}}\int dg_i dg_j \,\,D^\Gamma_{ab}(g_i^{-1}g_j)\ket{g_i,g_j}\bra{g_i,g_j},
\end{align}
which translates to the following contribution to the kernel $w$ in \eqref{eq:Exy_general}:
\begin{equation}
    \frac{1}{d_\Gamma}\sum_{a,b=1}^{d_\Gamma}  D^\Gamma_{ab}(l_i^{-1}l_j)D^{\Gamma,*}_{ab}(r_i^{-1}r_j)=\frac{1}{d_\Gamma}\chi_\Gamma(l_i^{-1}l_jr_j^{-1}r_i).
\end{equation}
To gain physical intuition on the action of the Kraus operator $\hat{K}^\Gamma_{ab;i,j}$, apply it to the ``empty state" $\ket{0}_i \ket{0}_j$. By employing the basis $\ket{\Gamma, a,b}$ and proper normalization, the resulting state is $\ket{\psi;a,b}=\frac{1}{\sqrt{d_\Gamma}}\sum_{c=1}^{d_\Gamma}\ket{\bar{\Gamma},c,a}_i\ket{\Gamma, c,b}_j$. 
Hence, $\hat{K}^\Gamma_{ab}$ indeed creates a pair of irreps $\Gamma$ and $\bar{\Gamma}$ in the trivial fusion channel (i.e., a singlet) on neighboring sites $i$ and $j$, which has a bipartite entanglement equal to
$\log(d_\Gamma)$. 
The local quantum channel acting on the reference vacuum state $\rho_0 = (\ket{0}\bra{0})^{\otimes V}$ then takes the form
\begin{equation}\label{eq:full_channel}
    \mathcal{E}_{ij}(\rho_0) = \sum_{\Gamma} p_\Gamma\sum_{a,b=1}^{d_\Gamma} \hat{K}^\Gamma_{ab;i,j}\, \rho_0\, \hat{K}^{\Gamma,\dagger}_{ab;i,j}\,,
\end{equation}
where the sum runs over \emph{all} irreps including the trivial one, and $\sum_\Gamma p_\Gamma = 1$ (the $\Gamma = 0$ term contributes $p\,\rho$ (no error), so $p = 1 - \sum_{\Gamma\neq 0}p_\Gamma$ plays the role of the no-error probability)~\footnote{The trace-preserving condition is guaranteed by the equation $\frac{1}{d_\Gamma}\sum_{a,b=1}^{d_\Gamma} D^{\Gamma, *}_{ab}(g_i^{-1}g_j)D^{\Gamma}_{ab}(g_i^{-1}g_j)=1$}. The decohered density matrix can then be explicitly written as 
\begin{equation}\label{eq:rho_explicit}
    \rho = \int \mathcal{D}l \mathcal{D}r \prod_{\langle i,j\rangle} w(l^{-1}_i l_j r^{-1}_j r_i)\; \ket{{\bf{l}}}\bra{{\bf{r}}}\,,
\end{equation}
with local weight
\begin{equation}\label{eq:local_weight}
    w(u) = \sum_{\Gamma} \frac{p_\Gamma}{d_\Gamma}\,\chi_\Gamma(u).
\end{equation}

By construction, the resulting $\rho$ is $G$-left strongly symmetric. However, the choice of Kraus operators~\eqref{eq:Kraus} also leads to an additional accidental \emph{local $G$-right weak symmetry,}. That is a symmetry under $l_v \to R_{g_v}l_v=l_v\, g_v^{-1}$, $r_v \to R_{g_v}r_v=r_v\, g_v^{-1}$ with a site dependent $g_v\in G$. 

We make a few remarks before proceeding
\begin{enumerate}[label=\roman*)]
\item As mentioned above, the channel in Eq.~\eqref{eq:Exy_general} is strongly symmetric only under $G$-left and not under $G$-right multiplication for a general choice of the kernel $w$.
\item We do not include strongly $G$-left symmetric Kraus operators of the form $R_{g,v}$. These would be the non-Abelian generalization of on site $\hat{X}$ Kraus operators in the $\mathbb{Z}_2$-symmetric setting and they complicate the analysis of both the Abelian and non-Abelian cases because they do not commute with the two-site Kraus operators. The latter are sufficient as a minimal model to bring out the novel features of the non-Abelian case. 
\item For groups whose irreps are not self-dual, i.e., $\bar{\Gamma}$ is not isomorphic to $\Gamma$ (e.g.\ cyclic $\mathbb{Z}_n$ with $n\geq 3$ or U$(1)$), the weight $w(u)$ in Eq.~\eqref{eq:local_weight} is not real-valued. In those situations, it can be made explicitly real by combining an irrep $\Gamma$ and its dual $\bar{\Gamma}$ with the same probability, as done in Eq.~\eqref{eq:U1_example_channel}. Then the condition $p \geq \sum_{\Gamma\neq 0}p_\Gamma$ is sufficient for the kernel $w$ to be non-negative for any group element. 
\item In App.~\ref{app:weakSym} we show that for a general channel of the form given in Eq.~\eqref{eq:Exy_general}, the weak global $G$-left symmetry cannot be spontaneously broken. Hence, the symmetry $G_{\textrm{ket}}\times G_{\textrm{bra}}$ can at most break to its diagonal subgroup $(G_{\textrm{ket}}\times G_{\textrm{bra}})_{\textrm{diag}}$.
\item Finally, for general $G$-left strongly symmetric quantum channels, the projector onto the global invariant subspace
\begin{equation} \label{eq:Haar_proj}
    \mathbb{P}_{\textrm{inv}} = \int dg\, {\bf{L}}_g = \int dg \bigotimes_{v} L_{g},
\end{equation}
is a stationary state. Refs.~\cite{li2025entanglement, Sahu_24, sahu2026entanglement} show that (for finite-dimensional local Hilbert spaces) $\mathbb{P}_{\textrm{inv}}$ has non-vanishing mixed-state entanglement when $G$ is non-Abelian. 

\end{enumerate}

\subsection{Block-diagonal structure of $\rho$}
We can gain more insight into the structure of the mixed state Eq.~\eqref{eq:rho_explicit} by expressing it in the irrep basis [Eq.~\eqref{eq:G_ab}]: 
\begin{equation} \label{eq:coherences}
\begin{split} 
      \rho &= \int \mathcal{D}g \,W({\bf{g}})\\ &\bigotimes_{v}\left(\sum_{\substack{\Gamma_v, b_v \\ a_v,a_v'}} D^{\Gamma_v}_{a_v',a_v}(g_v)\ket{\Gamma_v, a_v, b_v}\bra{\Gamma_v, a_v', b_v}\right) ,
\end{split}
\end{equation}
with $W({\bf{g}})=\prod_{\expval{xy}} w(g_x g_y^{-1})$ (see App.~\ref{app:coherences}). This expression shows that, in general, $\rho$ is only block-diagonal in the local irrep labels $\Gamma_v$. In particular, the fusion outcome of two irreps $\Gamma_v$, $\Gamma_w$ on vertices $v, w$, can be distinct for the bra and the ket.

When the group $G$ is Abelian, all irreps are one-dimensional (and equal to their characters), and $\rho$ is fully diagonal in the irrep basis:
\begin{equation} \label{eq:Abelian_rho}
    \begin{split} 
      \rho_{\textrm{Abe.}} = \sum_{\Gamma_v} \lambda(\{\Gamma_v\})\ket{\{\Gamma_v\}}\bra{\{\Gamma_v\}}, 
    \end{split}
\end{equation}
with the eigenvalues given by~\footnote{An eigenvalue $\lambda(\{ \Gamma_v\})$ corresponds to the partition function of a disordered statistical mechanics model with the disorder configuration given by $\{\Gamma_v\}$, see Sec.~\ref{sec:review}.}
\begin{equation}\label{eq:statmech}
    \lambda(\{\Gamma_v\}) = \int \mathcal{D}g \,W({\bf{g}})\prod_v \chi_{\Gamma_v}(g_v).
\end{equation}
It directly follows that, for Abelian symmetries, measuring the irrep on every site is the optimal measurement strategy.

On the other hand, for non-Abelian groups there always exists at least one irrep with dimension greater than one ($d_\Gamma>1$). $\rho$ is then not diagonal on the (left) representation index. This leads to off-diagonal coherences, making an explicit diagonalization of the decohered state, in general, intractable.

\section{Statistical Physics Mapping}\label{sec:syndrome}

We now turn to one of the central goals of this work: formulating the problem of distinguishing between corrupted quantum states from measurement data under strong non-Abelian symmetry.
The key to solving this inference problem is to express the exact syndrome probability distributions as partition functions whose local degrees of freedom are group elements $g\in G$.

The two states we want to distinguish are (i) $\rho=\mathcal{E}(\rho_0)$, the vacuum state corrupted by the strongly symmetric channel defined in section \ref{sec:channel}; (ii) $\rho^\Upsilon_{xy}=\mathcal{E}(\rho^\Upsilon_{0,xy})$, a state with a singlet Bell-pair, composed of irreps $\Upsilon$ and $\bar{\Upsilon}$ on distant sites $x$ and $y$, created from the vacuum and then corrupted by the same channel. Specifically:
\begin{equation} \label{eq:sigma_xy}
\rho^\Upsilon_{0,xy}=\sum_{a,b=1}^{d_\Upsilon} \hat{K}_{ab, x y }^{\Upsilon}\,\rho_0\, \hat{K}_{ab, x y }^{\Upsilon,\dagger}.
\end{equation}

A Bayesian decoder attempts to distinguish between these states by computing and then comparing the probabilities $p_\alpha(M_r)=\tr(\rho_\alpha M_r)$ for the states $\alpha=0,xy$ conditioned on the measurement outcome $M_r$. The states are perfectly distinguishable as long as the classical fidelity between the distributions
\begin{align} \label{eq:classical_fidelity}
     F^C(\rho,\rho^\Upsilon_{xy}) &=\sum_r \sqrt{p(M_r)p_{xy}(M_r)},
\end{align}
is vanishing. We will express the conditional probabilities as partition functions with quenched disorder configuration set by the measurement outcomes. The fidelity correlator \eqref{eq:classical_fidelity} will then be expressed as a disorder averaged correlation function. 

We begin with the simplest case in which the measured syndromes are only the local on-site irreps and analyze the symmetries of this problem. 
We then proceed to discuss a hierarchy of improved protocols that measure the fusion outcomes between syndromes at ascending levels.  

\subsection{Measuring on-site irreps}
\label{sec:Sym}

The simplest measurement protocol for attempting to distinguish $\rho$ from $\rho_{xy}^\Upsilon$ is measuring the irreps on sites. The elements of the POVM are, in this case, the products of projectors on the local irreps $\prod_v P_{\Gamma_v}$ given explicitly in   Eqs.~\eqref{eq:projectors} and \eqref{eq:PG_diag}. The probabilities of observing the given syndromes in the two candidate states are $p^{(1)}_\alpha(\{\Gamma_v\})=\tr\left(\rho_\alpha \prod_v P_{\Gamma_v}\right)$, where again $\alpha=0,xy$ and the superscript $\scriptsize{(1)}$ indicates that these probabilities correspond to observation of only local (``level 1") syndromes. 

Using the expressions for the density matrix ~\eqref{eq:rho_explicit} and the local projectors \eqref{eq:projectors} we derive the following expressions for the syndrome probabilities:
\begin{equation} \label{eq:p0}
\begin{aligned}
    p^{(1)}(\{\Gamma_v\}) &= \int\!\mathcal{D}g\, W({\bf{g}})\prod_v d_{\Gamma_v}\chi_{\Gamma_v}(g_v),\\
    p^{(1)}_{xy}(\{\Gamma_v\}) &= \int\!\mathcal{D}g\, W({\bf{g}})\prod_v d_{\Gamma_v}\chi_{\Gamma_v}(g_v)\tfrac{\chi_\Upsilon(g_x g_y^{-1})}{d_\Upsilon}\,,
\end{aligned}
\end{equation}
with~\footnote{Similarly, for the more general $G$-left strongly symmetric density matrix in Eq.~\eqref{eq:rho_general_B}, the syndrome distribution becomes $p^{(1)}(\{\Gamma_v\}) = \int\!\mathcal{D}l\mathcal{D}r\, W({\bf{l}},{\bf{r}})\prod_v d_{\Gamma_v}\chi_{\Gamma_v}(r_vl_v^{-1})$.}
\begin{equation} \label{eq:W_gigj}
     W({\bf{g}})=\prod_{\langle i,j\rangle}[\sum_{\Gamma_{ij}}\frac{p_{\Gamma_{ij}}}{d_{\Gamma_{ij}}}\Re\chi_{\Gamma_{ij}}(g_ig_j^{-1})]
\end{equation}
The factor $\tfrac{\chi_\Upsilon(g_x g_y^{-1})}{d_\Upsilon}$ follows from the definition of $\rho_{xy}^\Upsilon$ in Eq.~\eqref{eq:sigma_xy}. Here, the local group variables $g_v=r_vl_v^{-1}$ are the combination of ket and bra variables $l_v, r_v$ in Eq.~\eqref{eq:rho_explicit}. 
See Appendix~\ref{app:rho_derivation} for the detailed derivation.

These syndrome probability distributions can be viewed as partition functions of disordered 2D classical statistical mechanics models with local degrees of freedom that are the group elements of $G$. The syndromes play the role of quenched disorder whose configuration is set by the measured outcomes. We refer to the syndrome-dependent weight $S({\bf{g}}| \{\Gamma_v\})=\prod_v \chi_{\Gamma_v}(g_v)$ as the syndrome weight below.  

The classical fidelity \eqref{eq:classical_fidelity} can now be expressed as a disorder averaged correlation function in this statistical mechanics model. First, let us define the average in the disordered model with quenched syndromes $\{\Gamma_v\}$,
\begin{equation}
\langle O\rangle_{\{\Gamma_v\}} \equiv \frac{1}{p^{(1)}(\{\Gamma_v\})}
\int \mathcal{D}g\, W(\mathbf{g})\prod_v d_{\Gamma_v}\chi_{\Gamma_v}(g_v)\, O(g),
\end{equation}
Eq.~\eqref{eq:p0} gives $p^{(1)}_{xy}/p^{(1)} = \langle \chi_\Upsilon(g_xg_y^{-1})/d_\Upsilon\rangle_{\{\Gamma_v\}}$,
so the fidelity correlator becomes a disorder-averaged two-point function,
\begin{align}
F^C_\Upsilon &= \sum_{\{\Gamma_v\}} p^{(1)}(\{\Gamma_v\})
\sqrt{\Big\langle \tfrac{1}{d_\Upsilon}\chi_\Upsilon(g_xg_y^{-1})\Big\rangle_{\{\Gamma_v\}}}
\label{eq:FC_corr}
\end{align}
where the disorder distribution is the partition function itself (Nishimori-like condition).

We now analyze the symmetries of the statistical mechanics model. This will allow us to understand the possible phases and the corresponding long range behavior of the fidelity correlations associated with different irreps. 
Let us consider an observed syndrome configuration $\{\Gamma_v\}$. Since the initial states, the quantum channels and the measurements are invariant under the $G$-left strong symmetry, the fusion of all irreps is restricted to the trivial one. This also means that the statistical mechanics models are globally invariant under $G_{\text{ket}}\times G_{\text{bra}}$, corresponding to independent left and right multiplication by distinct group elements. 
The Hermiticity of $\rho$ and $\rho^\Upsilon_{xy}$ leads to an additional anti-unitary symmetry $\mathcal{J}$. It acts as $\mathcal{J}[f(g)]=f^*(g^{-1})$ on functions on the group, and exchanges left and right multiplications. Both symmetries act locally on the clean factor $W({\bf{g}})$, since in both cases the local bond weights $\chi_{\Gamma_{ij}}(g_ig_j^{-1})$ are invariant.

The syndrome contribution $S({\bf{g}}|\{\Gamma_v\})$ requires more attention. To gather some intuition, let us consider an Abelian group first. Using that for Abelian groups $\chi_{\Gamma_v}(ug_v)=\chi_{\Gamma_v}(u)\chi_{\Gamma_v}(g_v)$, we find that this symmetry transformation acts independently on each local degree of freedom
\begin{equation}
    S({\bf{g}}|\{\Gamma_v\})\to \prod_v \chi_{\Gamma_v}(hg_v) \overset{\text{Abe.}}{=} \chi_{\otimes_v \Gamma_v}(h) S({\bf{g}}| \{\Gamma_v\}),
\end{equation}
leading to an overall phase factor. For any allowed syndrome configuration, $\otimes_v\Gamma_v$ is trivial. Thus, when $G$ is Abelian, the statistical mechacnics model has $G_{\text{diag}}\times G\times \mathcal{J}$  (the weak $G_{diag}$ symmetry is obeyed trivially in the Abelian case).  

In contrast, for non-Abelian groups, a syndrome configuration $\{\Gamma_v\}$, in general leads to several distinct outcomes. For example, the two-dimensional irrep $\mathbf2$ of $S_3$ satisfies $\mathbf2 \otimes \mathbf 2 = \mathbf1 \oplus \mathbf1' \oplus \mathbf2$, so measuring two neighboring syndromes in the irrep $\mathbf2$ does not uniquely determine their global state. The consequences of this fact on the syndrome weight can be seen by how it transforms under left (or right) multiplication
\begin{equation}
\begin{aligned}
    S({\bf{g}}|\{\Gamma_v\})\to &\prod_v \tr\left(D^{\Gamma_v}(h)D^{\Gamma_v}(g_v)\right) \\
    &\not\propto  S({\bf{g}}| \{\Gamma_v\}).
\end{aligned}
\end{equation}
Therefore, unlike the clean weight $W({\bf{g}})$, the syndrome weight is not in itself invariant under the $(G\times G)\rtimes \mathcal{J}$ symmetry. Note that the complete Boltzmann weight is globally symmetric, but the symmetry is imposed only non-locally (more details in the following sections).  

That said, the syndrome weight is manifestly symmetric with respect to the following sub-group of $G\times G$. First, it is invariant under the diagonal symmetry $G_{\textrm{diag}}$ due to invariance of the characters under conjugation by group elements $g\to hgh^{-1}$.  Second, the syndrome weight is invariant under multiplication by elements in the center of $G$, i.e., $Z(G)=\{z\in G\,:\, zg=gz\,\,\forall \,g\in G\}$. Indeed, under the transformation $g_v\to zg_v$ with $z\in Z(G)$, on-site characters transform as $\chi_{\Gamma_v}(g_v)\to \chi_{\Gamma_v}(zg_v)=\lambda_{\Gamma_v}(z)\chi_{\Gamma_v}(g_v) $ with $\lambda_{\Gamma_v}(z)\in {\rm U}(1)$ a central character. Therefore, when $G$ is non-Abelian, the statistical mechanics models in Eq.~\eqref{eq:p0} are only symmetric under $G_{\textrm{diag}}\times Z(G)\times \mathcal{J}$. 

From this symmetry analysis we conclude that non-Abelian irreps appearing in 
the syndrome weight act as local symmetry breaking fields that can induce long range fidelity correlations in irreps of $G$ on which the center acts trivially. At the same time irreps with non-trivial central characters are expected to exhibit short range fidelity correlations at small $p$ and acquire long range order only above a critical error rate $p_c$ where the center symmetry is spontaneously broken.

\subsection{Hiding mechanism and finite fidelity correlators}
\label{sec:Hiding}

There is a simple mechanism generating the long range fidelity correlations in irreps with trivial central character that also allows to bound their saturation values. 
An irrep $\Upsilon$ can be absorbed by a non-Abelian irrep $R$ generated by the noise, namely $R\otimes \Upsilon \supset R$, if $\Upsilon$ transforms trivially under the action of the center $Z(G)$ (see App.~\ref{app:hidingMech}).  Inserting the irrep $\Upsilon$ on a site where a noise process has generated irrep $R$, can leave the syndrome configuration and the state unchanged. Hence, as long as the noise creates a non-zero local density of irreps that can absorb $\Upsilon$, the fidelity $F_\Upsilon^C(p^{(1)}, p_{xy}^{(1)})$ remains finite. 

In Appendix~\ref{app:hidingMech} we show, by computing the local densities of all irreps $R$ that are created by decoherence, that the fidelity can be lower bounded for any non-zero error rate below the percolation threshold by
\begin{equation}\label{eq:FboundMT}
    \lim_{|x-y|\to\infty}F_\Upsilon^C(p^{(1)},p_{xy}^{(1)})\ge \left[\sum_R\rho_R\sqrt{\frac{N^R_{R\Upsilon}}{d_\Upsilon}}\right]^2.
\end{equation}
where $\rho_R$ is the average density of irreps $R$ and $N_{R\Upsilon}^R$ are the fusion multiplicities.
The critical threshold associated to the sw-SSB of the strong $G$-left symmetry when measuring onsite irreps is hence exactly zero $p_{c, G}^{(1)}=0$. However, for irreps $\Upsilon$ whose central characters are not trivial, this lower bound does not hold and the fidelity is expected to vanish at sufficiently small error rates. Hence, the center $Z(G)$ can still be spontaneously broken at a finite critical error rate $p_{c, Z(G)}^{(1)}\neq 0$.

Let us consider the groups $S_3$ and SU$(2)$ as examples. In both cases, all irreps are self-dual, and the characters are real-valued, so the symmetry $\mathcal{J}$ becomes a unitary $\mathcal{Z}^{\mathcal{J}}_2$ symmetry. For $S_3$, all irreps can hide inside the two-dimensional irrep $\mathbf2$, leading to $F^C_\Upsilon(p^{(1)}, p^{(1)}_{xy})>0$ for all of them. Since $S_3$ has a trivial center, a finite fidelity implies that the symmetry $(S_3\times S_3)\rtimes\mathbb{Z}^{\mathcal{J}}_2$ is broken down to its subgroup $(S_3)_{\textrm{diag}}\times \mathbb{Z}^{\mathcal{J}}_2$ for any non-zero error rate. Similarly, for ${\rm SU}(2)$, integer spin representations $J$ can hide since $J\otimes J\supset J$. The center $Z({\rm SU}(2))$ is given by the $\mathbb{Z}_2$ symmetry $g_v\to -g_v$, and half-integer spins pick up a $-1$ under this transformation. Hence, we expect that $\lim_{|x-y|\to \infty}F^C_{\Upsilon}(p^{(1)}, p^{(1)}_{xy})=0$ for half-integer spins at sufficiently small error rate. In conclusion, in this case the $({\rm SU}(2)\times {\rm SU}(2))\rtimes \mathbb{Z}_2^{\mathcal{J}}$ symmetry is broken down to SU$(2)_{\textrm{diag}}\times \mathbb{Z}_2^{\mathcal{J}}\times \mathbb{Z}_2$, and half-integer-spin fidelities can only detect a $\mathbb{Z}_2$-SSB. A detailed analysis of these two cases can be found in Section~\ref{sec:Examples}.

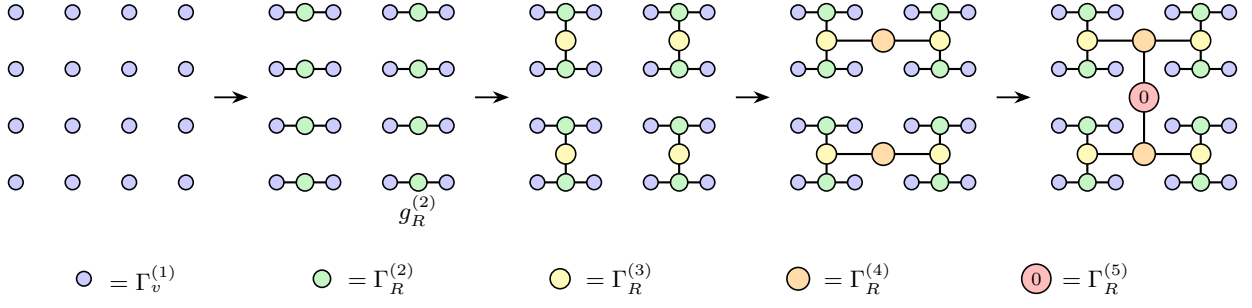
\begin{figure*}[t]
\begin{tikzpicture}[x=0.75cm, y=0.75cm, font=\small]
  \def\step{4.6}   
  \def\pw{3}       
  \def\al{0.6}     

  \foreach \lvl [evaluate=\lvl as \x using \step*(\lvl-1)] in {1,...,5} {
    \panel{\lvl}{\x}{0}
  }
  \node[font=\small, inner sep=0.5pt, below=4pt] at (\step+2.5,0) {$g_R^{(2)}$};
  \foreach \k [evaluate=\k as \xm using \step*\k-(\step-\pw)/2] in {1,...,4} {
    \draw[flow] (\xm-\al/2,1.5) -- (\xm+\al/2,1.5);
  }
 
  \begin{scope}[shift={(0,-1.7)}]
    \node[dotV]    at (1.2,0)  {}; \node[right] at (1.5,0)  {$=\Gamma_v^{(1)}$};
    \node[pentR]   at (5.4,0)  {}; \node[right] at (5.7,0)  {$=\Gamma_R^{(2)}$};
    \node[squareR] at (9.6,0)  {}; \node[right] at (9.9,0)  {$=\Gamma_R^{(3)}$};
    \node[triR]    at (13.8,0) {}; \node[right] at (14.1,0) {$=\Gamma_R^{(4)}$};
    \node[boxR]    at (18.0,0) {0}; \node[right] at (18.3,0) {$=\Gamma_R^{(5)}$};
  \end{scope}
\end{tikzpicture}
    \caption{\textbf{Hierarchical syndrome measurements.} Sequential measurement protocol starting with local irreps $\Gamma^{(1)}_v$. On each level of the measurement hierarchy, the measurement is performed on a pair of regions measured on the previous level.}
    \label{fig:HierarMeas}
\end{figure*}

\subsection{Syndrome fusion tree}\label{sec:syndprob}

To restore the global $G\times G$ symmetry of the statistical mechanics models in Eq.~\eqref{eq:p0}, we recall that since both $\rho$ and $\rho^\Upsilon_{xy}$ are fully supported on the globally $G$-invariant subspace, it follows that $\rho \,\mathbb{P}_{\mathrm{inv}}=\mathbb{P}_{\mathrm{inv}}\rho = \rho$ and similarly for $\rho^\Upsilon_{xy}$. The insertion of this global projector [as defined in Eq.~\eqref{eq:Haar_proj}] introduces a $G$-valued Lagrange multiplier $h$:
\begin{align}\label{eq:global_avg}
    \nonumber p^{(1)}(\{\Gamma_v\}) &= \tr\left(\rho \,\mathbb{P}_{\mathrm{inv}} \prod_v P_{\Gamma_v}\right)\\
    = \int\! \mathcal{D}g\, \, & W({\bf{g}}) \bigg[\int\!dh\prod_v d_{\Gamma_v}\,\chi_{\Gamma_v}(hg_v)\bigg]\,.
\end{align}

As discussed above, 
For a non-Abelian symmetry,  the Boltzmann weight is not manifestly invariant under left multiplication by any group element even for allowed syndrome configurations. 
To understand where the information about the symmetry is hidden and how to reveal it, it is instructive to  perform the above integral over  the $G$ valued Lagrange multiplier in \eqref{eq:global_avg}.

To this end we express the global projector $\mathbb{P}_{\mathrm{inv}}$ [Eq.~\eqref{eq:Haar_proj}] in the Peter-Weyl basis Eq.~\eqref{eq:G_ab}:
\begin{equation} \label{eq:P_inv_Gab}
    \mathbb{P}_{\mathrm{inv}}=\sum_{\{\substack{\Gamma_v, b_v,\\ a_v, a_v'}\}}\int dh \,\otimes_{v} D^{\Gamma_v}_{a_va'_v}(h)\,\ket{\{\Gamma_v, a_v,b_v\}}\bra{\{\Gamma_v, a'_v,b_v\}}.
\end{equation}
We can then resolve the entanglement structure between local degrees of freedom by explicitly accounting for different internal fusion channels of local irreps $\Gamma_v$. For example, let us consider the pairing of sites shown in Fig.~\ref{fig:HierarMeas}, and focus on a pair of two nearby irreps $\Gamma^{(1)}_x, \Gamma^{(1)}_y$ (blue dots). The intermediate fusion outcomes $\Gamma^{(2)}_{xy}$ (green dots in Fig.~\ref{fig:HierarMeas}) on non-overlapping size-$2$ regions can be decomposed as
 \begin{equation} \label{eq:twosites_fusion}
 \begin{aligned}
     &D^{\Gamma_x}_{a_xa'_x}(h)D^{\Gamma_y}_{a_ya'_y}(h) \\ &= \sum_{\Gamma^{(2)}_{xy}, a_{xy}, a'_{xy}} [C^{\Gamma^{(2)}_{xy}, a_{xy}}_{\Gamma_x, a_x; \Gamma_y, a_y}]^* D^{\Gamma^{(2)}_{xy}}_{a_{xy} a'_{xy}}(h)\,C^{\Gamma^{(2)}_{xy}, a'_{xy}}_{\Gamma_x, a'_x; \Gamma_y, a'_y},
\end{aligned}
\end{equation}
where $C^{\Gamma^{(2)}_{xy}, a_{xy}}_{\Gamma_x, a_x; \Gamma_y, a_y}$ are the Clebsch-Gordan (CG) coefficients given by $\langle \Gamma^{(2)}_{xy}, a_{xy}|\Gamma_x, a_x\rangle \otimes |\Gamma_y, a_y\rangle$. We assumed that no multiplicity factors appear in the fusion of two irreps, i.e., $\Gamma_x \otimes \Gamma_y= \sum_{\Gamma^{(2)}_{xy}} N^{\Gamma^{(2)}_{xy}}_{\Gamma_x, \Gamma_y}\Gamma^{(2)}_{xy}$ with $N^{\Gamma^{(2)}_{xy}}_{\Gamma_x, \Gamma_y}=0,1$.

We now iterate this procedure along the ``hierarchy'' shown in Fig.~\ref{fig:HierarMeas}. At the next step, two size-$2$ regions, for example $R^{(2)}_1=(x,y)$ and $R^{(2)}_2=(x',y')$, carrying intermediate irreps $\Gamma^{(2)}_{xy}$ and $\Gamma^{(2)}_{x'y'}$, are fused into a new intermediate irrep $\Gamma^{(3)}$. In an $L \times L$ system, this procedure produces a binary fusion tree $\mathcal T$ of the height $\log_2(L)$ with $L^2/2^n$ many sites on the $n$th level of the tree. The leaves of $\mathcal T$ are the measured on-site irreps $\Gamma_v$, the internal edges carry intermediate irreps, and the root is constrained to be the trivial irrep $0$. By using Eq.~\eqref{eq:ONB}, we can then fully integrate out the Lagrange multiplier $h$ which projects the two irreps $\Gamma_1$ and $\Gamma_2$ in the level just below the highest one onto the global trivial irrep.

Let us now denote by $\boldsymbol{\Gamma}$ all intermediate irreps on the edges of the fusion tree. We use the convention that the leaf labels are fixed by $\Gamma_v$ for each leaf $v$, and that the root label is fixed by $\Gamma_{\mathrm{root}}=0$. Each trivalent vertex of the tree, with one parent edge and two incoming children, involves a product of two CG coefficients as in Eq.~\eqref{eq:twosites_fusion}. Denoting by $\mathcal C^{\mathcal T,\boldsymbol{\Gamma}}_{{\boldsymbol{\Lambda}}}$ the product of the CG coefficients for a particular choice of irrep labels $\boldsymbol{\Gamma}$ and magnetic indices ${\boldsymbol{\Lambda}}$, we can write the global invariant projector as
\begin{equation}\mathbb P_{\mathrm{inv}}=\sum_{{b_v}}\sum_{\substack{\boldsymbol{\Gamma}:\Gamma_{\mathrm{root}}=0\\\boldsymbol{\Lambda},\boldsymbol{\Lambda'} }}\left(\mathcal C^{\mathcal T,\boldsymbol{\Gamma}}_{{\boldsymbol{\Lambda}}}\right)^*\mathcal C^{\mathcal T,\boldsymbol{\Gamma}}_{{\boldsymbol{\Lambda'}}}\ket{\{\Gamma_v,a_v,b_v\}}\bra{\{\Gamma_v,a'_v,b_v\}},
\label{eq:P_as_FT}\end{equation}
where the tree architecture $\mathcal T$ is fixed (for example, the binary tree structure as in Fig.~\ref{fig:HierarMeas}). Schematically, we can write this expression as a sum over all possible irrep confugirations on the fusion tree  $\mathcal T$: $\mathbb{P}_{\mathrm{inv}}=\sum_{\boldsymbol{\Gamma}}\ket{\mathcal T_{\boldsymbol{\Gamma}}}\bra{\mathcal T_{\boldsymbol{\Gamma}}}$, where the sum over magnetic indices is implicit. Eq.~\eqref{eq:P_as_FT} resolves the projector onto the global trivial representation
as a sum over internal fusion channels. As a result, the probability of an on-site syndrome configuration is given by [see Eqs.~\eqref{eq:P_as_FT} and \eqref{eq:coherences}]
\begin{align}\label{eq:fusion_tree_expansion}
    &p^{(1)}(\{\Gamma_v\}) = \prod_v d_{\Gamma_v}\int\mathcal{D}g  W({\bf{g}})  \sum_{\{{\boldsymbol{\Gamma}}\}} \bra{\mathcal T_{\boldsymbol{\Gamma}}}\bigotimes_v D^{\Gamma_v}(g_v)\ket{\mathcal T_{\boldsymbol{\Gamma}}}.
\end{align}

The internal fusion channels $\boldsymbol{\Gamma}$ introduce an auxiliary hierarchical structure that has no counterpart in the Abelian setting, where the global constraint is always additive.
In particular, this expression explicitly shows that in contrast to Abelian groups, performing on-site irrep measurements washes out the information regarding how local irreps $\{\Gamma_v\}$ globally fuse. This intuitively explains why the strong symmetry of the statistical mechanics model spontaneously breaks at any $p >0$.

The fusion tree $\mathcal{T}$ introduces an auxiliary direction playing the role \emph{analogous} to an extra spatial dimension in the statistical mechanics model. The on-site variables $g_v$ can be interpreted as the ``boundary'' degrees of freedom and the fusion channels $\boldsymbol{\Gamma}$ as the bulk degrees of freedom. As we will discuss later, multi-site measurements of fusion outcomes partially quench the ``bulk'' degrees of freedom but make the statistical mechanics model more non-local.

Before closing this section we notice that, because fusion is an associative operation ($(\Gamma_1\otimes \Gamma_2)\otimes \Gamma_3 = \Gamma_1\otimes (\Gamma_2\otimes \Gamma_3)$), we can fix any specific spatial architecture of the binary tree $\mathcal{T}$ with two incoming irreps and one outgoing (as in Fig.~\ref{fig:fusion_tree}). Nonetheless, any other tree architecture is permitted~\footnote{Different tree architectures are related to each other by so-called $F$-moves~\cite{simon2023topoquantum}.} and, in fact, may be better suited for applications in decoding.

\subsection{Threshold hierarchy from fusion}\label{sec:hierarchy}

The on-site syndrome distributions $p^{(1)}(\{\Gamma_v\})$, $p^{(1)}_{xy}(\{\Gamma_v\})$ capture only the first level of a richer structure. Here, we show that measuring the total irrep of successively larger regions (see Fig.~\ref{fig:HierarMeas}) defines a \emph{hierarchy of syndrome distributions}, with progressive refiniment of the distributions. 

We start by generalizing the on-site irrep projective measurements to any region $R$ including more than one site. This is done by replacing the single-site representation $L_{g}$ in Eq.~\eqref{eq:projectors} with the tensor product representation $\bigotimes_{v\in R}L_{g,v}$.
The projector measuring the total irrep $\Gamma_R$ of a region $R$ is given by
\begin{equation}\label{eq:P_region}
    P_{\Gamma_R} = d_{\Gamma_R}\int_G dg\, \chi_{\Gamma_R}(g)\bigotimes_{v\in R}L_{g,v}\,.
\end{equation}
This is a well-defined projector on the tensor product Hilbert space of sites in $R$.
We note that, given a subregion $A$ contained in $R$, $[P_{\Gamma_A}, P_{\Gamma_R}] = 0$ (see Appendix~\ref{app:transform} for the proof).
The same is not true for arbitrary regions: for partially overlapping regions $R,R'$ one finds that $[P_{\Gamma,R},P_{\Gamma',R'}]\neq 0$.

Consider a sequence of lattice partitions into non-overlapping regions $R$ of increasing sizes as in Fig.~\ref{fig:HierarMeas}. The measured regions on the level $N=1$ of the hierarchy correspond to single sites $R=\{v\}$ with on-site syndromes $\{\Gamma_v\}$; on the level $N=2$, the measured regions are pairs of sites $R=\{v,v'\}$ with syndromes $\{\Gamma^{(2)}_R\}$; and so on. At each level, we can simultaneously measure the irreps of all regions in the chosen partition \emph{as well as} all syndromes at lower subregions, since the nested regions commute. By doing so, we obtain a complete history record of the fusion outcomes in that region. Notice that, unlike on-site measurements, coarse regional measurements partially resolve coherences in $\rho$ [Eq.~\eqref{eq:coherences}].

%
At level $N=1$, the syndrome distribution is given by $p^{(1)}(\{\Gamma_v\})$ in Eq.~\eqref{eq:p0}.
For any higher hierarchy level $N>1$, the (joint) probability distribution for measurement outcomes on the first $N$ levels reads (App.~\ref{app:hierarchy_deriv})
\begin{widetext}
\begin{equation}\label{eq:level_general}
    p^{(N)}\Big(\big\{\Gamma^{(n)}_R\big\}_{n=1}^{N}\Big) = \int \mathcal{D}g\,W({\bf{g}}) \prod_{n=2}^{N}\left[\int \mathcal{D}g^{(n)}\prod_{{R}}  d_{\Gamma^{(n)}_{R}} {\chi_{\Gamma^{(n)}_{R}}}\left( g^{(n)}_{{R}}\right)\right]\prod_{v}  d_{\Gamma_v} {\chi_{\Gamma_v}}\left[  g_v(g^{(2)}_{{R}(v)})^{-1}\cdots (g^{(N)}_{R(v)})^{-1}\right],
\end{equation}
\end{widetext}
where $R(v)$ denotes the level-$n$ region containing vertex $v$, and $g^{(n)}_R$ is the group variable associated with that region (see Fig.~\ref{fig:HierarMeas}). Notice that the local coupling $W({\bf{g}})$ remains unchanged while the inter-level coupling enters only through the on-site syndrome insertions ${\chi_{\Gamma_v}}\left(   g_v(g^{(2)}_{{R}(v)})^{-1}\cdots (g^{(N)}_{R(v)})^{-1}\right)$. This new coupling guarantees that syndromes fuse according to a given configuration of the fusion tree.

It is natural to expect that having access to the higher levels of the fusion tree helps one to distinguish $\rho$ from $\rho_{xy}^\Upsilon$. Consider the hierarchy level $N$, where the distributions $p^{(N)}\Big(\big\{\Gamma^{(n)}_R\big\}_{n=1}^{N}\Big)$ and $p_{xy}^{(N)}\Big(\big\{\Gamma^{(n)}_R\big\}_{n=1}^{N}\Big)$ are specified by all the intermediate irreps up to the $N$th level. In particular, the distribution of syndromes up to level $N$ can be obtained as a  marginal of the distribution coming from syndromes up to level $M>N$:
\begin{equation} \label{eq:marginal}
    p^{(N<M)}_\alpha\!\Big(\big\{\Gamma^{(n)}_R\big\}_{n=1}^{N}\Big)=\sum_{n>N}\sum_{\{\Gamma_R^{(n)}\}}p^{(M)}\Big(\big\{\Gamma^{(n)}_R\big\}_{n=1}^{M}\Big),
\end{equation}

Therefore, higher levels of the hierarchy provide systematically more information: they refine lower levels, resolving previously ambiguous fusion channels, and progressively constraining the sum over fusion trees in Eq.~\eqref{eq:fusion_tree_expansion}. 

The full $G\times G$ symmetry is restored only at the level of the hierarchy where all the non-Abelian anyons fuse into Abelian ones. In non nilpotent groups like $S_3$ and $SU(2)$ this happens with certainty only when entire error clusters 
are fused. Achieving this requires an algorithm that locates the error clusters and constructs the tree adaptively by successively fusing syndrome irreps within those clusters. Below the percolation threshold such clusters have a maximal linear size of order $\log L$ ~\cite{Kesten1980}, therefore, while  the clustering algorithm succeeds it requires making measurements up to level $\sim \log(\log(L))$ of the fusion tree.  
Because the errors themselves are hidden, locating the error clusters requires solving an inference problem. The statistical mechanics models we derived for the probabilities at successive levels of the tree may allow to develop a maximum likelihood method for locating the clusters that improves on heuristic RG clustering approaches used previously \cite{ Brell_2014, Wootton_14, Dauphinais_2017, Schotte_22, schotte2022b, jing2026}.

\subsection{Improving thresholds}\label{sec:conjecture}

The importance of the increasing amount of information obtained from the higher levels of the hierarchy can be rigorously quantified. By virtue of the data-processing inequality~\cite{NielsenChuang2010},
\begin{equation} \label{eq:F_dpin}
F^C_\Upsilon(p^{(M)},p_{xy}^{(M)})\;\leq\;F^C_\Upsilon(p^{(N)},p_{xy}^{(N)}),
\end{equation}
for $M>N$. 
The equality holds if and only if all conditional distributions of $p$ and $p_{xy}$ agree, that is $F^C_\Upsilon\!\bigl(p(\cdot |X),p_{xy}(\cdot |X)\bigr)= 1$ for every set of syndromes $X$ on the levels N and below. Therefore, unless all conditionals exactly agree, one expects that having access to the whole joint distribution helps distinguish $\rho$ from $\rho_{xy}^\Upsilon$. 
For non-Abelian symmetry groups, one then finds the following hierarchy of distinguishability thresholds
\begin{equation} \label{eq:chain_ineq}
    p_c^{(1)}\leq p_c^{(2)} \leq \cdots \leq p_c^{(\log L)}\leq p_c^{\rm sw-SSB}.
\end{equation}
This sequence trivializes for Abelian states since then $ p_c^{(1)}= p^{\rm sw-SSB}_c$. 

For non-Abelian symmetries, this sequence should be considered separately for the center $Z(G)$ and $G$ itself. For a non-trivial center $Z(G)$, the hierarchy for $p_{c, Z(G)}^{(N)}$ can be non-trivial starting from level 1. We numerically demonstrate a significant improvement of the threshold of the $\mathbb{Z}_2$ center of ${\rm SU}(2)$ between levels 1 and 2 in Sec.~\ref{sec:numRes}. It is unclear whether the improvement between levels $N$ and $M$ for $N-M \sim \mathcal{O}(1)$ that we observe for SU(2) holds in general.

As long as $G \times G$ is not fully restored, its thresholds vanish  $p_{c, G}^{(N)} = 0$, and it remains so up to $N \sim \log \log L$. For $N \gtrsim \log \log L$, a finite threshold for $G \times G$ is expected to arise and non-trivially improve along the hierarchy. 
Finally, whether the hierarchy recovers the intrinsic sw-SSB threshold $p_c$ when measuring the entire fusion tree, i.e., $\lim_{N\to\log(L)}p_c^{(N)}=p_c$, remains an interesting open question. We leave further analysis on these questions to future work.

\subsection{Topological quantum memories under pure-charge noise}\label{sec:duality}

The strongly $G$-left symmetric open quantum dynamics we are considering in this work is dual to the effect of a pure-charge noise channel on a quantum double $\DG$ ground state based on a finite group $G$~\cite{kitaevTC2003}. This channel only nucleates anyons $a=(\{e\},\Gamma)$ labeled by a trivial conjugacy class of $G$ and an irrep $\Gamma$. A detailed formulation of this duality can be found in App.~\ref{app:equiv_QD}.
This duality KW$_G$, which generalizes the Kramers-Wannier transformation to finite non-Abelian groups~\cite{Tantivasadakarn_2023, yoshida2017gapped, lootens2023dualities}, follows from the mapping between the strongly-symmetric setup (with $G$-qudits on vertices) to the topological ordered system (with $G$-qudits on edges).
Within the subspace of $G$-left symmetric states, this unitary maps the symmetric product state $\ket{\psi_0}$ in Eq.~\eqref{eq:init_state}, to a ground state $\ket{\DG}$ within a fixed logical sector. Moreover, the strongly-symmetric channel $\mathcal{E}_{ij}$ in Eq.~\eqref{eq:full_channel} maps under $\textrm{KW}_G$ to a pure-charge decoherence channel acting on bonds (with a particular orientation on the relative insertion of $\Gamma$ and $\bar{\Gamma}$ when irreps are not self-dual). With probability $p_\Gamma$ it nucleates an anyon $a=(\{e\},\Gamma)$ anti-anyon $\bar{a}=(\{e\},\bar{\Gamma})$ pair labeled by a trivial conjugacy class of $G$ and the irrep $\Gamma$. Moreover, this duality implies that the expectation value of $G$-left symmetric operators $\mathcal{O}_G$ evaluated on the strongly-symmetric decohered density matrix $\rho$ equals the expectation value of corresponding gauge invariant operator $\mathcal{O}_{\DG}$ under KW$_G$, evaluated on the decohered mixed state $\rho_{\DG}$ derived from $\ket{\DG}$. In particular, the quantum fidelity between the decohered density matrices $\rho$ and $\rho_{xy}^\Upsilon$, correspond to the fidelity between the decohered ground state $\rho_{\DG}$, and $\rho_{\DG, xy}^a$. The second includes an additional anyon anti-anyon pair inserted on sites $x, y$ within the vacuum fusion outcome. 

Through this duality the  insights from the previous sections imply that as long as an anyon $a$ can be absorbed by another anyon without changing the absorbing anyon's type, the quantum fidelity between $\rho_{\DG}$, and $\rho_{\DG, xy}^a$ is finite, independently of the distance between $x$ and $y$. This then suggests that $a$ has condensed and that at best, the encoded information can only be partially recovered using local syndrome information. This is precisely the scenario for all pure-charge anyons in $S_3$ topological order (TO), since all irreps can be absorbed by $\mathbf 2$. This raises the question of whether a nonzero recovery threshold can nevertheless exist. Our previous discussion suggests that to fully restore the  $G\times G$~\footnote{Note that the same $G\times G$ symmetry (or $G\times \mathrm{Rep}(G)$ in the dual picture) remains when including all anyon types beyond pure-charge ones~\cite{Vadali_26}. } symmetry, and thereby prevent a subset of anyons from ``condensing'', the fusion outcome of regions of linear size of at least $\log L$, need to be resolved.  

This prediction is consistent with previous results in the literature. In particular the $D_4$ quantum double provides an example of a non-cyclic theory~\cite{Dauphinais_2017}, in which all anyons become Abelian after a finite number of fusion steps. Since Abelian anyons can be absorbed by non-Abelian anyons, there is no threshold when measuring only local syndromes. However, the non-cyclic property implies that the full symmetry can be restored at a finite level of the hierarchy where all anyons fuse into Abelian ones. Indeed Ref.~\cite{jing2026} showed numerically that a finite recovery threshold is achieved if one measures local syndromes, pairs up all non-Abelian anyons, and then measures again to correct the remaining Abelian anyons. More generally, Ref.~\cite{Dauphinais_2017} proved a non-vanishing recovery threshold (even in the presence of measurement errors) for non-cyclic theories. 

The $S_3$ quantum double is an example of a non cyclic theory and its decoding under pure-charge decoherence was investigated in Ref.~\cite{Wootton_14, lyons_26}. As expected from our analysis, to achieve a non-vanishing threshold, it was essential to locate complete error clusters which requires going up to level $\log\log L$ of the fusion tree. The clusters were identified via a heuristic scheme giving a lower bound to the optimal threshold~\cite{Wootton_14}.
A similar clustering scheme has been numerically implemented for more general non-Abelian TOs~\cite{Brell_2014, Schotte_22, schotte2022b}, also leading to finite recovery thresholds. Finally it is worth noting that the recently proposed just-in-time decoders for $D_4$~\cite{Davydova_26,delafuente2026} and $S_3$~\cite{lyons_26} non-Abelian quantum codes are dynamical fault tolerant versions of such adaptive clustering schemes and it too requires fusing clusters of maximal linear size $\sim \log(L)$.

\section{Illustrative examples} \label{sec:Examples}

In this section, we study the sw-SSB of $S_3$ and ${\rm SU}(2)$ in 2D, and the exactly solvable case of sw-SSB in 1D.

\subsection{$S_3$ strong-to-weak SSB} \label{sec:S3}

Consider the finite non-Abelian group $S_3\cong \mathbb Z_3\rtimes\mathbb Z_2$, namely the permutation group of three elements. This group of order $6$ counts with three irreps: the trivial representation $\mathbf1$, the sign representation $\mathbf1'$, and the two-dimensional representation $\mathbf2$. Their (non-trivial) fusion rules are
\begin{equation} \label{eq:S3_fusion}
\mathbf1'\otimes\mathbf2=\mathbf2,\quad
\mathbf2\otimes\mathbf2=\mathbf1\oplus\mathbf1'\oplus\mathbf2,\quad \mathbf1'\otimes\mathbf1'=\mathbf1
\end{equation}

For the noise model, we consider a quantum channel that with probability $p_{1'}$ it inserts a pair of $\mathbf1'$ irrep, with probability $p_2$ the $\mathbf 2$ irreps, and the trivial irrep otherwise.
Since the center of $S_3$ is trivial, but its characters are real, the only symmetry of the statistical mechanics models corresponding to the first level of the hierarchy are the diagonal $S_3$, which cannot be broken, together with the $\mathbb{Z}_2^P$ symmetry described in Sec.~\ref{sec:Sym}. This implies that the fidelity $F^C_\Upsilon(p^{(0)}, p_{xy}^{(1)})$ picks up a finite value for any irrep $\Upsilon$ of $S_3$. Let us start by considering the one-dimensional irrep $\mathbf1'$. Whenever each of the sites $x$ and $y$ hosts a $\mathbf2$, the two syndrome distributions coincide, and then $F^C_{\mathbf1'} \geq  \text{prob(both sites read } \mathbf2)$. Assuming the sites $x$ and $y$ are far apart this probability is given by the square of the local density $\rho_{\bm{2}}$ of irreps $\mathbf2$ on a vertex. Using the same reasoning and accounting for the non-deterministic fusion, one finds a similar lower bound for the fidelity of inserting a pair of the irreps $\mathbf2$. The resulting expressions read:
\begin{equation}\label{eq:S3fidBounds}
\begin{split} 
    F^C_{\bm{1}'}(p^{(0)}, p_{xy}^{(1)}) &\geq \rho_{\bm{2}}^2,\\
    F^C_{\bm{2}}(p^{(0)}, p_{xy}^{(1)}) &\geq \frac{\rho_{\bm{2}}^2}{2}.
\end{split}
\end{equation}
with $\rho_{\bm{2}}=\frac{2}{3}\qty[1-(1-\frac{3}{2}p_2)^4]$ given an error rate $p_2$. 
Details of the derivation can be found in App.~\ref{app:hidingMech}.

\subsection{SU$(2)$ strong-to-weak SSB} \label{sec:SU2}

We now move on to the analysis of the ${\rm SU}(2)$ strong-to-weak SSB. Although the most general channel inserts any irrep $J\in \tfrac{1}{2} \mathbb{Z}$ with probabilities $p_J$,  we restrict ourselves to the local channel inserting pairs of spins $\tfrac{1}{2}$ on neighboring sites $\expval{i,j}$ with probability $p$. Given this channel, the local on-site syndrome probability distribution on the resulting decohered state $\rho$ reads
\begin{align}\label{eq:SU2_synd_sec}
    &p^{(1)}(\{J_v\}) = \int_{\textrm{SU}(2)}\mathcal{D}g\;\prod_{\langle i,j\rangle}\bigg[1-p +\frac{p}{2}\chi_{\tfrac{1}{2}}(g_ig_j^{-1})\bigg]\nonumber\\
    &\hspace{3cm}\times\prod_v d_{J_v}\, \chi_{J_v}(g_v)\,,
\end{align}
with $d_J=2J+1$. Notice that the local weights on a bond are guaranteed to be non-negative only for $p\in [0,1/2]$. 

While this statistical mechanics model is globally $[{\rm SU}(2)\times {\rm SU}(2)]\rtimes \mathbb{Z}_2^{\mathcal{J}}$ symmetric, only the quotient $\left([{\rm SU}(2)\times {\rm SU}(2)]/\{(e, e),(-e, -e)\}\right)\rtimes \mathbb{Z}_2^{\mathcal{J}}\cong {\rm O}(4)$ acts faithfully. To make this symmetry explicit, let us consider the no-syndrome configuration, i.e., $J_v=0$ for all vertices $v$. It is then convenient to express group elements $g\in \textrm{SU}(2)$ in the quaternionic representation: $g = n^0 \mathbbm{1}_2 + i\mathbf{n}\cdot\mathbf{\sigma}$, where $\mathbf{N} = (n^0,\mathbf{n})$ is a unit 4-vector on the $3$-sphere $S^3$.
In this parametrization, the characters are given by Chebyshev polynomials $\chi_J(g)=U_{2J}(n^0)$~\cite{dlmf}; note that $\chi_J(-g)=(-1)^{2J}\chi_J(g)$.
Using that $\chi_{\tfrac{1}{2}}(g_x g_y^{-1}) = 2 \mathbf{N}_x \cdot \mathbf{N}_y$, we express the model in a manifestly O$(4)$-symmetric form: 
\begin{align}\label{eq:SU2_synd_clean}
     &p^{(1)}(\{J_v=0\}) = \int_{S^3}\mathcal{D}\mathbf{N}\;\prod_{\langle i,j\rangle}\bigg[1-p + p\mathbf{N}_i\cdot\mathbf{N}_j\bigg].
 \end{align} 
On the honeycomb lattice, it becomes the usual O$(4)$ loop model with tension $t=\frac{p}{4(1-p)}$ and loop fugacity $n=4$~\cite{Chayes_2000}.

Once syndromes are present, $\{J_v\}\neq\{0\}$, and the ${\rm O}(4)$ symmetry of Eq.~\eqref{eq:SU2_synd_sec} is spontaneously broken down to ${\rm O}(3)\times\mathbb{Z}_2$ for any $p>0$. This breaking is diagnosed by the integer-spin fidelities $F^C_J$, which are bounded from below by the hiding mechanism (see Appendix~\ref{app:hidingMech}). In the small-$p$ limit, these bounds read
\begin{equation}\label{eq:su2fidBounds}
\begin{split}
    F^C_{1}(\rho, \rho_{xy}) &\geq \frac{1}{3}(4p)^2,\\
    F^C_{2}(\rho, \rho_{xy}) &\geq \frac{1}{5}\qty(\frac{9}{2} p^2)^2.
\end{split}
\end{equation}
The residual $\mathbb{Z}_2$ symmetry acts as $g_v\to -g_v$, or equivalently $\mathbf{N}\to-\mathbf{N}$, so the integer-spin characters are invariant under its action.

We now focus on the residual $\mathbb{Z}_2$ symmetry. Since the half-integer spins are charged under it, long-range order in their fidelities detects its SSB.  We argue that this transition belongs to the universality class of the 2D RBIM on the Nishimori line~\cite{Nishimori_81}. Once
${\rm O}(4)$ is broken down to ${\rm O}(3)\times\mathbb{Z}_2$, the remaining order parameter, the sign of $n^0$, transforms exactly as an Ising spin, and the disorder is drawn from the partition function itself. The only difference lies in the additional degrees of freedom $\mathbf{n}$, which are absent in the RBIM. However, these fluctuations are gapped by the syndromes that locally break the full ${\rm SU}(2)$ symmetry and therefore they are irrelevant at the $\mathbb{Z}_2$ SSB transition. They cannot change the universality but can shift the critical point. Because the ${\rm SU}(2)$ syndromes carry more information, we expect the critical $p$ to be higher than that of the plain vanilla Nishimori RBIM~\cite{honecker2001universalityNishimori, wan2026Nishimori}. We numerically verify these conclusions in Section~\ref{sec:numRes}.

Finally, we consider the case in which the measurements resolve the fusion outcomes on entire error clusters, so that the full ${\rm O}(4)$ symmetry of the corresponding statistical mechanics model is restored. The partition function Eq.~\eqref{eq:level_general} then describes a system of size $L\times L\times M$, where $M$ is the minimal height of the tree for a given syndrome. As long as an algorithm can construct a tree of height $M \ll \log L$, the system is effectively two-dimensional on the scale $L$, and the HMW theorem~\cite{hohenberg1967LRO, mermin1966absence} rules out ${\rm O}(4)$ ordering. Hence, the error threshold of any such tree-constructing algorithm provides a lower bound on $p_{c,{\rm SU}(2)}^{\rm sw\text{-}SSB}$. When the syndromes become dense, however, we expect the minimal tree height itself to grow as $\log L$, at which point the effectively two-dimensional description, and with it the HMW theorem, breaks down. We leave the construction of such an algorithm and the resulting bound to future work.

\subsection{Absence of non-Abelian sw-SSB in 1D}
\label{sec:Cayley}

We close this section by discussing an example where the sw-SSB can be diagnosed exactly. 
To simplify the calculation, here we use the Rényi-$1$ correlator:
\begin{equation} \label{eq:R1}
    \mathcal{R}_1\qty(x,y) = \sum_{a,b=1}^{d_\Upsilon}\tr\qty[\sqrt{\rho} K^\Upsilon_{ab, xy}\sqrt{\rho} K^{\Upsilon,\dagger}_{ab, xy}],
\end{equation}
which also detects the sw-SSB transition for the channels we consider~\cite{weinstein2025efficient, Whightman_corr}. Specifically, the quantum fidelity $F^Q(\rho, \rho_{xy}^\Upsilon)$ both lower and upper bounds the Rényi-$1$ correlator:
\begin{equation} \label{eq:R1_vs_FQ_bounds}
    \frac{1}{d_\Upsilon} \qty[F^{Q}\qty(\rho, \mathcal{E}_{xy}^\Upsilon(\rho))]^2\leq \R^\Upsilon_1(x, y) \leq F^{Q}\qty(\rho, \mathcal{E}_{xy}^\Upsilon(\rho)).
\end{equation}
The bounds from Refs.~\cite{weinstein2025efficient, Whightman_corr} are recovered for Abelian groups where $d_\Upsilon=1$ for every irrep $\Upsilon$. 
The derivation can be found in App.~\ref{app:R1bound}, and it applies to both finite and compact Lie groups $G$ as long as one inserts finite-dimensional irreps. 

This simplification permits an exact calculation of $p_c$ for systems defined on lattices without any loops, e.g., on a 1D chain with open boundary conditions. Specifically, the square root of the decohered density matrix Eq.~\eqref{eq:rho_explicit} is given by~\footnote{This result can be easily verified using the character convolution identity
\begin{equation}
    \int_G du\, \chi_\Gamma(a u)\chi_{\Gamma'}(u^{-1}b) =\frac{\delta_{\Gamma,\Gamma'}}{d_\Gamma}\chi_\Gamma(ab).
\end{equation}}
\begin{equation}
\label{eq:sqrtRho}
    \sqrt{\rho} = \int \mathcal{D}l \mathcal{D}r\, \prod_{\expval{xy}} w^{(1/2)}\qty(l_x^{-1} l_y r_y^{-1} r_x) \ket{\qty{l_v}}\bra{\qty{r_v}}
\end{equation}
with the new local weight $ w^{(1/2)}(u) = \sum_\Gamma \sqrt{p_\Gamma} \,\chi_\Gamma(u)$. By a straightforward calculation, the Rényi-$1$ correlator can be mapped onto a two-point correlation function in a \emph{clean} statistical mechanics model
\begin{equation}
    \label{eq:cleanTwoPoint}\mathcal{R}^\Upsilon_1\qty(x,y) = \expval{\frac{1}{d_\Upsilon} \chi_{\Upsilon}\qty(g_x g_y^{-1})}_{ \textrm{clean}}\sim e^{-|x-y|/\xi_\Upsilon},
\end{equation}
with Boltzmann weights $\prod_{\expval{xy}} |w^{(1/2)}(g_x g_y^{-1})|^2$.

The resulting two-point correlation functions~\eqref{eq:cleanTwoPoint} always decay exponentially in 1D, and the state is strongly symmetric at any finite error rate. 
This calculation explicitly shows that the coherences introduced by the non-Abelian nature of the symmetry, are insufficient to break down the mapping to one-dimensional statistical mechanics in 1D. 
The generalization of this result to Cayley trees will be addressed in a future work.

\section{Numerical methods} 
\label{sec:decoders}

In this Section, we develop a numerical method that allows us to verify our analytical claims and address several questions: 
\begin{itemize}
    \item How does the threshold of the $Z(G)$ SSB depend on the hierarchy level $n$? 
    \item Does the symmetry $G$ get restored when higher levels of the hierarchy are measured? Is it spontaneously broken at any finite $p$?
    \item Is there a finite $p_c$ for the sw-SSB of ${\rm SU}(2)$?
\end{itemize}

We begin with the error threshold of the optimal decoder that has access only to on-site syndrome measurements. This threshold corresponds to a phase transition in the statistical mechanics model Eq.~\eqref{eq:p0} and can be probed by the classical fidelity Eq.~\eqref{eq:FC_corr}, which is given by the disorder-averaged two-point correlation function in the statistical mechanics model:
\begin{equation}\label{eq:CF_N1}
    F^C_\Gamma(p^{(1)},p^{(1)}_{xy}) = \!\sum_{\{\Gamma_v\}} p^{(1)}(\{\Gamma_v\})\sqrt{\frac{p^{(1)}_{xy}(\{\Gamma_v\})}{p^{(1)}(\{\Gamma_v\})}}.
\end{equation}
Numerical evaluation of this expression requires (i) importance sampling of syndrome configurations according to $p^{(1)}(\{\Gamma_v\})$, and (ii) computing the probability ratio $p^{(1)}_{xy}(\{\Gamma_v\})/p^{(1)}(\{\Gamma_v\})$. While for Abelian groups the syndrome sampling can be easily performed by just sampling a reference error configuration which completely fixes the syndromes, for non-Abelian groups, non-trivial correlations among onsite syndromes need to be handled differently. Moreover, standard Monte Carlo methods do not straightforwardly generalize to arbitrary groups $G$ given that local weights are not always non-negative. To circumvent these obstacles, we formulate the statistical mechanics model~\eqref{eq:p0} as the contraction of a two-dimensional tensor network (TN), and we provide a general recipe for computing the classical fidelity correlator~\eqref{eq:CF_N1}.

\begin{figure*}[t]
    \centering
 \begin{tikzpicture}[>=stealth,line cap=round,line join=round]
 
\begin{scope}[shift={(-6.2,1.5)}]
\node[font=\bfseries,left] at (-2.35,2.45) {(a)};
 
\def\hs{0.78}   
\def\ll{1.15}   
 
\tikzset{midarrow/.style={postaction={decorate},
  decoration={markings,mark=at position 0.55 with {\arrow{Stealth[length=5pt,width=4pt]}}}}}
\draw[blue!45,thick,midarrow] ({-\hs-\ll},0) -- (-\hs,0);   
\draw[blue!45,thick,midarrow] ( \hs,0) -- ({\hs+\ll},0);    
\draw[blue!45,thick,midarrow] (0, \hs) -- (0,{\hs+\ll});    
\draw[blue!45,thick,midarrow] (0,{-\hs-\ll}) -- (0,-\hs);   
 
\draw[red!65!black,thick] (0.55,0.55) -- (1.30,1.45);
\filldraw[fill=red!20,draw=red!65!black,thick] (1.30,1.45) circle (2.0pt);
\node[font=\scriptsize,red!50!black,right] at (1.40,1.50) {$\Gamma_v$};
 
\filldraw[fill=blue!12,draw=blue!65!black,thick,rounded corners=2pt]
  (-\hs,-\hs) rectangle (\hs,\hs);
\node[font=\small] at (0,0) {$\mathbf{T}(\Gamma_v)$};
 
\node[font=\scriptsize,above] at (-1.45, 0.05) {$\Gamma_{v,-\hat x}$};
\node[font=\scriptsize,above] at ( 1.45, 0.05) {$\Gamma_{v,+\hat x}$};
\node[font=\scriptsize,left]  at (-0.05, 1.70) {$\Gamma_{v,+\hat y}$};
\node[font=\scriptsize,left]  at (-0.05,-1.45) {$\Gamma_{v,-\hat y}$};
 
\node[font=\scriptsize,text=blue!65!black,below] at (-1.45,-0.08) {$\sqrt{\omega_{\Gamma_{v,-\hat x}}}$};
\node[font=\scriptsize,text=blue!65!black,below] at ( 1.45,-0.08) {$\sqrt{\omega_{\Gamma_{v,+\hat x}}}$};
\node[font=\scriptsize,text=blue!65!black,left] at (-0.05, 1.15) {$\sqrt{\omega_{\Gamma_{v,+\hat y}}}$};
\node[font=\scriptsize,text=blue!65!black,right] at ( 0.05,-1.45) {$\sqrt{\omega_{\Gamma_{v,-\hat y}}}$};
\end{scope}
 
\begin{scope}[shift={(-1.1,0)}]
\node[font=\bfseries,left] at (-1.25,3.95) {(b)};

\def\a{1.0}     
\def\out{0.6}   

\foreach \y in {0,...,3}{
  \draw[thick,blue!25] ({-\out*\a},\y*\a) -- ({(3+\out)*\a},\y*\a);
}
\foreach \x in {0,...,3}{
  \draw[thick,blue!25] (\x*\a,{-\out*\a}) -- (\x*\a,{(3+\out)*\a});
}

\node[font=\scriptsize,blue!65!black,above]
  at ({(3+\out-0.1)*\a},0.03) {$D$};

\foreach \x in {0,...,3}{
  \foreach \y in {0,...,3}{
    \pgfmathsetmacro{\legtip}{0.35+0.07*mod(2*\x+\y+floor(\x/2),4)}
    \coordinate (Btip\x\y) at ({\x*\a+\legtip},{\y*\a+\legtip});
    \draw[red!65!black,thick,preaction={draw=white,line width=2.2pt}]
      ({\x*\a+0.20},{\y*\a+0.20}) -- (Btip\x\y)
      node[circle,fill=red!20,draw=red!65!black,thick,
           minimum size=4pt,inner sep=0pt] {};
    \node[rectangle,rounded corners=2pt,
          fill=blue!12,draw=blue!65!black,thick,
          minimum size=14pt,inner sep=0pt,font=\scriptsize]
      at (\x*\a,\y*\a) {};
  }
}

\node[font=\scriptsize,red!65!black,right=1pt] at (Btip33) {$\Gamma_v$};

\draw[->,thick] (-0.90,-0.15) -- (-0.90,3.30);
\node[font=\scriptsize,text=black,rotate=90] at (-1.15,1.575) {time};
\end{scope}

\draw[->,thick,>=stealth] (2.85,1.5) -- (3.55,1.5);

\begin{scope}[shift={(4.6,0)}]
\node[font=\bfseries,left] at (-0.55,3.95) {(c)};

\def\a{1.0}
\def\m{0.32}
\def\out{0.6}
\tikzset{
  sgl/.style={green!55!black,line width=0.7pt,line cap=butt},
  dbl/.style={sgl,double,double distance=2pt}
}

\foreach \y in {0,...,3}{
  \draw[sgl] ({(1+\m)*\a},\y*\a) -- ({(2-\m)*\a},\y*\a);
  \draw[sgl] ({-\out*\a},\y*\a) -- ({-\m*\a},\y*\a);
  \draw[sgl] ({(3+\m)*\a},\y*\a) -- ({(3+\out)*\a},\y*\a);
}

\foreach \c in {0.5,2.5}{
  \foreach \y in {0,...,2}{
    \draw[dbl] (\c*\a,{(\y+\m)*\a}) -- (\c*\a,{(\y+1-\m)*\a});
  }
  \draw[dbl] (\c*\a,{-\out*\a}) -- (\c*\a,{-\m*\a});
  \draw[dbl] (\c*\a,{(3+\m)*\a}) -- (\c*\a,{(3+\out)*\a});
}

\foreach \x in {0,2}{
  \foreach \y in {0,...,3}{
    \filldraw[fill=green!12,draw=green!55!black,thick,rounded corners=4pt]
      ({(\x-\m)*\a},{(\y-\m)*\a})
      rectangle ({(\x+1+\m)*\a},{(\y+\m)*\a});
  }
}

\foreach \x in {0,2}{
  \foreach \y in {0,...,3}{
    \draw[sgl] ({(\x-\m)*\a},\y*\a) -- ({(\x-0.22)*\a},\y*\a);
    \draw[sgl] ({(\x+0.22)*\a},\y*\a) -- ({(\x+1-0.22)*\a},\y*\a);
    \draw[sgl] ({(\x+1+0.22)*\a},\y*\a) -- ({(\x+1+\m)*\a},\y*\a);

    \foreach \s in {0,1}{
      \draw[blue!40,line width=0.65pt,rounded corners=1.4pt]
        ({(\x+\s-0.22)*\a},{(\y-0.22)*\a})
        rectangle ({(\x+\s+0.22)*\a},{(\y+0.22)*\a});
    }

    \coordinate (Fuse\x\y) at ({(\x+0.82)*\a},{(\y+0.48)*\a});

    \draw[red!65!black,line width=0.65pt]
      ({(\x+0.205)*\a},{(\y+0.205)*\a})
      .. controls ({(\x+0.46)*\a},{(\y+0.205)*\a})
              and ({(\x+0.64)*\a},{(\y+0.30)*\a})
      .. (Fuse\x\y);
    \draw[red!65!black,line width=0.65pt]
      ({(\x+1)*\a},{(\y+0.22)*\a}) -- (Fuse\x\y);

    \pgfmathsetmacro{\lt}{0.82+0.015*mod(2*\x+\y+floor(\x/2),4)}
    \pgfmathsetmacro{\rt}{0.68+0.02*mod(2*(\x+1)+\y+floor((\x+1)/2),4)}
    \pgfmathsetmacro{\lefttipx}{(1-\lt)^3*0.205+3*(1-\lt)^2*\lt*0.46+3*(1-\lt)*\lt^2*0.64+\lt^3*0.82}
    \pgfmathsetmacro{\lefttipy}{(1-\lt)^3*0.205+3*(1-\lt)^2*\lt*0.205+3*(1-\lt)*\lt^2*0.30+\lt^3*0.48}
    \coordinate (Localtip\x\y0)
      at ({(\x+\lefttipx)*\a},{(\y+\lefttipy)*\a});
    \coordinate (Localtip\x\y1)
      at ({(\x+1-0.18*\rt)*\a},{(\y+0.22+0.26*\rt)*\a});
    \foreach \s in {0,1}{
      \filldraw[fill=red!20,draw=red!65!black,line width=0.55pt]
        (Localtip\x\y\s) circle (1.15pt);
    }

    \pgfmathsetmacro{\leglen}{0.17+0.02*mod(\x/2+3*\y+floor(\y/2),4)}
    \coordinate (Ctip\x\y)
      at ({(\x+0.82+\leglen)*\a},{(\y+0.48+0.38*\leglen)*\a});
    \draw[green!55!black,thick] (Fuse\x\y) -- (Ctip\x\y);
    \fill[green!55!black] (Fuse\x\y) circle (0.9pt);
    \filldraw[fill=green!20,draw=green!55!black,thick]
      (Ctip\x\y) circle (2.0pt);
  }
}
\node[font=\scriptsize,green!35!black,right=1pt]
  at (Ctip23) {$\Gamma_R^{(2)}$};

\node[font=\scriptsize,green!35!black,above]
  at ({(3+\out-0.1)*\a},0.03) {$D$};
\node[font=\scriptsize,green!35!black,right]
  at (0.58,{-0.5*\a}) {$D^2$};
\end{scope}

\end{tikzpicture}

\caption{
\textbf{Tensor network structure of the syndrome probability.}
\textbf{(a)} Local tensor ${\bf{T}}(\Gamma_v)$ at vertex $v$. Four bond legs carry indices $(j_{v,\hat\mu}, m_{v,\hat\mu}, n_{v,\hat\mu})$ with bond dimension $D = \sum_{\Gamma=0}^{\Gamma_{\textrm{max.}}} d_\Gamma^2$; the ``physical'' index (red) is postselected on the syndrome $\Gamma_v$. Internally, opposite link pairs fuse into intermediate irreps $\ell^\pm$ (green dashed), recoupled with $\Gamma_v$; each leg carries a weight $\sqrt{\omega_\Gamma}$ with $\omega_\Gamma = p_\Gamma/d_\Gamma$. \textbf{(b)}~Full 2D tensor network when measuring local irreps $\Gamma_v$: a boundary MPS evolves along the vertical direction to give $P(\{\Gamma_v\}) = \mathrm{tr}(\prod_v {\bf{T}}(\Gamma_v))$. Different length of sticking-out legs represent distinct irrep measurements.
\textbf{(c)}~At the second hierarchy level, the tensor network becomes a bilayer: level-1 tensors ${\bf{T}}(\Gamma_v)$ are coupled to coarser level-2 degrees of freedom via the local tensor ${\bf{I}}^{(2)}(\Gamma_R)$ in Eq.~\eqref{eq:tensor_I2}.
}
\label{fig:TN}
\end{figure*}
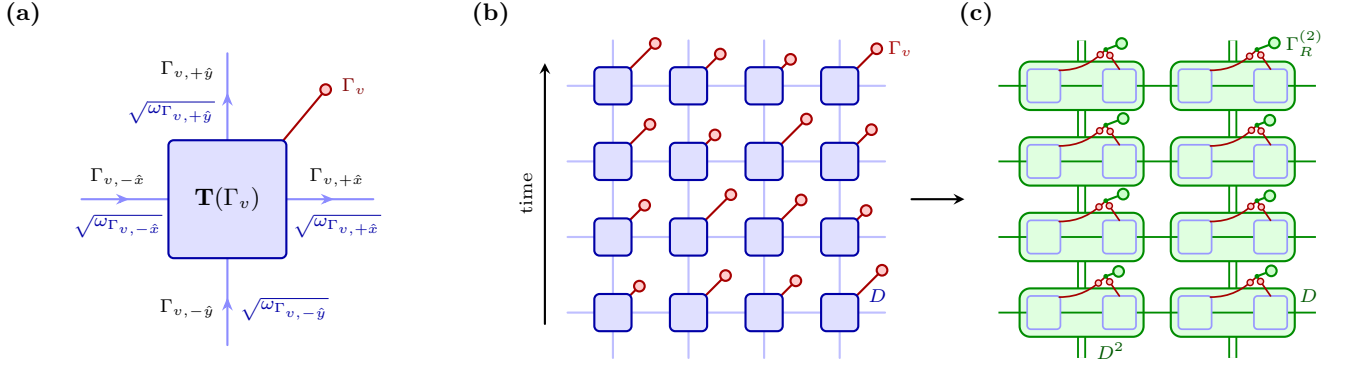

\subsection{Tensor network representation}
\label{sec:TN}

The TN representation rests on a simple observation. Consider the group average over $g_v$ in Eq.~\eqref{eq:p0}, which is a Haar integral for compact Lie groups and a normalized sum for finite groups. Combined with the Clebsch--Gordan decomposition of products of representation matrices, Eq.~\eqref{eq:twosites_fusion}, this average produces a local tensor at each vertex.

For concreteness, we consider a square lattice and the probabilities $p^{(1)}, p_{xy}^{(1)}$ defined in Eq.~\eqref{eq:p0}, with local weights $w(g_ig_j^{-1})=\sum_{\Gamma} w_\Gamma \Re \chi_\Gamma(g_ig_j^{-1})$ where $w_\Gamma = p_\Gamma/d_\Gamma$. Instead of retaining the two terms $\chi_\Gamma(g_ig_j^{-1}),\chi_\Gamma(g^{-1}_ig_j)$, which are equal for groups with real-valued characters, we will keep only $\chi_\Gamma(g_ig_j^{-1})$ and fix an orientation along the edges of the lattice. Our convention is that a bond $\langle i, j\rangle$ points from left $(i)$ to right $(j)$ and from bottom $(i)$ to top $(j)$. Each vertex $v$ is then connected to four oriented links $\hat\mu = \pm\hat x,\pm\hat y$. Each link carries an irrep $\Gamma_{v,\hat\mu}$ appearing in the noise channel with internal (``magnetic'') quantum numbers $(a_{v,\hat\mu}, b_{v,\hat\mu})$ [see Fig.~\ref{fig:TN}(a)]. As in previous sections, we restrict to multiplicity-free groups, for which the fusion of any two irreps contains each irrep at most once.

To construct a local tensor for each vertex, we first split the link weights into two factors, one for each endpoint of the bond: 
\begin{equation}\label{eq:link_split}
    \chi_\Gamma(g_ig_j^{-1})=\sum_{a_{i,j}, b_{i,j}=1}^{d_{\Gamma}}D_{a_{i,j} b_{i,j}}^{\Gamma}(g_i)\, D_{a_{i,j} b_{i,j}}^{\Gamma,*}(g_j).
\end{equation}
Each oriented link $(v,\hat\mu)$ therefore carries a triple $\alpha_{v,\hat\mu} \equiv (\Gamma_{v,\hat\mu}, a_{v,\hat\mu}, b_{v,\hat\mu})$, with $a_{v,\hat\mu}, b_{v,\hat\mu} = 1,\dots,d_{\Gamma_{v,\hat\mu}}$. The vertex at the tail of the link receives the factor $D^{\Gamma}$, and the vertex at its head receives $D^{\Gamma,*}$. Further, we collect all factors that depend on $g_v$ into a single tensor with four link legs and an additional syndrome leg $\qty(A_v, B_v)$:
\begin{equation}\label{eq:tensor_I}
    \mathbf{I}(\Gamma_v)^{A_v B_v}_{\{\alpha_{v,\hat\mu}\}} = \int \! dg_v \left[\prod_{\hat\mu = \pm\hat x, \pm\hat y} D^{\Gamma_{v,\hat\mu},\, s_{\hat\mu}}_{a_{v,\hat\mu} b_{v,\hat\mu}}(g_v)\right] D^{\Gamma_v}_{A_v B_v}(g_v),
\end{equation}
where $A_v, B_v = 1,\dots,d_{\Gamma_v}$. The superscript $s_{\hat\mu} = \pm$ distinguishes outgoing links ($\hat\mu = +\hat x, +\hat y$) from incoming ones ($\hat\mu = -\hat x, -\hat y$), with $D^{\Gamma,+} \equiv D^{\Gamma}$ and $D^{\Gamma,-} \equiv D^{\Gamma,*}$.

To integrate out $g_v$, we repeatedly apply the fusion rule~\eqref{eq:twosites_fusion} to the product in square brackets in Eq.~\eqref{eq:tensor_I}. Each application replaces a product of two representation matrices by a sum over fusion channels, so that after three steps every term contains a single representation matrix of $g_v$. The orthogonality relation~\eqref{eq:ONB} then evaluates the integral. Concretely, we fuse the two incoming links $\Gamma_{v,-\hat x}$ and $\Gamma_{v,-\hat y}$ into an intermediate irrep $\ell^v_{-x,-y}$, the two outgoing links $\Gamma_{v,+\hat x}$ and $\Gamma_{v,+\hat y}$ into $\ell^v_{+x,+y}$, and finally fuse the two intermediate irreps with each other. By orthogonality, the integral vanishes unless $\Gamma_v \in \ell^v_{-x,-y} \otimes \ell^v_{+x,+y}$. This selection rule is how the vertex tensor locally enforces the $G\times G$ symmetry of the model.

Next, we absorb the link weights into the vertex tensors. Since each link is shared by two vertices, we assign a factor $\sqrt{w_\Gamma}$ to each of its endpoints and define
  \begin{equation}\label{eq:tensor_T}
       \mathbf{T}(\Gamma_v)^{A_v B_v}_{\{\alpha_{v,\hat\mu}\}} = d_{\Gamma_v}\, \mathbf{I}(\Gamma_v)^{A_v B_v}_{\{\alpha_{v,\hat\mu}\}} \prod_{\hat\mu = \pm\hat x, \pm\hat y} \sqrt{w_{\Gamma_{v,\hat\mu}}}.
   \end{equation}
The probability of a syndrome configuration $\{\Gamma_v\}$ on the state $\rho$ is given by the contraction of the TN:
\begin{equation}\label{eq:TN}
    p^{(1)}(\{\Gamma_v\}) = \mathrm{tTr}\bigg[\prod_v T(\Gamma_v)\bigg], \quad T(\Gamma_v) = \sum_{M_v=1}^{d_{\Gamma_v}} \mathbf{T}(\Gamma_v)^{M_v M_v},
\end{equation}
where $\mathrm{tTr}$ denotes the contraction of all bond indices shared by neighboring vertices, with open boundary conditions. Each bond index runs over the triples $\alpha = (\Gamma, a, b)$ with $p_\Gamma \neq 0$, i.e., over the irreps created by the noise channel in Eq.~\eqref{eq:full_channel}, so the bond dimension is $D = \sum_{\Gamma\:p_\Gamma \neq 0} d_\Gamma^2$. For finite groups, $D$ remains small even when all irreps are included, since $D \leq |G| = \sum_\Gamma d_\Gamma^2$: for instance, $D = 2$ for $\mathbb{Z}_2$, $D = 6$ for $S_3$, and $D = 8$ for $D_4$.

Computing $p^{(1)}_{xy}(\{\Gamma_v\})$ requires inserting an additional weight $\chi_{\Upsilon}(g_x g_y^{-1})/d_\Upsilon$ between the vertices $x$ and $y$. We split this weight as in Eq.~\eqref{eq:link_split} and assign a factor $1/\sqrt{d_\Upsilon}$ to each endpoint. The vertex tensors at $x$ and $y$ thereby acquire a sixth leg, carrying the magnetic indices $a, b = 1,\dots,d_\Upsilon$:
\begin{equation}\label{eq:tensor_I_mod}
\begin{split}
    \mathbb{I}(\Gamma_v)^{A_v B_v;\,ab}_{\{\alpha_{v,\hat\mu}\}}
    = \frac{1}{\sqrt{d_\Upsilon}} \int \! dg_v
    &\left[\prod_{\hat\mu} D^{\Gamma_{v,\hat\mu},\,s_{\hat\mu}}_{a_{v,\hat\mu} b_{v,\hat\mu}}(g_v)\right] \\
    &\times D^{\Gamma_v}_{A_v B_v}(g_v)\, D^{\Upsilon,\,\sigma_v}_{ab}(g_v),
\end{split}
\end{equation}
where $v \in \{x, y\}$, with $\sigma_x = +$ and $\sigma_y = -$: the tail $x$ of the inserted link receives $D^{\Upsilon}$ and its head $y$ receives $D^{\Upsilon,*}$. 

Evaluating $p^{(1)}_{xy}(\{\Gamma_v\})$ then maps to the contraction
\begin{equation}\label{eq:TN_p1}
    p^{(1)}_{xy}(\{\Gamma_v\}) = \mathrm{tTr}\bigg[\prod_{v \neq x,y} T(\Gamma_v) \sum_{a,b=1}^{d_\Upsilon} \mathbb{T}(\Gamma_x)^{ab}\, \mathbb{T}(\Gamma_y)^{ab}\bigg],
\end{equation}
where the tensor $\mathbb{T}\qty({\Gamma_v})$ is defined through Eq.~\eqref{eq:tensor_T} by exchanging ${\bf{I}}$ by $\mathbb{I}$. This long-range coupling can be numerically implemented by increasing the bond dimension of bonds connecting $x$ and $y$. Bonds on this path have an enlarged bond dimension $D\to D\times d_\Upsilon^2$, and the tensors at intermediate vertices act as the identity on the additional indices. 

The same construction yields the joint distribution of syndromes on the first $N$ levels of the hierarchy, Eq.~\eqref{eq:level_general}. Consider, for example, $N = 2$, and partition the lattice into non-overlapping regions $R = \{i, j\}$ of nearest-neighbor sites [see Fig.~\ref{fig:TN} (c)]. The probability $p^{(2)}_0\big(\{\Gamma_v\} \cup \{\Gamma^{(2)}_R\}\big)$ is obtained by contracting the first-level tensors $\mathbf{T}(\Gamma_v)^{A_v B_v}_{\{\alpha_{v,\hat\mu}\}}$ of Eq.~\eqref{eq:tensor_T}, with their syndrome indices left open, with one second-level tensor per region,
\begin{multline}\label{eq:tensor_I2}
    \mathbf{I}^{(2)}\big(\Gamma^{(2)}_R\big)^{A^{(2)}_R B^{(2)}_R}_{\{\Gamma_v, A_v, B_v\}_{v\in R}} \\
    =d_{\Gamma^{(2)}_R} \int \! dg^{(2)}_R \left[\prod_{v\in R} D^{\Gamma_v,*}_{A_v B_v}\big(g^{(2)}_R\big)\right] D^{\Gamma^{(2)}_R}_{A^{(2)}_R B^{(2)}_R}\big(g^{(2)}_R\big).
\end{multline}
The lower indices of $\mathbf{I}^{(2)}$ are contracted with the syndrome indices $(A_v, B_v)$ of the two first-level tensors in $R$. Its upper indices $(A^{(2)}_R, B^{(2)}_R)$ are either traced over, for $N = 2$, or contracted with the tensors of the next level. Higher levels follow analogously, with a tensor $\mathbf{I}^{(n)}\big(\Gamma^{(n)}_R\big)$ introduced for each region at level $n$ [see Fig.~\ref{fig:TN}(c)]. Explicit expressions for the tensors for $G = {\rm SU}(2)$ in terms of Wigner 3-j symbols are provided in App.~\ref{app:tensorsSU2}.

We notice that the resulting TN~\eqref{eq:TN} corresponds to a \emph{spin network}~\cite{Ilkka_19, Schmoll_2021,Freidel_2010,Bianchi, asymptotic_free_heisenberg,Borisenko_2005}, where representation-theoretic data (irreps on edges, intertwiners at vertices) are contracted into a symmetric-invariant way. In our context, the spin-network structure arises naturally from the non-Abelian fusion rules of the syndrome distribution, and the intertwiner at each vertex is precisely ${\bf{T}}\qty({\Gamma_v})$.

\subsection{Syndrome sampling}
\label{sec:SyndromeSampling}

Below, we discuss two complementary algorithms for syndrome sampling. The first is the perfect sampling algorithm for projected entangled pair states~\cite{ferris2010perfect, vieijra2021direct} which is applicable for any $p$ in our setting. The second algorithm specifically utilizes the structure of the error channel and provides significant speedup at $p$ well below the percolation threshold $p_{\rm perc}$. However, as will be discussed below, this algorithm fails at $p$ close to $p_{\rm perc}$ and beyond.

The perfect sampling algorithm~\cite{ferris2010perfect} proceeds by sampling the syndromes on each vertex of the one-dimensional boundary matrix product state (MPS)~\cite{verstraete2008mps}. Ordering the vertices row by row, from left to right within each row, we use the chain rule
\begin{equation}
    p^{(1)}(\{\Gamma_v\}) = p(\Gamma_1)\, p(\Gamma_2 | \Gamma_1)\, p(\Gamma_3 | \Gamma_1, \Gamma_2) \cdots,
\end{equation}
and sample each syndrome conditioned on those already sampled. Each conditional probability follows from a marginal of $p^{(1)}$, which is a contraction of the TN~\eqref{eq:TN} in which the syndromes of the already-sampled vertices are fixed to their sampled values, while those of the remaining vertices are summed over (we assume that there is a finite number of possible syndromes on each vertex). Evaluating this marginal for every possible value of the current syndrome and normalizing gives its conditional distribution.

Alternatively, we can leverage the structure of our model and construct a more efficient sampling algorithm for $p \ll p_{\rm perc}$. We recall that, using the conditional probabilities, we can write $p^{(1)}(\{\Gamma_v\})$ as a sum over physical error clusters $E$ compatible with the syndrome configuration:
\begin{equation}
p^{(1)}(\{\Gamma_v\}) = \sum_E p(E)\, p(\{\Gamma_v\}|E),
\end{equation}
where $p(E)$ factorizes over the bonds $\langle ij\rangle$ of the lattice, and an error cluster $E$ is a configuration of bond irreps $\Gamma_{ij}$ sampled according to the probabilities $p_\Gamma$ in Eq.~\eqref{eq:full_channel}. The sampling therefore proceeds in two steps: first, we sample an error cluster $E$ from $p(E)$; second, given $E$, we sample a syndrome configuration $\{\Gamma_v\}$ from $p(\{\Gamma_v\}|E)$. 

The first step is efficient, since the errors can be sampled independently on each link. The sampling of the syndromes on a given error cluster can also be done efficiently and \emph{exactly} as long as the error cluster is not too wide. This is done by performing independent sampling on the tree component of the cluster and subsequent exact conditional sampling on the vertices belonging to a minimal feedback vertex set (FVS)---a set of vertices breaking the cycles of the graph---of the error cluster. An approximately minimal FVS can be found efficiently~\cite{becker1996optimization, bafna1999APXFVS}. The conditional sampling of the syndromes on the FVS vertices is done by an exact TN contraction, which becomes impossible when the error clusters span the entire system. This corresponds to the bond percolation threshold $p_{\rm perc}$, beyond which this algorithm should not be used. See Appendix~\ref{app:FVSsampler} for the detailed construction of the algorithm.

\subsection{Numerical results}
\label{sec:numRes}

\begin{figure}[t]
     \centering
        \includegraphics[width=0.48\textwidth]{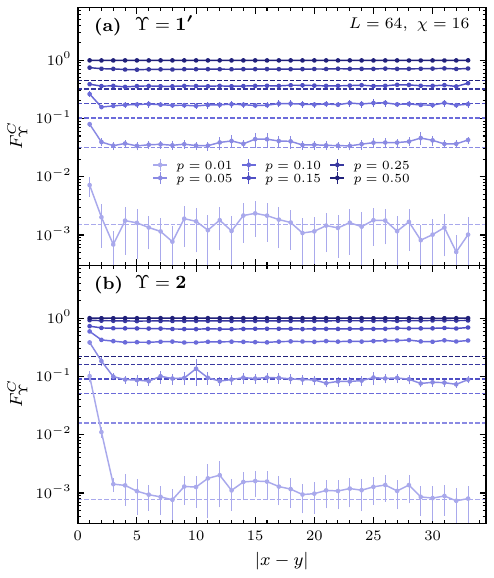}
        \caption{\textbf{$S_3$ sw-SSB.} Numerical results for the $S_3$ model obtained using the TN representation with open boundary conditions from Section~\ref{sec:TN}. The channel is chosen with $p_{1'} = 0$ and $p_{2} = p$, and only the local on-site syndrome information is known, i.e., the first level of the hierarchy. The $L\times L$ TN is contracted using the boundary MPS method with a finite bond dimension $\chi = 16$. 
        Fidelity correlators~\eqref{eq:CF_N1} (solid lines) for $\Upsilon = \bm{1}', \bm{2}$ in the $L=64$ system for various $p$. Horizontal dashed lines depict the lower bounds from Eq.~\eqref{eq:S3fidBounds} for the solid lines of the same color.}
        \label{fig:S3numerics}
\end{figure}

We start by validating the predicted absence of a phase transition at the first level of the hierarchy for $S_3$, using the $S_3$-symmetric decoherence channel with $p_{1'}=0$ and $p_{2}=p$ on the square lattice. For a given syndrome configuration $\{\Gamma_v\}$, we numerically compute the probabilities $p^{(1)}(\{\Gamma_v\})$ and $p^{(1)}_{xy}(\{\Gamma_v\})$ by expressing the two-dimensional TN as the time evolution of an MPS~\cite{vidal2004efficient}. As shown in Fig.~\ref{fig:S3numerics}, the classical fidelity correlators~\eqref{eq:CF_N1} for the irreps $\bm{1}'$ and $\bm{2}$ do not decay to $0$ for any $p>0$. As predicted in Section~\ref{sec:S3}, they are bounded from below by the finite values in Eq.~\eqref{eq:S3fidBounds}, signaling SSB of $S_3\times S_3$ down to
$S_3^{\rm diag}$. 

\begin{figure*}[t]
     \centering
        \includegraphics[width=\textwidth]{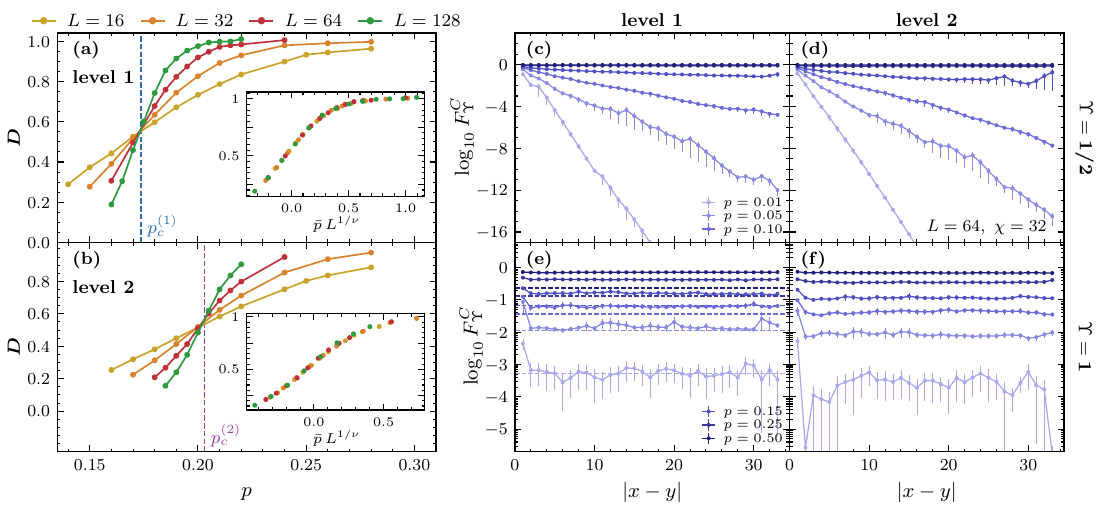}
        \caption{\textbf{${\rm SU}(2)$ sw-SSB.} Numerical results for the ${\rm SU}(2)$ model~\eqref{eq:SU2_synd_sec} using the boundary MPS contraction of the $L\times L$ tensor networks with open boundary conditions. The virtual bond dimensions are $\chi=32$ for $L > 64$, and $\chi=24$ and $32$ for $L \leq 64$. \textbf{Panels (a) and (b):} scale-invariant ratio $\bm{D}$ of the typical correlator defined in Eq.~\eqref{eq:Dtypical}. Locations of the crossings correspond to the critical points on each level. Insets: finite-size scaling collapses of $D$ with $\nu = \nu_{\rm RBIM}$.  \textbf{Panels (c) and (d):} fidelity correlators of spin-$\tfrac{1}{2}$ in system size $L=64$ for various $p$. The correlators decay exponentially at small $p$ and become long ranged for $p > p_c$. \textbf{Panels (e) and (f):} same as (c) and (d) but with spin-1. The correlators are long-ranged for any finite $p$. The horizontal dashed lines in panel (e) correspond to the lower bounds [Eq.~\eqref{eq:su2fidBounds}] for the solid lines of the same color.}
        \label{fig:SU2numerics}
\end{figure*}

The ${\rm SU}(2)$ case is qualitatively different, since its non-trivial center $\mathbb{Z}_2$ allows for a finite-$p$ transition. At the first level of the hierarchy, the integer-spin fidelity with $\Upsilon=1$ is long-ranged for any $p>0$, in agreement with the bounds in Eq.~\eqref{eq:su2fidBounds}, whereas the half-integer fidelity with $\Upsilon=\tfrac12$ decays exponentially at small $p$ and develops long-range order only above a critical error rate $p_c^{(1)}$. This is precisely the breaking pattern ${\rm O}(4)\to{\rm O}(3)\times\mathbb{Z}_2$ followed by the SSB of the residual $\mathbb{Z}_2$.

To locate the transition, we study the typical correlators of half-integer spins through the dimensionless ratio
\begin{equation}\label{eq:Dtypical}
    \bm{D}(p) = \frac{\exp \qty[\frac{1}{2}\overline{\log \langle \frac{1}{d_\Upsilon} \chi_{\Upsilon}(g_{L/2} g_0^{-1})\rangle}]}{\exp \qty[\frac{1}{2}\overline{\log \langle \frac{1}{d_\Upsilon} \chi_{\Upsilon}(g_{L/4} g_0^{-1})\rangle}]},
\end{equation}
which is related to the difference of Kullback-Leibler divergences between the corrupted states [see App.~\ref{app:addNumerics}]. To reduce statistical errors, the overline in Eq.~\eqref{eq:Dtypical} denotes an average over both the syndromes and several placements of the probe pair. The curves $\bm{D}(p)$ for different $L$ cross near $p\approx0.173$, with a slow drift of the crossing point with system size. A finite-size scaling (FSS) fit that includes the leading correction to scaling gives $p_c^{(1)}=0.1736(4)$ and $\nu=1.51(2)$, consistent with the Nishimori-Ising value $\nu_{\rm RBIM}=1.534(5)$~\cite{wan2026Nishimori}. The inset of Fig.~\ref{fig:SU2numerics}(a) shows the data collapse with $\nu$ fixed to $\nu_{\rm RBIM}$.

At the second level of the hierarchy, where the total irrep is measured on the pairs of neighboring sites in addition to the on-site irreps, we use the tensors from Eq.~\eqref{eq:tensor_I2} and follow the same procedure. We find a transition at $p_c^{(2)}=0.2024(6)$ with exponent $\nu = 1.47(5)$ [see Fig.~\ref{fig:SU2numerics}(b)]. The threshold lies above $p_c^{(1)}$, confirming that higher levels of the hierarchy have higher thresholds. The exponent is again consistent with Nishimori-Ising universality, as shown by the FSS collapse with $\nu = \nu_{\rm RBIM}$ in the inset of Fig.~\ref{fig:SU2numerics}(b).

We present additional numerical results and details in Appendix~\ref{app:addNumerics}. In particular, we further verify the Nishimori-Ising universality of the transitions by directly comparing the numerical data with the scaling function of the Nishimori RBIM. Additionally, we verify that the critical scaling of the MPS entanglement entropy is consistent with the Nishimori-Ising central charge $c_{\rm RBIM}=0.464(4)$~\cite{honecker2001universalityNishimori}.

\section{Discussion}\label{sec:discussion}

In this work, we have introduced a framework to study strong-to-weak spontaneous symmetry breaking for general compact Lie group or finite group $G$. The construction uses commuting two-site quantum channels that preserve the strong symmetry by creating pairs of conjugate irreps in their trivial fusion channel. While commutativity greatly simplifies the Abelian case, here pairs of higher-dimensional irreps carry entanglement, and the resulting density matrix is only block-diagonal in the local irrep basis. Consequently, unlike for Abelian groups, the quantum fidelity between two different initial symmetric states generally cannot be computed exactly.

Instead, we exploit the operational definition of the fidelity and construct a family of measurement protocols tailored to non-Abelian symmetries. Because the fusion of non-Abelian irreps is not unique, the total irrep of a region carries information beyond that of its local charges. Measuring it on progressively larger, nested regions (Fig.~\ref{fig:HierarMeas}) therefore extracts more information about the state, organizing the measurement protocols into a hierarchy. The corresponding distinguishability thresholds form a non-decreasing sequence of lower bounds on the intrinsic sw-SSB threshold [Eq.~\eqref{eq:chain_ineq}].

For every level $n$ of the hierarchy, we have expressed the syndrome probability distribution $p^{(n)}_\alpha$ for two initial states $\alpha=0,xy$ exactly as the partition function of a disordered statistical mechanics model [Eq.~\eqref{eq:level_general}]. The global singlet constraint resolves into a sum over internal fusion channels [Eq.~\eqref{eq:fusion_tree_expansion}], making explicit the information that is absent from the on-site syndrome record. Strikingly, the fluctuating fusion channels act as an effective additional dimension of the statistical mechanics model, making non-Abelian inference problems qualitatively different from the Abelian ones.

For groups like $S_3$ and ${\rm SU}(2)$, at any fixed level $n$ which doesn't scale with system size, we identify a hiding mechanism by which an inserted irrep can hide within higher-dimensional irreps created by the decoherence. This results in provably finite classical fidelities between $p^{(n)}_0$ and $p^{(n)}_1$ at any non-zero error rate, reflecting the spontaneous breaking of $G\times G$ down to
$Z(G)\times G_{\rm diag}$. We have analyzed this in detail for $S_3$ and SU$(2)$, combining analytical results with tensor-network simulations. For $S_3$, whose center is trivial, the fidelity for any non-trivial irrep insertion is lower bounded by a quantity proportional to the density of $2$-dimensional irreps, so there is no finite-$p$ transition. For ${\rm SU}(2)$, we similarly prove finite fidelities for integer spin insertions, while half-integer spins detect the spontaneous breaking of the center $\mathbb{Z}_2$ at a finite threshold, consistent with the Nishimori Ising universality. Simulating the first two levels of the hierarchy, we find that this threshold improves from $p_c^{(1)} \approx 0.173$ to $p_c^{(2)} \approx 0.202$, confirming the hierarchy.

\subsection{Open questions}

The central open question is whether the additional information obtained through fusion measurements at higher levels of the hierarchy provides tighter lower bounds on $p_c$. In particular, it is unclear whether the sequence $p_c^{(n)}$ saturates below the intrinsic transition or reaches it when the complete fusion tree is measured. For ${\rm SU}(2)$ and other continuous non-Abelian groups, even the existence of a finite threshold remains an open question. 

A related question is how the fusion tree becomes populated as the error rate increases. Our percolation argument relates nontrivial fusion channels at large scales to large physical error clusters. However, the height of an occupied fusion channel in the tree depends on the choice of tree. For example, a fixed partition into spatial regions can split a nearby entangled pair at successive levels, leaving nontrivial irreps high in the tree even at small error rates. Hence, nontrivial irreps high in a fixed tree do not by themselves signal large error clusters. The relevant construction should instead optimize the tree around the locations of measured nontrivial local irreps. Understanding how the occupied depth of such an adaptive tree scales with error rate and system size would clarify the proposed mechanism for $\mathrm{SU}(2)$ sw-SSB. Constructing the optimal tree requires a maximum-likelihood clustering, which poses another interesting challenge. Such an algorithm could use the output of the optimal decoders at lower levels to choose the geometry of the higher ones.

Resolving these questions requires contracting the relevant tensor networks at different levels of the hierarchy. At the first two levels, we found that the entanglement entropy of the boundary MPS obeys an area law away from criticality and grows logarithmically at the Nishimori critical points. It is unclear whether a transition to a volume-law phase could occur once the full $G\times G$ symmetry is restored. An intriguing possibility is that such a transition occurs at an error rate $p^\ast$ below the decodability threshold. In that case, $\rho$ and $\rho^\Upsilon_{xy}$ would remain distinguishable in principle given the syndrome information, but the optimal decoder could no longer be implemented efficiently using our tensor network contraction. Related entanglement obstructions to decoding have been found for the surface code under coherent unitary errors~\cite{bao2026surfacecode}. A similar mechanism is plausible here, since the creation of pairs of higher-dimensional irreps already generates coherences within local irrep sectors.

Another interesting direction is to construct and understand suboptimal decoders obtained from the zero-temperature limit of the optimal decoder, in which the sum over error configurations is replaced by the most likely one. For Abelian models, this procedure yields minimum-weight perfect matching~\cite{dennisTQM2002}. Higher-dimensional irreps, in contrast, impose additional constraints from the fusion rules on the resulting optimization problem~\cite{jing2026}. Such decoders are also of interest in their own right: they provide additional lower bounds on the threshold, and at high error rates they correspond to strongly disordered zero-temperature phases of the associated statistical mechanics models~\cite{dennisTQM2002, temkin2025charge}.

For Abelian topological quantum memories, faulty measurements introduce an additional temporal dimension, turning the two-dimensional statistical mechanics model into a three-dimensional one~\cite{dennisTQM2002}. In non-Abelian settings, in contrast, it has been argued that a successful strategy requires correcting errors on the fly, applying a correction as soon as sufficiently many measurements have accumulated to do so confidently~\cite{Wootton_14, Davydova_26, lyons_26}. The problem then no longer maps onto an equilibrium statistical mechanics model but becomes a genuine non-equilibrium control problem. Whether our hierarchical construction can be adapted to such dynamical decoding strategies remains an open question.

Finally, it would be interesting to extend the $\mathrm{U}(1)$ sw-SSB experiment of Wang et al.~\cite{wang2026swssb} to $\mathrm{SU}(2)$. Their quantum-gas-microscope experiment estimates nonlinear correlators from occupation snapshots and a classical Gaussian model of the dephased Fermi gas. In the non-Abelian setting, local irrep measurements leave quantum coherences unresolved, so extending this strategy requires identifying which measurements can diagnose the transition. Based on our framework, it is natural to investigate the effects of strongly $\mathrm{SU}(2)$-symmetric noise together with measurements of local irreps and regional fusion outcomes. Establishing which levels of the hierarchy are experimentally accessible would connect the distinguishability bounds to an implementable measurement protocol. 

\emph{Note added:} While preparing this manuscript, we became aware of a related work~\cite{sajith26} by Rohith Sajith, Yabo Li, Zijian Song, Tsung-Cheng Lu, and Carolyn Zhang, which appears in the same arXiv posting.

\section{Acknowledgments}

We acknowledge useful discussions with Jason Alicea, Ignacio Cirac, Matthew Fisher, Tarun Grover, Dian Jing, Anasuya Lyons, Michael J. Ragone, Nathaniel Selub, Avinash Vadali,  Robijn Vanhove, and Ruben Verresen. We also acknowledge usage of ChatGPT Sol-5.6 and Astra as well as Claude Opus 5 and 5.5 for assistance in numerical simulations and symbolic technical derivations. The authors take full responsibility for the results. This research was supported in part by a Simons Investigator award (E.A.), NSF QLCI program through Grant No. OMA-2016245 (E.A. and V.T.), U.S. Department of Energy, Office of Science, Office of High Energy Physics, under QuantISED Award DE-SC0019380 (E.A. and P.S.),
Gordon and Betty Moore Foundation Grant GBMF8688 (D.P.).
D.P. acknowledges support from the Israel Science Foundation (ISF), grant No. 2005/23.
C.X. acknowledges support from the Simons Foundation International through the Simons Investigator program. 
This research was done using services provided by the OSG Consortium \cite{osg1, osg2, osg3, osg4}, which is supported by the National Science Foundation awards \#2030508 and \#2323298.


%

\let\oldaddcontentsline\addcontentsline
\renewcommand{\addcontentsline}[3]{}

\appendix
\onecolumngrid

\section{Properties of the weights w} \label{app:weightW}

In this Appendix, we elaborate on the conditions imposed on the form of the decohered density matrix. In particular, using the density matrix $\rho$ given by
\begin{equation}\label{eq:rho_general_A}
    \rho = \int \mathcal{D}l \mathcal{D}r \prod_{\langle x,y\rangle} w(l_x, l_y,r_x,r_y)\; \ket{{\bf{l}}}\bra{{\bf{r}}}\,,
\end{equation}
we address the properties of the weight $w(l_x, l_y,r_x,r_y)$. Below, we derive constraints on the weight $w$ dictated by the properties of the decohered density matrix and a generic CPTP map. 

\begin{enumerate}
    \item Global strong $G$-left symmetry. 

    The desired density matrix is invariant under global transformations taking $l_v\to gl_v$ and $r_v\to hr_v$ for any $g,h\in G$. This is automatically satisfied if the kernel $w(l_x, l_y,r_x,r_y)$ is a function of only two arguments: $w(l_x, l_y,r_x,r_y) = w(l_x^{-1} l_y,r_x^{-1}r_y)$.

    \item The channel $\mathcal{E}$ preserves the trace. 
 
    The trace of the density matrix after application of the channel $\mathcal{E}$ is equal to 1 if $w(g, g) = 1$. This condition also implies \emph{unitality} of the channel, i.e., $\mathcal{E}(\mathds{1})=\mathds{1}$.

    \item $\rho$ is a \emph{self-adjoint} operator which corresponds to $w(l,r)^*=w(r,l)$.

    \item The channel $\mathcal{E}$ is completely positive.

    The complete positivity of the channel can be seen by applying the Choi-Jamio\l kowski isomorphism~\cite{choi1975map, jamiolkowski1972map}. Using the maximally entangled state between the system $S$ and the reference $R$
    \begin{equation}
        \ket{\Phi} = \int d(l_x, l_y)\,\ket{l_x, l_y}_S \otimes \ket{l_x, l_y}_R
    \end{equation}
    the channel gets mapped onto
    \begin{equation}
        \qty(\mathcal{E} \otimes \mathbbm{1}) \qty(\ket{\Phi} \bra{\Phi}) = \int d(l_x, l_y) \,d(r_x, r_y)\, w(l_x^{-1} l_y, r_x^{-1} r_y) \ket{l_x, l_y} \bra{r_x, r_y}_S \otimes \ket{l_x, l_y}\bra{r_x, r_y}_R.
    \end{equation}
    The complete positivity of the channel now corresponds to the positive semi-definiteness of this operator. The kernel is thus 
    \begin{equation}
        \int d(l_x, l_y)\, d(r_x, r_y) \,v(l_x, l_y)^*w(l_x^{-1} l_y,r_x^{-1} r_y)v(r_x, r_y)\geq 0
    \end{equation}
    for any complex vector $v$. This leads to the following Kraus form of the kernel:
    \begin{equation}
        w(l_x^{-1} l_y, r_x^{-1} r_y) = \sum_{\alpha} k_{\alpha}(l_x^{-1} l_y) k_{\alpha}^*(r_x^{-1} r_y), \qquad \sum_{\alpha} |k_\alpha(g)|^2 = 1,
    \end{equation}
    where $\alpha$ can take infinitely many values for continuous groups and at most $|G|$ for finite groups. 
\end{enumerate}

\section{Absence of spontaneous breaking of weak symmetry \label{app:weakSym}}

Since the local quantum channels $\mathcal{E}_{ij}$ are mutually commuting, the global channel $\mathcal{E}$ admits a local finite-depth purification by inserting an ancilla degree of freedom close to each bond. Hence, the $G$-left weak symmetry cannot be spontaneously broken at finite error rate $p$~\cite{fanDiagnostics2024}. Nonetheless, we will explicitly show that all standard two-point correlation functions of local operators that are charged under the $G$-left weak symmetry vanish. Indeed, notice that the density matrix $\rho$
\begin{equation}\label{eq:rho_general_B}
    \rho = \int \mathcal{D}l \mathcal{D}r \prod_{\langle x,y\rangle} w(l^{-1}_x l_y, r^{-1}_y r_x)\; \ket{{\bf{l}}}\bra{{\bf{r}}}\,,
\end{equation}
is invariant under the action of the weak symmetry $G_{\rm diag}$, i.e., ${\bf{L_g}}\rho {\bf{L_g^{-1}}}=\rho$. For any irrep $\Gamma \in \hat{G}$, the weak symmetry order parameter is given by the diagonal operator $Z_{ab}^\Gamma$:
\begin{equation} \label{eq:Z_Gamma}
    Z^\Gamma_{ab}=\sum_{g\in G}\, D^{\Gamma}_{ab}(g)\ket{g}\bra{g},\quad Z^\Gamma_{ab}=\int_G d\mu(g)\, D^{\Gamma}_{ab}(g)\ket{g}\bra{g}.
\end{equation} 
For any $g\in G$, under conjugation by the left action $L_g$, this operator transforms as
\begin{equation}\label{eq:Z_transform}
    L_{g} Z^\Gamma_{ab}\, L_{g^{-1}} = \sum_{c=1}^{d_\Gamma}D^\Gamma(g^{-1})_{ac}  Z^\Gamma_{cb}\,.
\end{equation}
For example, for $G=\mathbb{Z}_2$, the only non-trivial $Z^\Gamma$ operator is the Pauli $Z$, while for $G={\rm U}(1)$, these correspond to $e^{ik\hat{\theta}}=\int \frac{d\theta}{2\pi}e^{ik\theta}|\theta\rangle \langle \theta |$ for any $k\in \mathbb{Z}$. 

It is easy to see that the two-point correlation function of $Z^{\Gamma}$ vanishes:
\begin{align}
    \tr \qty[\rho \sum_{c=1}^{d_\Gamma}(Z^{\bar{\Gamma}}_X)_{ca}(Z^{{\Gamma}}_Y)_{cb}]=\sum_{c=1}^{d_\Gamma}\int dg_X D^\Gamma_{ac}(g_X^{-1})\int dg_Y D^\Gamma_{cb}(g_Y)=0.
\end{align}
In the first equality, we used that  the kernel satisfies $w(g,g)=1$ for any $g\in G$, and that $\langle D^\Gamma_{ab}| D^{\Gamma'}_{a'b'}\rangle=\delta_{\Gamma, \Gamma'}\delta_{a,a'}\delta_{b,b'}$. This result implies that the weak-symmetry order parameter never develops a finite expectation value and the weak symmetry cannot be broken spontaneously. Hence, $\rho$ can at most showcase a strong-to-weak spontaneous symmetry breaking transition, from $G_{\textrm{ket}}\times G_{\textrm{bra}}$ to its diagonal subgroup $(G_{\textrm{ket}}\times G_{\textrm{bra}})_{\textrm{diag}}$.

\section{Orthogonality and commutativity of \texorpdfstring{$P_\Gamma$}{P\_Gamma}}\label{app:transform}

In this Appendix, we explicitly verify the properties of the projectors $P_\Gamma=d_\Gamma\int dg\,\chi_\Gamma(g)L_{g}$ defined in Eq.~\eqref{eq:projectors}. First, we show that they satisfy $P_\Gamma P_{\Gamma'}= \delta_{\Gamma\Gamma'}P_\Gamma$. Using $L_{g}L_{h}=L_{gh}$ and the change of variables $k=gh$,
\begin{align}
    P_\Gamma P_{\Gamma'} &= d_\Gamma d_{\Gamma'}\!\int_G dg\,dh\;\chi_\Gamma(g)\,\chi_{\Gamma'}(h)\,L_{gh}\nonumber= d_\Gamma d_{\Gamma'}\!\int_G dk\,dh\;\chi_\Gamma(k h^{-1})\,\chi_{\Gamma'}(h)\,L_{k}\,.
\end{align}
Expanding the first character as $\chi_\Gamma(k h^{-1})=\sum_{m,n} D^{\Gamma}_{mn}(k)\,D^{\Gamma,*}_{mn}(h)$, we perform the integral over $h$: $\int dh\,D^{\Gamma, *}_{mn}(h)\,D^{\Gamma'}_{l l}(h) =\frac{\delta_{\Gamma\Gamma'}}{d_\Gamma}\delta_{mn}$. Plugging this result back into the original expression leads to
\begin{equation}
    P_\Gamma P_{\Gamma'}    =\delta_{\Gamma\Gamma'}\,d_\Gamma\!\int_G dk\,\chi_\Gamma(k)\,L_{k}=\delta_{\Gamma\Gamma'}\,P_\Gamma\,.
\end{equation}
The completeness relation $\sum_\Gamma P_\Gamma=\mathbbm{1}$ follows from the character completeness identity $\sum_\Gamma d_\Gamma\,\chi_\Gamma(g)=|G|\,\delta_{g,e}$, upon noting that $L_{e}=\mathbbm{1}$.

It is also easy to see that $P_\Gamma$ commutes with the $G$-left symmetry generator. Indeed, for any $h\in G$ the rearrangement theorem gives
\begin{equation}
    L_{h}\,P_\Gamma = d_\Gamma\!\int dg\,\chi_\Gamma(g)\,L_{hg} \overset{g\to h^{-1}gh}{=} d_\Gamma\!\int dg\, \chi_\Gamma(h^{-1}gh)\,L_{gh} = P_\Gamma\,L_{h}\,,
\end{equation}
where in the last equality we used the character property $\chi_\Gamma(h^{-1}gh)=\chi_\Gamma(g)$. Hence, $[P_\Gamma,L_{h}]=0$ for all $h\in G$ and $P_\Gamma$ is a $G$-symmetric observable.

Finally, we prove that two ``region''-irrep projectors commute if the domain of action of one of them is completely contained in the other's domain of action. That is, for $A\subseteq B$,
\begin{equation}\label{eq:appK_commute}
    [P_{\Gamma_A},P_{\Gamma_B}]=0\,,
\end{equation}
with $P_{\Gamma_R}=d_{\Gamma_R}\!\int dg\,\chi_{\Gamma_R}(g)\,\prod_{v\in R}L_{g,v}$ defined in Eq.~\eqref{eq:P_region}. Using the explicit form of $P_{\Gamma_R}$,
\begin{align}
    P_{\Gamma_A} P_{\Gamma_B} &=d_{\Gamma_A} d_{\Gamma_B}\!\int dg_A\,dg_B\; \chi_{\Gamma_A}(g_A)\,\chi_{\Gamma_B}(g_B)\, \prod_{a\in A}L_{g_A g_B,\,a} \prod_{b\in B\setminus A}L_{g_B,\,b}\,,
\end{align}
where we used $L_{g_A,a}L_{g_B,a}=L_{g_A g_B,a}$ for sites in $A$. We now perform the change of variables $g_A=g_B h_A g_B^{-1}$ (so that $g_A g_B=g_B h_A$) at fixed $g_B$. The Haar measure is invariant, so $dg_A=dh_A$, and the same character property $\chi_{\Gamma_A}(g_B h_A g_B^{-1})=\chi_{\Gamma_A}(h_A)$ as before allows us to drop the conjugation:
\begin{align}
    P_{\Gamma_A} P_{\Gamma_B} &=d_{\Gamma_A} d_{\Gamma_B}\!\int dh_A\,dg_B\; \chi_{\Gamma_A}(h_A)\,\chi_{\Gamma_B}(g_B)\, \prod_{a\in A}L_{g_B h_A,\,a}\prod_{b\in B\setminus A}L_{g_B,\,b}\nonumber\\
    &=d_{\Gamma_A} d_{\Gamma_B}\!\int dh_A\,dg_B\; \chi_{\Gamma_A}(h_A)\,\chi_{\Gamma_B}(g_B)\, \prod_{b\in B}L_{g_B,\,b}\,\prod_{a\in A}L_{h_A,\,a}\nonumber\\
    &= P_{\Gamma_B}\,P_{\Gamma_A}.
\end{align}

Importantly, for two regions $R,R'$ that \emph{partially} overlap (neither is contained in the other), $[P_{\Gamma_R},P_{\Gamma_{R'}}]\neq 0$ in general. Take, for example, two two-site regions $R=\{1,2\}$ and $R'=\{2,3\}$ sharing site $2$ on a chain. The product $P_{\Gamma_R}P_{\Gamma_{R'}}$ involves $L_{g_{R} g_{R'},2}L_{g_R,1}L_{g_{R'},3}$, which cannot be resolved by the same change of variables as before. 
This is why the syndrome hierarchy of Sec.~\ref{sec:hierarchy} is defined on partitions of the lattice into non-overlapping regions at each level $N$: the projectors at the same level cover disjoint regions and trivially commute, while the nested-region result \eqref{eq:appK_commute} guarantees compatibility across levels.

\section{On-site syndrome distribution}
\label{app:rho_derivation}

This Appendix details the derivation of the on-site syndrome probability [Eq.~\eqref{eq:p0}]. We compute $P(\{\Gamma_v\})=\mathrm{tr}(\rho\prod_v P_{\Gamma_v})$ by inserting $P_{\Gamma_v}=d_{\Gamma_v}\!\int dh_v\,\chi_{\Gamma_v}(h_v)L_{h_v,v}$. By $L_{h_v}\!\ket{l_v}=\ket{h_v l_v}$ and the orthonormality $\braket{r_v}{h_v l_v}=\delta(h_v - r_v l_v^{-1})$, the trace produces a delta function on each vertex, fixing $h_v=r_v l_v^{-1}$. Hence,
\begin{equation}\label{eq:appD_Psynd1}
    P(\{\Gamma_v\}) = \int\!\mathcal{D}l\,\mathcal{D}r\; \prod_{\langle xy\rangle}w\left(l_x^{-1}l_y r_y^{-1}r_x\right)\, \prod_v d_{\Gamma_v}\,\chi_{\Gamma_v}(r_v l_v^{-1})\,.
\end{equation}
After introducing $g_v = r_v l_v^{-1}$, we can rewrite the argument of each link weight $w$
\begin{equation}
    l_x^{-1}l_y r_y^{-1}r_x = l_x^{-1}l_y l_y^{-1}g_y^{-1}g_xl_x = l_x^{-1}(g_y^{-1}g_x)l_x\,.
\end{equation}
Thus the character is evaluated on an element conjugate to $g_y^{-1}g_x$. Since characters are class functions, this is equivalently $\chi_\Gamma(g_xg_y^{-1})$ with the orientation convention used in the Main Text. The remaining $l$-integrals factorize and integrate to unity, leaving
\begin{equation}\label{eq:appD_Psynd2}
    P(\{\Gamma_v\}) = \int\!\mathcal{D}g\;W({\bf{g}})\, \prod_v d_{\Gamma_v}\,\chi_{\Gamma_v}(g_v)\,,\qquad W({\bf{g}})=\prod_{\langle xy\rangle}w(g_x g_y^{-1})\,,
\end{equation}
which is the first line of Eq.~\eqref{eq:p0}.

By construction, $\rho^\Upsilon_{xy}=\sum_{a,b} K^\Gamma_{ab,\langle xy\rangle}\rho K^{\Gamma\,\dagger}_{ab,\langle xy\rangle}$, where the Kraus operators include the normalization in Eq.~\eqref{eq:Kraus}. Following the same steps as above, but with one extra Kraus operator acting on the sites $x$ and $y$, we obtain a similar expression with the only modification of an extra factor of $\frac{1}{d_\Gamma}\chi_\Gamma(g_x g_y^{-1})$ inside the integrand:
\begin{equation}
    p^{(1)}_{xy}(\{\Gamma_v\}) = \int\!\mathcal{D}g\;W({\bf{g}})\, \prod_v d_{\Gamma_v}\chi_{\Gamma_v}(g_v)\; \frac{\chi_\Gamma(g_x g_y^{-1})}{d_\Gamma}\, = \expval{\frac{1}{d_\Gamma} \chi_\Gamma(g_x g_y^{-1})},
\end{equation}
which is the second line of Eq.~\eqref{eq:p0}.

\section{Density matrix in the Peter--Weyl basis}\label{app:coherences}

Here we derive Eq.~\eqref{eq:coherences} of the main text starting from \eqref{eq:rho_explicit} and changing variables to the irrep basis. This result explicitly demonstrates the coherent block-diagonal structure of the density matrix.

We start by writing the decohered density matrix Eq.~\eqref{eq:rho_explicit} as
\begin{equation}\label{eq:appE_rho_pre}
    \rho = \int\!\mathcal{D}l\,\mathcal{D}g\; \prod_{\langle xy\rangle}w(g_x g_y^{-1})\; \ket{\{l_v\}}\!\bra{\{g_v l_v\}}\,.
\end{equation}
Next, we expand the states in the irrep basis using
\begin{equation}
    \ket{l_v}=\sum_{J,m,n}\sqrt{d_J}\,D^{J\,*}_{m n}(l_v)\,\ket{J,m,n}\,, \quad \bra{g_v l_v}=\sum_{J',m',n'}\sqrt{d_{J'}}\, D^{J'}_{m'n'}(g_v l_v)\,\bra{J',m',n'}\,.
\end{equation}
Combined with $D^{J'}_{m'n'}(g_v l_v)=\sum_k D^{J'}_{m' k}(g_v)\,D^{J'}_{k n'}(l_v)$, Eq.~\eqref{eq:appE_rho_pre} becomes
\begin{align}
    \rho =\!\int\!\mathcal{D}g\;W({\bf{g}})\;\bigotimes_v\,\left[\!\sum_{J,m,n}\sum_{J',m',n',k}\!\sqrt{d_J d_{J'}}\,D^{J'}_{m'k}(g_v)\, \left(\!\int dl_v\,D^{J\,*}_{mn}(l_v)D^{J'}_{k n'}(l_v)\right) \ket{J,m,n}\!\bra{J',m',n'}\right]\,.
\end{align}
The $l_v$ integral gives $\delta_{J,J'}\,\delta_{m,k}\,\delta_{n,n'}/d_J$:
\begin{equation}\label{eq:appE_rho_PW}
    \rho =\!\int\!\mathcal{D}g\;W({\bf{g}})\,\bigotimes_v\,\Big[\!\sum_J\sum_{m,m',n}D^J_{m'm}(g_v)\,\ket{J,m,n}\!\bra{J,m',n}\Big]\,.
\end{equation}

Equation \eqref{eq:appE_rho_PW} can be re-organized by separating the irrep labels (which are diagonal) from the magnetic indices on the left (which carry the residual coherences) and the right (which are also diagonal):
\begin{equation}\label{eq:appE_main}
    {\rho = \!\int\!\mathcal{D}g\;W({\bf{g}})\, \sum_{\{J_v\}}\,\ket{\{J_v\}}\!\bra{\{J_v\}} \bigotimes_v\!\bigg(\!\sum_{m_v,m'_v} D^{J_v}_{m'_v m_v}(g_v)\,\ket{m_v}\!\bra{m'_v}\bigg)\,,}
\end{equation} reproducing Eq.~\eqref{eq:coherences} of the main text (the right magnetic indices $n_v$ have been absorbed into a degenerate identity, since the kernel does not depend on them). The off-diagonal coherences $D^J_{m'm}(g)$ with $m\neq m'$ vanish for Abelian groups (where $d_J=1$ and $D^J(g)=\chi_J(g)$ is a phase), and this is why for Abelian symmetries $\rho$ is fully diagonal in the irrep basis. For non-Abelian groups the coherences are intrinsic and make exact diagonalization intractable; this is the technical reason why we resort to evaluating the syndrome distribution \eqref{eq:appD_Psynd2} rather than the spectrum of $\rho$ itself.

Using \eqref{eq:appE_main} together with the on-site projector $P_{J_v}=\ket{J_v}\!\bra{J_v}\otimes\sum_{m,n}\ket{m,n}\!\bra{m,n}$ (Appendix~\ref{app:transform}), we find for the syndrome probability:
\begin{equation}\label{eq:appE_probJ}
    \mathrm{tr}\left(\rho\prod_v P_{J_v}\right) = \!\int\!\mathcal{D}g\;W({\bf{g}})\, \prod_v d_{J_v}\chi_{J_v}(g_v)\,,
\end{equation}
in agreement with \eqref{eq:appD_Psynd2}. The factor of $d_{J_v}$ arises from $\mathrm{tr}_{m,n}\big(\sum_{m,m'}D^{J_v}_{m'm}(g_v) \ket{m_v}\!\bra{m'_v}\big)=\chi_{J_v}^*(g_v)\sum_n 1=d_{J_v}\chi^*_{J_v}(g_v)$.

\section{Hiding and a lower bound on the fidelity correlator}
\label{app:hidingMech}

An irrep $\Upsilon$ {\em hides} in an irrep $R$ if $R\otimes \Upsilon\supset R$ or, equivalently, if $N_{R\Upsilon}^R\ge 1$.  Thus, a site in irrep $R$ can absorb an inserted $\Upsilon$ through a fusion channel that leaves its measured irrep unchanged.  The relevant fusion multiplicities satisfy
\begin{equation}
    N^{\Upsilon}_{R\bar{R}}=\int dg\, |\chi_{R}(g)|^2\chi^*_\Upsilon(g)=(N_{R\Upsilon}^R)^*=N^{R}_{R\Upsilon},
\end{equation}
where the last equality follows because fusion multiplicities are real. Hence,  $\Upsilon$ hides in $R$ if and only if $\Upsilon\subset R\otimes \bar{R}$.

Two consequences follow. (i) Every irrep with $d_R>1$ hides at least one nontrivial irrep. Indeed, $R\otimes\bar{R}$ has dimension $d_R^2>1$, while the trivial irrep $\mathbf{1}$ occurs exactly once, by Schur's lemma, so $R\otimes \bar{R}$ contains at least one nontrivial irrep, which hides in $R$. (ii) The center $Z(G)$ acts trivially on every irrep that hides somewhere.  For $z\in Z(G)$, Schur's lemma gives $D^R(z)=\lambda_R(z) 1_R$, with $|\lambda_R(z)|=1$, where $1_R$ is the identity on $R$. Therefore,
\begin{equation}
    D^{R\otimes\bar{R}}(z)=\lambda_R(z)\qty(\lambda_R(z))^* \,1_{R\otimes\bar{R}}=1_{R\otimes\bar{R}},
\end{equation}
so $z$ acts trivially on every irreducible component of ${R\otimes\bar{R}}$.

We now use the hiding mechanism to bound the fidelity correlator from below.  The probe goes unnoticed when both $x$ and $y$ carry irreps in which $\Upsilon$ hides, and both absorb it (for a non-self-dual $\Upsilon$, the endpoint at $y$ carries $\bar{\Upsilon}$, which hides in $R$ equally often, since $N^R_{R\bar{\Upsilon}}=N^R_{R\Upsilon}$).  Let $\rho_R=p(\Gamma_x=R)$ be the probability that a site carries the irrep $R$.  For distant $x$ and $y$, the irreps measured at the two sites are independent.  Given that site $x$ carries $R$, the inserted $\Upsilon$ fuses with it back into $R$ with probability $N^R_{R\Upsilon}d_R/(d_R d_\Upsilon)=N^R_{R\Upsilon}/d_\Upsilon$, the fraction of the states of $R\otimes\Upsilon$ that lie in $R$.  
An endpoint that carries $R$ with probability $\rho_R$ without the probe therefore carries it with probability at least $\rho_R N^R_{R\Upsilon}/d_\Upsilon$ with the probe, so the fidelity of the syndrome at that endpoint is at least $\sum_R\rho_R\sqrt{N^R_{R\Upsilon}/d_\Upsilon}$.  When $x$ and $y$ are far apart, the fidelities at the two endpoints multiply, giving
\begin{equation}\label{eq:FboundGen}
    \lim_{|x-y|\to\infty}F_\Upsilon^C(p^{(1)},p_{xy}^{(1)})\ge \left[\sum_R\rho_R\sqrt{\frac{N^R_{R\Upsilon}}{d_\Upsilon}}\right]^2.
\end{equation}
The right-hand side is positive for every $p>0$ whenever $\Upsilon$ hides in an irrep that occurs with nonzero probability.  We justify the counting below, where we also show that it holds as long as the error bonds do not percolate, which includes all small $p$.

To make the counting rigorous, we first condition on the error configuration $E$, i.e., the irreps inserted on the bonds. By the Cauchy--Schwarz inequality, $F_\Upsilon^C\ge\sum_E p(E)F_\Upsilon(E)$, where $F_\Upsilon(E)$ is the fidelity between the syndrome distributions conditioned on $E$.  For fixed $E$, the state factorizes over error clusters (connected components of error bonds), and both the state of each cluster and its syndrome projectors are $G$-invariant. Clusters containing neither $x$ nor $y$ are unaffected by the probe and contribute a factor $1$ to the conditional fidelity. Two facts then follow from Schur's lemma.  First, conditioned on the other syndrome outcomes, the state of site $x$ is maximally mixed within each irrep space, and the same holds separately at $y$.  Thus, if either endpoint carries the irrep $R$, fusing in $\Upsilon$ returns it to $R$ with probability $N^R_{R\Upsilon}/d_\Upsilon$.  Second, if $x$ and $y$ belong to different clusters, then for each syndrome outcome on one cluster the operator left on its end of the probe singlet is a multiple of the identity, so the singlet does not correlate the two clusters.  The counting therefore holds for every $E$ in which $x$ and $y$ lie in different clusters.  At large separation, the probability that $x$ and $y$ belong to the same cluster vanishes whenever the error bonds do not percolate, i.e., when $\sum_{\Gamma\ne\mathbf{1}}p_\Gamma$ does not exceed the bond-percolation threshold (1/2 for the square lattice).  In this limit, the two cluster contributions also become independent, and averaging over $E$ gives  Eq.~(\ref{eq:FboundGen}).  In particular, the bound holds for arbitrarily small $p>0$, which is sufficient to establish that the threshold vanishes, $p_c^{(1)}=0$.

The densities $\rho_R$ follow directly from the syndrome distribution.  Summing over the syndromes of all sites other than $x$ and using $\sum_\Gamma d_\Gamma\chi_\Gamma(g)=\delta_{g,e}$ pins their group elements to the identity, leaving
\begin{equation}\label{eq:rhoR}
    \rho_R=\int dg\, [w(g)]^z d_R \chi_R(g),
\end{equation}
where $w(g)=\sum_\Gamma (p_\Gamma/d_\Gamma)\,\mathrm{Re}\,\chi_\Gamma(g)$ is the bond weight and $z=4$ is the coordination number of the square lattice.  The same argument shows that the irreps of two non-adjacent sites are independent.

\begin{table}[t]
    \centering
    \renewcommand{\arraystretch}{1.2}
    \begin{tabular}{c|ccc}
        & $\{e\}$ & $\{s,rs,r^2s\}$ & $\{r,r^2\}$ \\
        \hline
        Class size & 1 & 3 & 2 \\
        \hline
        $\chi_{1}$ & 1 &  1 &  1 \\
        $\chi_{1'}$ & 1 & -1 &  1 \\
        $\chi_2$     & 2 &  0 & -1
    \end{tabular}
    \caption{Character table of $S_3$~\cite{etingof2011representation}, with generators $r=(123)$ and $s=(23)$.}
    \label{tab:s3-character-table}
\end{table}

For $S_3$ with $p_{\mathbf{1}'}=0$ and $p_{\mathbf{2}}=p$, Eq.~(\ref{eq:rhoR}) and the character table of $S_3$ (Table~\ref{tab:s3-character-table}) give $\rho_{\mathbf{2}}=\frac{2}{3}\left[1-(1-\frac{3p}{2})^4\right]$.  The sign irrep $\mathbf{1}'$ hides only in $\mathbf{2}$, and since $\chi_{\mathbf{2}}\chi_{\mathbf{1}'}=\chi_{\mathbf{2}}$, inserting $\mathbf{1}'$ at a site in irrep $\mathbf{2}$ leaves the integrand of the syndrome probability unchanged.  Hence $p_{xy}^{(1)}=p^{(1)}$ for every syndrome with $\Gamma_x=\Gamma_y=\mathbf{2}$, and for any non-adjacent $x$, $y$ and any $p$,
\begin{equation}
    F_{\mathbf{1}'}^C\ge \sum_{\{\Gamma_v\}:\Gamma_x=\Gamma_y=\mathbf{2}}p^{(1)}(\{\Gamma_v\})=\rho_\mathbf{2}^2\simeq 16p^2.
\end{equation}
The irrep $\mathbf{2}$ also hides only in $\mathbf{2}$, with $N^{\mathbf{2}}_{\mathbf{2}\mathbf{2}}=1$, so Eq.~(\ref{eq:FboundGen}) gives $F_{\mathbf{2}}^C\ge\frac{1}{2}\rho_\mathbf{2}^2\simeq 8p^2$.

For SU(2) with spin-$\frac{1}{2}$ errors, each of the $n$ error bonds touching a site contributes half of a singlet, so the site has total spin $R$ with probability given by the multiplicity of $R$ in $(\frac{1}{2})^{\otimes n}$, times $d_R/2^n$.  Summing over $n$ with binomial weights gives
\begin{equation}
    \rho_{1/2}=4p(1-p)^3+2p^3(1-p),\quad \rho_1=\frac{9}{2}p^2(1-p)^2+\frac{9}{16}p^4,\quad \rho_{3/2}=2p^3(1-p),\quad \rho_2=\frac{5}{16}p^4.
\end{equation}
Half-integer spins never hide, while an integer spin $\Upsilon$ hides in every spin $R\ge\Upsilon/2$, with $N^R_{R\Upsilon}=1$.  Eq.~(\ref{eq:FboundGen}) then gives $F_\Upsilon^C\ge\frac{1}{2\Upsilon+1}\left[\sum_{R\ge\Upsilon/2}\rho_R\right]^2$, i.e.,
\begin{equation}
    F_1^C\ge \frac{1}{3}\left[4p-\frac{15}{2}p^2+7p^3-\frac{21}{8}p^4\right]^2\simeq\frac{16}{3}p^2,\qquad F_2^C\ge \frac{1}{5}\left[\frac{9}{2}p^2-7p^3+\frac{27}{8}p^4\right]^2\simeq\frac{81}{20}p^4.
\end{equation}
At small $p$, the dominant events in which the probe goes unnoticed are a single error bond at $x$ and another one at $y$, which are exactly the events counted above.  The bounds on both $S_3$ fidelities and on $F_1^C$ for SU(2) are therefore saturated to leading order in $p$.

The bounds on $F_{\mathbf{2}}^C$ for $S_3$ and on $F_1^C$, $F_2^C$ for SU(2) follow from Eq.~(\ref{eq:FboundGen}) and therefore hold for $p\le 1/2$, while the bound on $F_{\mathbf{1}'}^C$ holds for all $p$.

\section{Example supporting strict information gain.} 
\label{app:Example_4sites}

As discussed in the Main Text, the classical fidelity of the joint probability distributions of the measurement outcomes on 2 levels of the hierarchy $p_{0, 1}(X, Y)$ is less or equal to the classical fidelity of the probability distributions of a single-layer measurement outcome $p_{0, 1}(X)$. The equality is only achieved when the conditional distributions are the same, that is, $p(Y|X) = p_{xy}(Y|X)$. In this Appendix, we construct an explicit example where this doesn't hold, implying a strict inequality of the fidelities. 

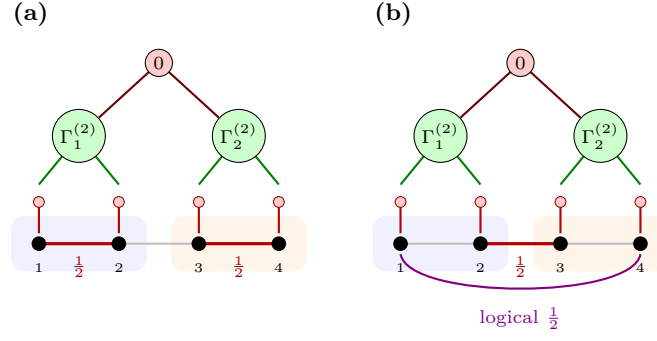
\begin{figure}[t]
    \centering
    \begin{tikzpicture}[>=stealth, scale=0.92,
        site/.style={circle, draw, fill=black, inner sep=1.8pt},
        errlink/.style={very thick, red!70!black},
        nolink/.style={thick, black!25},
        fnode/.style={circle, draw, fill=green!20, inner sep=1pt, font=\scriptsize},
        rnode/.style={circle, draw, fill=red!20, inner sep=1.5pt, font=\scriptsize\bfseries}
    ]
    \def\sp{1.15}  
 
    \begin{scope}[shift={(0,0)}]
    \node[font=\bfseries\small, anchor=north west] at (-0.5,3.6) {(a)};
 
    \fill[blue!6, rounded corners=5pt] (-0.4,-0.4) rectangle (1*\sp+0.4,0.4);
    \fill[orange!8, rounded corners=5pt] (2*\sp-0.4,-0.4) rectangle (3*\sp+0.4,0.4);
 
    \draw[errlink] (0,0) -- node[font=\tiny, red!60!black, below=1pt] {$\frac{1}{2}$} (\sp,0);
    \draw[nolink] (\sp,0) -- (2*\sp,0);
    \draw[errlink] (2*\sp,0) -- node[font=\tiny, red!60!black, below=1pt] {$\frac{1}{2}$} (3*\sp,0);
 
    \foreach \i in {0,...,3}{ \node[site] (S\i) at (\i*\sp, 0) {}; }
    \node[font=\tiny, below=4pt] at (S0) {1};
    \node[font=\tiny, below=4pt] at (S1) {2};
    \node[font=\tiny, below=4pt] at (S2) {3};
    \node[font=\tiny, below=4pt] at (S3) {4};
 
    \foreach \i in {0,...,3}{
        \draw[thick, red!70!black] (\i*\sp, 0.15) -- (\i*\sp, 0.6);
        \filldraw[fill=red!20, draw=red!60!black] (\i*\sp, 0.6) circle (2.2pt);
    }
 
    \draw[thick, green!50!black] (0*\sp, 0.85) -- (0.5*\sp, 1.55);
    \draw[thick, green!50!black] (1*\sp, 0.85) -- (0.5*\sp, 1.55);
    \draw[thick, green!50!black] (2*\sp, 0.85) -- (2.5*\sp, 1.55);
    \draw[thick, green!50!black] (3*\sp, 0.85) -- (2.5*\sp, 1.55);
    \node[fnode] (FR0) at (0.5*\sp, 1.55) {$\Gamma_1^{(2)}$};
    \node[fnode] (FR1) at (2.5*\sp, 1.55) {$\Gamma_2^{(2)}$};
    \draw[thick, red!40!black] (FR0) -- (1.5*\sp, 2.6);
    \draw[thick, red!40!black] (FR1) -- (1.5*\sp, 2.6);
    \node[rnode] at (1.5*\sp, 2.6) {$0$};
    \end{scope}
 
    \begin{scope}[shift={(5.2,0)}]
    \node[font=\bfseries\small, anchor=north west] at (-0.5,3.6) {(b)};
 
    \fill[blue!6, rounded corners=5pt] (-0.4,-0.4) rectangle (1*\sp+0.4,0.4);
    \fill[orange!8, rounded corners=5pt] (2*\sp-0.4,-0.4) rectangle (3*\sp+0.4,0.4);
 
    \draw[nolink] (0,0) -- (\sp,0);
    \draw[errlink] (\sp,0) -- node[font=\tiny, red!60!black, below=1pt] {$\frac{1}{2}$} (2*\sp,0);
    \draw[nolink] (2*\sp,0) -- (3*\sp,0);
 
    \foreach \i in {0,...,3}{ \node[site] (S\i) at (\i*\sp, 0) {}; }
    \node[font=\tiny, below=4pt] at (S0) {1};
    \node[font=\tiny, below=4pt] at (S1) {2};
    \node[font=\tiny, below=4pt] at (S2) {3};
    \node[font=\tiny, below=4pt] at (S3) {4};
 
    \draw[thick, violet] (0, -0.15)
        to[out=-80, in=-100, looseness=0.5] (3*\sp, -0.15);
    \node[font=\scriptsize, violet, below=1pt] at (1.5*\sp, -0.75) {logical $\frac{1}{2}$};
 
    \foreach \i in {0,...,3}{
        \draw[thick, red!70!black] (\i*\sp, 0.15) -- (\i*\sp, 0.6);
        \filldraw[fill=red!20, draw=red!60!black] (\i*\sp, 0.6) circle (2.2pt);
    }
 
    \draw[thick, green!50!black] (0*\sp, 0.85) -- (0.5*\sp, 1.55);
    \draw[thick, green!50!black] (1*\sp, 0.85) -- (0.5*\sp, 1.55);
    \draw[thick, green!50!black] (2*\sp, 0.85) -- (2.5*\sp, 1.55);
    \draw[thick, green!50!black] (3*\sp, 0.85) -- (2.5*\sp, 1.55);
    \node[fnode] (FR0) at (0.5*\sp, 1.55) {$\Gamma_1^{(2)}$};
    \node[fnode] (FR1) at (2.5*\sp, 1.55) {$\Gamma_2^{(2)}$};
    \draw[thick, red!40!black] (FR0) -- (1.5*\sp, 2.6);
    \draw[thick, red!40!black] (FR1) -- (1.5*\sp, 2.6);
    \node[rnode] at (1.5*\sp, 2.6) {$0$};
    \end{scope}
    \end{tikzpicture}
    \caption{
    \textbf{Four-site chain: dominant error configurations and fusion trees.}
    Sites $\{1,2,3,4\}$ on an open chain with all on-site syndromes $\Gamma_v = \frac{1}{2}$ (red legs).
    Level-2 blocks $\{1,2\}$ (blue shading) and $\{3,4\}$ (orange shading).
    Active error links are shown in red; the fusion tree above shows the level-2 irreps $\Gamma_1^{(2)}$, $\Gamma_2^{(2)}$ (green) and the singlet root (red).
    \textbf{(a)}~Vacuum $\rho_0$: at small $p$, the dominant error has two within-block active links.
    \textbf{(b)}~Inserted state $\rho_1$: an additional irrep pair (violet line, below) is inserted between sites~1 and~4.
    The dominant error at small~$p$ has one active link on $(2,3)$.
    }
    \label{fig:4site}
\end{figure}

Let us consider an ${\rm SU}(2)$-left strongly symmetric channel as described in Sec.~\ref{sec:SU2} on a one-dimensional system of size $L=4$ with open boundary conditions [see Fig.~\ref{fig:4site}]. Consider the configuration for level-$1$ of the hierarchy with all on-site irreps $\Gamma_v = \frac{1}{2}$ (red legs), and the level-$2$ blocks $R_1=\{1,2\}$ (blue shading) and $R_2 = \{3,4\}$ (orange shading).

The conditional probabilities of the syndrome on the second level for the logical state $\rho_0$ are straightforward to find. Since each of the fusion blocks $R_1$ and $R_2$ contains an entire Bell pair, their measurements always return $\Gamma_1^{(2)} = 0$, $\Gamma_2^{(2)} = 0$, that is,
\begin{equation}
    p\qty(\Gamma^{(2)} = \qty(0, 0)\left| \Gamma^{(1)} = \qty(\frac{1}{2}, \frac{1}{2}, \frac{1}{2}, \frac{1}{2})\right. ) = 1, \quad p\qty(\Gamma^{(2)} = \qty(1, 1)\left| \Gamma^{(1)} = \qty(\frac{1}{2}, \frac{1}{2}, \frac{1}{2}, \frac{1}{2})\right. ) = 0.
\end{equation}
The probability of $\Gamma^{(2)} = (0,1)$ is trivially $0$ since this combination doesn't transform as a singlet under the action of ${\rm SU}(2)$.

The probabilities for logical $\rho_1$ can be computed using a simple geometric transformation. On the second level, sites 1 and 2 (3 and 4) effectively merge into a site $1^\prime$ ($2^\prime$) with 2 incident errors. Thus, the probabilities are given by the probabilities of observing different syndromes on the following error-graph:

\begin{equation}
    \raisebox{-12.5pt}{$
    \vcenter{\hbox{
    \begin{tikzpicture}[>=stealth, scale=0.92,
        site/.style={circle, draw, fill=black, inner sep=1.8pt},
        errlink/.style={very thick, red!70!black},
        nolink/.style={thick, black!25}
    ]
        \def\sp{1.15}
        
        \fill[blue!6, rounded corners=5pt] (-0.4,-0.4) rectangle (1*\sp+0.4,0.4);
        \fill[orange!8, rounded corners=5pt] (2*\sp-0.4,-0.4) rectangle (3*\sp+0.4,0.4);

        \foreach \i in {0,...,3}{ \node[site] (S\i) at (\i*\sp, 0) {}; }
        \node[font=\tiny, below=4pt] at (S0) {1};
        \node[font=\tiny, below=4pt] at (S1) {2};
        \node[font=\tiny, below=4pt] at (S2) {3};
        \node[font=\tiny, below=4pt] at (S3) {4};

        \draw[nolink] (S0) -- (S1);
        \draw[errlink] (S1) -- node[font=\tiny, red!60!black, below=1pt] {$\frac{1}{2}$} (S2);
        \draw[nolink] (S2) -- (S3);

        \draw[thick, violet] (S0) to[out=-80, in=-100, looseness=0.5] (S3);
        \node[font=\scriptsize, violet, below=1pt] at (1.5*\sp, -0.75) {logical $\frac{1}{2}$};
    \end{tikzpicture}
    }}$}
    \quad \rightarrow \quad
    \vcenter{\hbox{
    \begin{tikzpicture}[>=stealth, scale=0.92,
        site/.style={circle, draw, fill=black, inner sep=1.8pt},
        errlink/.style={very thick, red!70!black}
    ]
        \def\sp{1.5}
        
        \node[site] (S1p) at (0, 0) {};
        \node[site] (S2p) at (\sp, 0) {};
        \node[font=\tiny, below=4pt] at (S1p) {$1'$};
        \node[font=\tiny, below=4pt] at (S2p) {$2'$};

        \draw[errlink] (S1p) to[bend left=45] node[font=\tiny, red!60!black, above=2pt] {$\frac{1}{2}$} (S2p);
        \draw[thick, violet] (S1p) to[bend right=45] node[font=\scriptsize, violet, below=2pt] {logical $\frac{1}{2}$} (S2p);
    \end{tikzpicture}
    }}
\end{equation}

The probabilities for the fusion outcomes on the sites $1^\prime$ and $2^\prime$, given this error configuration $E^\prime$, can be found as
\begin{equation}
\begin{split}
    p(\Gamma_{1^\prime} = 0, \Gamma_{2^\prime} = 0 \left| E^\prime \right.) &= \int dg_{1^\prime} dg_{2^\prime} \frac{1}{2^2}\qty[\chi_{12}(g_{1^\prime} g_{2^\prime}^{-1})]^2 = \frac{1}{4}, \\
    p(\Gamma_{1^\prime} = 1, \Gamma_{2^\prime} = 1 \left| E^\prime \right.)  &= \frac{3}{4},
\end{split}  
\end{equation}
which results in
\begin{equation}
    p_{xy}\qty(\Gamma^{(2)} = \qty(0, 0)\left| \Gamma^{(1)} = \qty(\frac{1}{2}, \frac{1}{2}, \frac{1}{2}, \frac{1}{2})\right. ) = \frac{1}{4}, \quad p_{xy}\qty(\Gamma^{(2)} = \qty(1, 1)\left| \Gamma^{(1)} = \qty(\frac{1}{2}, \frac{1}{2}, \frac{1}{2}, \frac{1}{2})\right. ) = \frac{3}{4}.
\end{equation}
Thus, the conditional probabilities for the two logical states differ and the fidelity decays on the second level of the hierarchy. However, we do not claim that this is the typical case---while we provide an example where the fidelity inequality is strict, we emphasize that this is a specific fine-tuned example which is not guaranteed to describe the typical scenario.

\section{Equivalence to quantum double under pure-charge noise} \label{app:equiv_QD}

A strongly $G$-left symmetric open quantum dynamics, is dual to a pure-charge noise channel (i.e., creating pure-charge anyons $a=(\{e\},\Gamma)$ labeled by a trivial conjugacy class of $G$ and an irrep $\Gamma$) on the quantum double $\DG$ ground state for a symmetry group $G$. This duality KW$_G$, which generalizes the Kramers--Wannier transformation to non-Abelian groups~\cite{Tantivasadakarn_2023, yoshida2017gapped, lootens2023dualities}, follows from the mapping between the strongly-symmetric setup (with $G$-qudits on vertices) to the topologically ordered system (with $G$-qudits on edges). Within the duality, the right action of the group is dual to the vertex operator of the corresponding quantum double:
\begin{equation}\label{eq:KW_R_to_A}
    R_{k,v}\overset{\mathrm{KW}_G}{\longrightarrow}A_v^k, \qquad A_v^k= \prod_{e:v\to }L_{k,e} \prod_{e:\to v}R_{k,e}.
\end{equation}

With the definitions in Eq.~\eqref{eq:left_reg}, the central projectors constructed from the left and right regular actions obey
\begin{equation}\label{eq:PL_equals_PR}
\begin{split}
    P_{\Gamma,v}^{L}\equiv \frac{d_\Gamma}{|G|}\sum_{g\in G}\chi_\Gamma(g)L_{g,v}=\frac{d_\Gamma}{|G|}\sum_{g\in G}\chi^*_\Gamma(g)R_{g,v}
    \equiv P_{\Gamma,v}^{R}.
\end{split}
\end{equation}
Both sides act as the identity on the irrep block labeled by $\Gamma$ and vanish on all other blocks. The complex conjugation in the right-action expression is essential: using $\chi_\Gamma(g)$ instead would give $P_{\bar\Gamma}^{L}$.  Combining Eqs.~\eqref{eq:KW_R_to_A} and \eqref{eq:PL_equals_PR}, the on-site irrep measurements map to the pure-charge measurement at the corresponding quantum-double vertex,
\begin{equation}\label{eq:KW_charge_projector}
\begin{split}
    P_{\Gamma,v}^{L} &\overset{\mathrm{KW}_G}{\longrightarrow}P_{(\{e\},\Gamma),v},\\
    P_{(\{e\},\Gamma),v} &=\frac{d_\Gamma}{|G|}\sum_{k\in G}\chi^*_\Gamma(g)A_v^g.
\end{split}
\end{equation}
This expression remains valid for irreps with complex characters and fixes the possible $\Gamma\leftrightarrow\bar\Gamma$ ambiguity. On the other hand, for an oriented link $e:x\to y$, the $G$-left symmetric Kraus operator $K^\Gamma_{ab,\langle xy\rangle}$ can be written as
\begin{equation}
    \frac{1}{\sqrt{d_\Gamma}} \sum_{c=1}^{d_\Gamma} (Z^{\bar{\Gamma}}_x)_{ca} (Z^\Gamma_y)_{cb},
\end{equation}
where $(Z_x^{\Gamma})_{ab}$ is defined in Eq.~\eqref{eq:Z_Gamma}. Under, the KW$_G$ this transforms to
\begin{equation}\label{eq:KW_Z}
    \sum_{c=1}^{d_\Gamma}
    (Z_x^{\bar\Gamma})_{ca}(Z_y^\Gamma)_{cb}
    \overset{\mathrm{KW}_G}{\longrightarrow}
    (Z_{\langle xy\rangle}^\Gamma)_{ab},
\end{equation}
where $(Z_{\langle xy \rangle}^\Gamma)_{ab}$ multiplies the link basis state $\ket{h_e}$ by $D^\Gamma_{ab}(h_e)$. This follows immediately from $h_e=g_x^{-1}g_y$.  If the arrow is reversed, the matrix element instead contains $D^\Gamma_{ab}(h_e^{-1})=D^{\bar\Gamma}_{ba}(h_e)$: reversing the arrow exchanges $\Gamma$ and $\bar\Gamma$ and transposes the matrix indices. Table~\ref{tab:comparison} summarizes the 1-to-1 mapping between the strong-to-weak SSB for a finite non-Abelian group $G$, and the quantum double formulation for the same group $G$ under pure-charge noise. 

\begin{table*}[t]
\centering
\begin{tikzpicture}[
  corner/.style ={fill=black!10, text width=2.25cm, align=left},
  hdrS/.style   ={fill=blue!15,  text width=7.25cm, align=center, font=\bfseries},
  hdrT/.style   ={fill=orange!22,text width=7.25cm, align=center, font=\bfseries},
  rlabel/.style ={fill=black!6,  text width=2.25cm, align=left, font=\bfseries},
  cellS/.style  ={fill=blue!4,   text width=7.25cm, align=left},
  cellT/.style  ={fill=orange!5, text width=7.25cm, align=left},
]
\matrix (T) [
  matrix of nodes,
  nodes={anchor=center, outer sep=0, inner xsep=3pt, inner ysep=2pt,
         execute at begin node={\small \thinmuskip=3mu \medmuskip=4mu \thickmuskip=5mu}},
  row sep=2pt, column sep=2pt,
  row 1/.style={nodes={minimum height=0.62cm}},
  row 2/.style={nodes={minimum height=0.68cm}},
  row 3/.style={nodes={minimum height=0.78cm}},
  row 4/.style={nodes={minimum height=1.04cm}},
  row 5/.style={nodes={minimum height=1.04cm}},
  row 6/.style={nodes={minimum height=0.68cm}},
  ]{
  |[corner]| {}
    & |[hdrS]| Strong-to-weak SSB
    & |[hdrT]| Topological quantum memory \\
  |[rlabel]| Degrees of freedom
    & |[cellS]| $G$-qudits on \emph{vertices}
    & |[cellT]| $G$-qudits on \emph{edges} \\
  |[rlabel]| Initial state
    & |[cellS]| Product state $\ket{+}_G^{\otimes V}$; 
                $\ket{+}_G = \int_G dg\,\ket{g}$
    & |[cellT]| Ground state $\ket{\mathcal{D}(G)}$; 
                $\ket{+}_G^{\otimes V} \overset{\mathrm{KW}_G}{\leftrightarrow} \ket{\mathcal{D}(G)}$ \\
  |[rlabel]| Noise channel
    & |[cellS]| Edge-correlated (Eq.~\eqref{eq:full_channel}): \newline
                $K^\Gamma_{ab,\langle xy\rangle} = \sum_{c=1}^{d_\Gamma} (Z^{\bar{\Gamma}}_x)_{ca} (Z^\Gamma_y)_{cb}/\sqrt{d_\Gamma}$
    & |[cellT]| Pure-charge on edge $\langle xy\rangle$ (Eq.~\eqref{eq:QD_channel}): \newline
                $K^\Gamma_{ab, \langle xy \rangle} = (Z^\Gamma_{\langle xy \rangle})_{ab}/\sqrt{d_\Gamma}$ \\
  |[rlabel]| Syndrome \newline measurement
    & |[cellS]| Irrep projector on vertex: \newline
                $P_{\Gamma_v} =\frac{d_\Gamma}{|G|} \sum_{g\in G}\,\chi_\Gamma(g)\,L_{g,v}$
    & |[cellT]| Anyon charge at vertex: \newline
                $P_{(\{e\},\Gamma)} = \frac{d_\Gamma}{|G|}\,\sum_{g\in G}\,\chi^*_\Gamma(g)\,A^g_v$ \\
  |[rlabel]| Charge detection
    & |[cellS]| $2$-local on sites $x,y$: $\sum_{c=1}^{d_\Gamma} (Z^{\bar{\Gamma}}_x)_{ca} (Z^\Gamma_y)_{cb}$
    & |[cellT]| Wilson string along $\gamma$: $W^\Gamma_{ab}(\gamma)=\big(\prod_{e\in\gamma}Z^\Gamma_e\big)_{ab}$ \\
};
\end{tikzpicture}
\caption{\textbf{Duality between sw-SSB and topological quantum memories.} Comparison between the strong-to-weak SSB setup with strongly $G$-left symmetric channels, and the topological quantum memory $\DG$ under pure-charge decoherence, related by the generalized Kramers-Wannier duality $\mathrm{KW}_G$. See main text for details.}
\label{tab:comparison}
\end{table*}

Putting everything together, the strongly-symmetric channel $\mathcal{E}_{xy}$ in Eq.~\eqref{eq:full_channel} maps under $\textrm{KW}_G$ to a channel acting on the edges of the quantum double,
\begin{equation}\label{eq:QD_channel}
     \sum_{\Gamma} \frac{p_\Gamma}{2d_\Gamma} \sum_{a,b} \left[(Z^\Gamma_e)_{ab}\,\rho\, [(Z^\Gamma_e)_{ab}]^\dagger + (Z^{\bar{\Gamma}}_e)_{ab}\,\rho\, [(Z^{\bar{\Gamma}}_e)_{ab}]^\dagger\right],
\end{equation}
that only creates pairs of pure-charge anyons: with probability $p_\Gamma$, it creates a pair of pure charge anyons labeled by irreps $\Gamma, \bar{\Gamma}$ at the endpoints of an edge. To avoid an orientation-dependent noise model for groups for which $\bar{\Gamma}\neq \Gamma$, we have included the second symmetrizing term.

\section{FVS Syndrome Sampler}
\label{app:FVSsampler}

We recall that by Bayes' theorem, $p^{(1)}(\{\Gamma_v\})$ can be written as a sum over physical error chains $E$ compatible with the syndrome configuration:
\begin{equation}
p^{(1)}(\{\Gamma_v\}) = \sum_E p(E)\, p(\{\Gamma_v\}|E),
\end{equation}
where $p(E)$ factorizes over the bonds $\langle x y\rangle$ of the lattice, and an error chain $E$ is a configuration of bond irreps $\Gamma_{xy}$ sampled according to the probabilities $p_\Gamma$ in Eq.~\eqref{eq:full_channel}. The sampling therefore proceeds in two steps: first, we sample an error chain $E$ from $p(E)$; second, given $E$, we sample a syndrome configuration $\{\Gamma_v\}$ from $p(\{\Gamma_v\}|E)$. The second step is the nontrivial one, and we now describe how to carry it out.

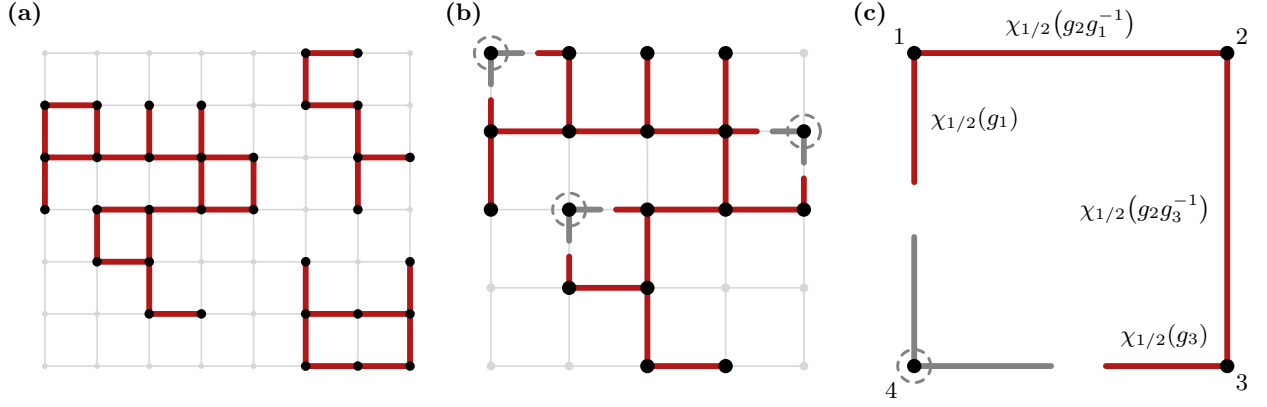
\begin{figure*}[t]
\begin{tikzpicture}
 
\begin{scope}[scale=0.69, line cap=round]
  \node[anchor=west] at (-0.9,6.75) {\textbf{(a)}};
 
  \foreach \x in {0,...,7} \draw[clatt, line width=0.6pt] (\x,0) -- (\x,6);
  \foreach \y in {0,...,6} \draw[clatt, line width=0.6pt] (0,\y) -- (7,\y);
 
  \foreach \x in {0,...,7} \foreach \y in {0,...,6}
    \fill[clatt] (\x,\y) circle (1.6pt);
 
  \begin{scope}[shift={(0,1)}]
    \foreach \a/\b in {%
        {0,4}/{1,4}, {0,4}/{0,3}, {4,3}/{3,3}, {4,3}/{4,2},
        {1,2}/{1,1}, {1,2}/{2,2},
        {1,4}/{1,3}, {1,3}/{0,3}, {0,3}/{0,2},
        {1,3}/{2,3}, {2,3}/{2,4}, {2,3}/{3,3}, {3,3}/{3,4},
        {3,3}/{3,2}, {3,2}/{4,2}, {3,2}/{2,2},
        {2,2}/{2,1}, {2,1}/{1,1}, {2,1}/{2,0}, {2,0}/{3,0}}
      \draw[cred, line width=2.2pt] (\a) -- (\b);
    \foreach \p in {{0,4},{1,4},{1,3},{0,3},{0,2},{2,3},{2,4},{3,3},{3,4},
                    {4,3},{3,2},{4,2},{1,2},{2,2},{2,1},{1,1},{2,0},{3,0}}
      \fill[black] (\p) circle (2.6pt);
  \end{scope}
 
  \foreach \a/\b in {%
      {5,5}/{5,6}, {5,6}/{6,6}, {5,5}/{6,5},
      {6,5}/{6,4}, {6,4}/{7,4}, {6,4}/{6,3}}
    \draw[cred, line width=2.2pt] (\a) -- (\b);
  \foreach \p in {{5,5},{5,6},{6,6},{6,5},{6,4},{7,4},{6,3}}
    \fill[black] (\p) circle (2.6pt);
 
  \foreach \a/\b in {%
      {5,0}/{6,0}, {6,0}/{7,0}, {7,0}/{7,1}, {7,1}/{6,1},
      {6,1}/{5,1}, {5,1}/{5,0},
      {7,1}/{7,2}, {5,1}/{5,2}}
    \draw[cred, line width=2.2pt] (\a) -- (\b);
  \foreach \p in {{5,0},{6,0},{7,0},{7,1},{6,1},{5,1},{7,2},{5,2}}
    \fill[black] (\p) circle (2.6pt);
\end{scope}
 
\begin{scope}[xshift=5.9cm, scale=1.035, line cap=round]
  \node[anchor=west] at (-0.7,4.5) {\textbf{(b)}};
 
  \foreach \x in {0,...,4} \draw[clatt, line width=0.6pt] (\x,0) -- (\x,4);
  \foreach \y in {0,...,4} \draw[clatt, line width=0.6pt] (0,\y) -- (4,\y);
 
  \foreach \a/\b in {%
      {0,4}/{1,4}, {0,4}/{0,3},   
      {4,3}/{3,3}, {4,3}/{4,2},   
      {1,2}/{1,1}, {1,2}/{2,2}}   
  {
    \draw[cgray, line width=2.2pt] (\a) -- ($(\a)!0.40!(\b)$);
    \draw[cred,  line width=2.2pt] ($(\a)!0.60!(\b)$) -- (\b);
  }
 
  \foreach \a/\b in {%
      {1,4}/{1,3}, {1,3}/{0,3}, {0,3}/{0,2},
      {1,3}/{2,3}, {2,3}/{2,4}, {2,3}/{3,3}, {3,3}/{3,4},
      {3,3}/{3,2}, {3,2}/{4,2}, {3,2}/{2,2},
      {2,2}/{2,1}, {2,1}/{1,1}, {2,1}/{2,0}, {2,0}/{3,0}}
    \draw[cred, line width=2.2pt] (\a) -- (\b);
 
  \foreach \p in {{0,0},{0,1},{1,0},{3,1},{4,0},{4,1},{4,4}}
    \fill[clatt] (\p) circle (1.6pt);
 
  \foreach \p in {{1,4},{1,3},{0,3},{0,2},{2,3},{2,4},{3,3},{3,4},
                  {3,2},{4,2},{2,2},{2,1},{1,1},{2,0},{3,0}}
    \fill[black] (\p) circle (2.6pt);
 
  \foreach \p in {{0,4},{4,3},{1,2}}
  {
    \fill[black] (\p) circle (2.6pt);
    \draw[cgray, dashed, line width=1.1pt] (\p) circle (6pt);
  }
\end{scope}
 
\begin{scope}[xshift=11.5cm, scale=1.035, line cap=round]
  \node[anchor=west] at (-0.9,4.5) {\textbf{(c)}};
 
  \coordinate (v1) at (0,4);
  \coordinate (v2) at (4,4);
  \coordinate (v3) at (4,0);
  \coordinate (v4) at (0,0);
 
  \draw[cred,  line width=2.2pt] (v1) -- (v2);                    
  \draw[cred,  line width=2.2pt] (v2) -- (v3);                    
  \draw[cred,  line width=2.2pt] (v1) -- (0,2.35);                
  \draw[cgray, line width=2.2pt] (0,1.65) -- (v4);                
  \draw[cgray, line width=2.2pt] (v4) -- (1.75,0);                
  \draw[cred,  line width=2.2pt] (2.45,0) -- (v3);                
 
  \foreach \p in {v1,v2,v3} \fill[black] (\p) circle (2.6pt);
  \fill[black] (v4) circle (2.6pt);
  \draw[cgray, dashed, line width=1.1pt] (v4) circle (6pt);
  \node[above left]  at (v1) {$1$};
  \node[above right] at (v2) {$2$};
  \node[below right] at (v3) {$3$};
  \node[below left=3pt]  at (v4) {$4$};
 
  \node[above=2pt] at (2,4)    {$\chi_{1/2}\!\left(g_2 g_1^{-1}\right)$};
  \node[left=3pt]  at (4,2)    {$\chi_{1/2}\!\left(g_2 g_3^{-1}\right)$};
  \node[right=3pt] at (0,3.15) {$\chi_{1/2}\!\left(g_1\right)$};
  \node[above=3pt] at (3.2,0)  {$\chi_{1/2}\!\left(g_3\right)$};
\end{scope}
 
\end{tikzpicture}
\caption{\textbf{(a)} Error configuration consisting of three disconnected error clusters; the syndrome on each of the clusters is sampled independently. The upper right cluster doesn't contain any loops and the syndrome distribution factorizes over the vertices of this cluster. \textbf{(b)} An example of a cluster with 3 cycles. Sites from the feedback vertex set (FVS) are circled by a dashed gray line. \textbf{(c)} The simplest non-trivial error cluster containing one cycle. The site 4 is removed on the first step. The syndrome on the sites 1, 2, and 3 is sampled independently. The site 4 is then added and sampled conditioned on the fixed syndrome on the sites 1, 2, and 3.}
\label{fig:errors}
\end{figure*}

Given an error configuration $E$, we identify its disconnected components $E_C$ using a union--find data structure (for example, the error configuration in Fig.~\ref{fig:errors}(a) contains three error clusters). This can be done in almost linear time \cite{cormen2022algo}.
The syndrome is sampled on each disconnected cluster separately, since the conditional syndrome distribution factorizes over clusters, $p(\{\Gamma_v\}|E)=\prod_C p(\{\Gamma_v\}|E_C)$. Each of the error clusters, being disconnected from the rest, must fuse to vacuum separately. Thus, there are no correlations between the syndromes on disconnected error clusters. If a cluster is a tree (i.e., contains no cycles), then $p(\{\Gamma_v\}| E_C)$ factorizes further over its vertices:
\begin{equation} \label{eq:p_tree}
p(\Gamma_v | E_C)=\frac{d_{\Gamma_v}}{\sum_{\Gamma'_v\in S_v} d_{\Gamma'_v}},
\end{equation}
where $S_v$ denotes the set of fusion outcomes at vertex $v$ compatible with $E_C$. For vertices where the same total irrep appears with multiplicity, the local probability should include the multiplicity factor $m_v(J_v)$; the displayed formula is correct only when the local fusion outcome is multiplicity-free. The proof of this fact can be found in Appendix~\ref{app:tree_sampling}. Sampling on tree-like error clusters is therefore fast.

In general, however, error clusters contain cycles. A \emph{feedback vertex set} (FVS) $\mathcal{V}$ is a set of vertices whose removal renders the cluster acyclic [circled vertices in Figs.~\ref{fig:errors}(b,c)]; we refer to the remaining vertices as the \emph{tree component} $\mathcal{T}$ of the cluster. Notice that in general, removing vertices from an error cluster can lead to disconnected tree-like components, and hence to a forest rather than a tree graph. Starting from the joint distribution of syndromes on a given error cluster, $p(\{\Gamma_v\}|E_C)$, we obtain the marginal distribution over the tree component via:
\begin{align}
   \nonumber & p(\{\Gamma^{\mathcal{T}}_v\}|E_C)=\sum_{\{\Gamma_v\,: v\in \mathcal{V}\}}p(\{\Gamma_v\}|E_C)  \\
   &=\int \prod_{v\in \mathcal{T}} dg_v \,W({\bf{g}}\,:\,g_v=e \text{ if } v\notin \mathcal{T})\prod_{v\in \mathcal{T}}d_{\Gamma_v}\chi_{\Gamma_v}(g_v).
\end{align}
Here the local weights $w(g_xg_y^{-1})$ appearing in $W({\bf{g}}\,:\,g_v=e \text{ if } v\notin \mathcal{T})$ are modified such that $g_v$ is set equal to the identity element $e\in G$ for every $v\in \mathcal{V}$. This turns $E_C$ into a forest with ``dangling'' legs---edges of $E_C$ incident to FVS sites, which now terminate on a fixed identity element [see Figs.~\ref{fig:errors}(c)]. 

The sampling on a cluster with cycles then proceeds in two steps. First, having removed the FVS sites, we sample $\{\Gamma^{\rm tree}_v\}$ from $p(\{\Gamma^{\rm tree}_v\}|E_C)$ independently on each vertex of the forest component using Eq.~\eqref{eq:p_tree}. Second, utilizing Bayes' theorem once more, we add the FVS sites back to the error cluster one by one. Each time an FVS site is added, all of its ``dangling'' legs are returned to the cluster along with it, and we compute the probabilities of all possible fusion outcomes at that site, conditioned on the syndromes fixed so far on the rest of the cluster. The number of possible fusion outcomes on a lattice with finite coordination number and a channel inserting a finite number of different irreps is always finite. For the square lattice and the channel inserting pairs of spin-$\frac{1}{2}$, the maximum number of different fusion outcomes on a vertex is 3.  Calling $\mathcal{T}_n$ the set of vertices belonging to tree union those in $\mathcal{V}$ already added back at the $n$th iteration of this protocol with a fixed irrep, the probability of inserting the $(n+1)$th vertex $x\in \mathcal{V}$ in an irrep $\Gamma^{n+1}_x$ is given by 
\begin{align}
   p(\Gamma_x^{n+1}|E_C)=\frac{p(\{\Gamma_v\,:\, v\in \mathcal{T}_{n+1}\}|E_C)}{p(\{\Gamma_v\,:\, v\in \mathcal{T}_{n}\}|E_C)},
\end{align}
with 
\begin{align}
    \nonumber &p(\{\Gamma_v\,:\, v\in \mathcal{T}_{1}\}|E_C)=\int \prod_{v\in \mathcal{T}_1} dg_v \,W({\bf{g}}\,:\,g_v=e \text{ if } v\notin \mathcal{T}_1)\prod_{v\in \mathcal{T}_1}d_{\Gamma_v}\chi_{\Gamma_v}(g_v),
\end{align}
where $\mathcal{T}_1=\mathcal{T}\cup \{x\}$. Given these probabilities, we sample one of the possible fusion outcomes and record it. The procedure continues until every FVS site has been reinserted and a syndrome has been assigned to every vertex of the cluster. 

The complexity of this procedure scales linearly with the number of FVS sites: for every inserted FVS site, one needs to contract a constant number of tensor networks for each possible fusion outcome on this site. This number is determined by the coordination number of the inserted FVS site in the error cluster and is independent of the system size. 

Although finding a minimum FVS on a general graph is NP-complete~\cite{karp1972NP}, on undirected graphs the problem admits a polynomial-time constant-factor approximation~\cite{becker1996optimization, bafna1999APXFVS}. Moreover, if the coordination number of the lattice is less than or equal to three, the problem can be solved exactly in polynomial time \cite{ueno1988FVSd3}. Thus, this step of the sampling algorithm can also be performed efficiently. 

\section{Local fusion probabilities on tree error clusters}
\label{app:tree_sampling}

In this Appendix, we prove that the fusion probabilities on tree-like error clusters factorize, which significantly simplifies syndrome sampling. The syndrome probability on a given error cluster $E$ is
\begin{equation}
    p(\{J_1, \dots J_N\} | E) = \tr\qty[\rho_E \prod_{v} P_{J_v}] = \int dg_N\dots dg_2 dg_1 \prod_{\expval{xy} \in E} \frac{\chi_{\Gamma_{xy}}\qty(g_x g_y^{-1})}{d_{\Gamma_{xy}}} \prod_{v\in E} d_{\Gamma_v} \chi_{\Gamma_v}(g_v),
\end{equation}
where $\Gamma_{xy}$ stands for the error on the link, and $\Gamma_v$ is the on-site syndrome. The proof below relies on the following identity:
\begin{equation}
\label{eq:inductionTree}
    \int dg_v \qty[\prod_{u} \frac{\chi_{\Gamma_{uv}}(g_v)}{d_{\Gamma_{uv}}}] d_{\Gamma_v} \chi_{\Gamma_v}(g_v) \frac{1}{d_{\Gamma_{vw}}}\chi_{\Gamma_{v w}} (g_v g_w^{-1}) = \frac{d_{\Gamma_v} m_{\Gamma_v}(\otimes_{i = \{u, \omega\}}\Gamma_{iv}) }{\prod_{i = \{ u, w\}} d_{\Gamma_{iv}}} \frac{1}{d_{\Gamma_{vw}}}\chi_{\Gamma_{vw}}(g_w), 
\end{equation}
where $u$ is the set of all the leaves connected to the site $v$, $w$ is the next non-leave site in the cluster, and  $m_\Gamma(\otimes_i \Gamma_i)$ is the multiplicity of the outcome $\Gamma$ in the fusion of $\otimes_i \Gamma_i$. To derive this result, we first fuse all of the characters except the last one $\chi_{\Gamma_{uw}}$:
\begin{equation}
    \qty[\prod_{u} \chi_{\Gamma_{uv}}(g_v)] \chi_{\Gamma_v}(g_v) = \sum_\Gamma m_{\Gamma}\qty(\otimes_u \Gamma_{ub} \otimes \Gamma_v)\, \chi_{\Gamma}(g_v).
\end{equation}
The integral can now be computed using the Schur orthogonality relation, leading to
\begin{equation}
    \int dg_v  \frac{d_{\Gamma_v}}{\prod_u d_{\Gamma_{uv}}} \qty[\sum_\Gamma m_{\Gamma}\qty(\otimes_u \Gamma_{ub} \otimes \Gamma_v) \chi_{\Gamma}(g_v)]  \chi_{\Gamma_{v w}} (g_v g_w^{-1}) =  \frac{d_{\Gamma_v} m_{\Gamma_{vw}}(\otimes_u \Gamma_{ub} \otimes \Gamma_v) }{\prod_{i = \{ u, w\}}d_{\Gamma_{iv}}}  \frac{\chi_{\Gamma_{vw}}(g_w)}{d_{\Gamma_{vw}}}.
\end{equation}
Finally, we use the symmetry of the multiplicity factor and exchange $\Gamma_v$ with $\Gamma_{vw}$, reproducing the result in Eq.~\eqref{eq:inductionTree}. 

Using this result as the elementary step of induction, it is now simple to see how the proof proceeds. Given a tree-like error cluster, we can first identify the leaves and integrate them out. Afterwards, we can remove the integrated out leaves from the cluster and identify the leaves of the reduced cluster. By proceeding this way we can reach the root of the tree and, therefore, integrate over the entire error cluster. On each step, as was show in Eq.~\eqref{eq:inductionTree}, we get a factor corresponding to the independent fusion probability:
\begin{equation}
    p_{\rm ind}(\Gamma_v) = \frac{d_{\Gamma_v} m_{\Gamma_v} \qty(\otimes_{u: \expval{uv}} \Gamma_{uv})}{\prod_{u: \expval{uv}} d_{\Gamma_{uv}}},
\end{equation}
where $u: \expval{uv}$ stands for the nearest neighbors of $v$.

\section{Equivalence of quantum fidelity and R\'enyi-1 correlator}
\label{app:R1bound}

In this Appendix, we prove equivalence between the R\'enyi-1 correlator and quantum fidelity for the class of channels considered in this paper, generalizing the proofs from Refs.~\cite{weinstein2025efficient, Whightman_corr}.

The first part of the proof proceeds by using (1) Jensen's operator inequality for unital channels~\cite{hansen2003HPJ}, and (2) the inequality for the trace of bounded operators: $|\tr X| \leq \| X\|_1$~\cite{simon2010trace}. Combining these inequalities and using $\mathcal{E}_{xy}^{\Upsilon}(\rho) = \sum_{a, b =1}^{d_\Upsilon} K^{\Upsilon}_{ab, xy} \rho K^{\Upsilon, \dagger}_{ab, xy}$, we find
\begin{equation}
    \R_1(x, y) = \tr\qty[\sqrt{\rho} \mathcal{E}^{\Upsilon}_{xy} (\sqrt{\rho})] \overset{(1)}{\leq}  \tr\qty[\sqrt{\rho} \sqrt{\mathcal{E}^{\Upsilon}_{xy} (\rho)}] \overset{(2)}{\leq} \| \sqrt{\rho} \sqrt{\mathcal{E}^{\Upsilon}_{xy} (\rho)}\|_1 = F^Q\qty(\rho, \mathcal{E}^{\Upsilon}_{xy} (\rho)).
\end{equation}

To prove the lower bound on the R\'enyi-1 correlator, we first rewrite the quantum fidelity as
\begin{equation}
    F^Q\qty(\rho, \mathcal{E}^{\Upsilon}_{xy} (\rho)) = \tr \sqrt{M M^\dagger} = \| M\|_1, 
\end{equation}
with 
\begin{equation}
    V = \sum_{a, b = 1}^{d_\Upsilon} K^\Upsilon_{ab, xy} \otimes \bra{ab},\qquad M = \sqrt{\rho}\, V\, \qty(\sqrt{\rho} \otimes \mathbbm{1}_{d_{\Upsilon}^2})
\end{equation}
We further split the operator $M$ into $ M = \rho^{1/4} \, \qty[\rho^{1/4}\, V\, \qty(\rho^{1/4}\otimes \mathbbm{1})] \, \qty(\rho^{1/4} \otimes \mathbbm{1})$,  and apply H\"older's inequality~\cite{simon2010trace}:
\begin{equation}
    \| M\|_1 \leq \| \rho^{1/4}\|_4 ~\| \rho^{1/4}\, V\, \qty(\rho^{1/4}\otimes \mathbbm{1})\|_2 ~\| \rho^{1/4} \otimes \mathbbm{1}\|_4.
\end{equation}
Each factor from the right hand side can be computed explicitly to give
\begin{equation}
\begin{split}
    \| \rho^{1/4}\|_4 &= \qty(\tr \rho)^{1/4} = 1, \\
    \| \rho^{1/4}\, V \,\qty(\rho^{1/4} \otimes \mathbbm{1})\|_2 &= \sqrt{\tr \sqrt{\rho}\, V \qty(\rho^{1/2} \otimes \mathbbm{1})\, V^\dagger} = \sqrt{\R_1(x, y)}, \\
    \| \rho^{1/4} \otimes \mathbbm{1}\|_4 &= \qty(\tr \rho \otimes \mathbbm{1})^{1/4} = \sqrt{d_{\Upsilon}}.
\end{split}
\end{equation}
Hence, all together, we find
\begin{equation}
    \frac{1}{d_\Upsilon} \qty[F^{Q}(\rho, \mathcal{E}_{xy}^\Upsilon(\rho))]^2\leq \R_1(x, y) \leq F^{Q}(\rho, \mathcal{E}_{xy}^\Upsilon(\rho)),
\end{equation}
for any finite dimensional representation $d_\Upsilon$.

\section{Hierarchy distributions at general level $N$}
\label{app:hierarchy_deriv}

Similarly to App.~\ref{app:rho_derivation}, we derive the joint syndrome probability distribution $p^{(N)}\!\big(\{\Gamma^{(n)}_R\}_{n=1}^N\big)$ at general hierarchy level $N$, Eq.~\eqref{eq:level_general} of the main text, together with the marginal property that connects level $N$ to level $N-1$.

We fix a sequence of nested
partitions of the lattice into non-overlapping regions of increasing size: at level $n=1$ each region is a single vertex, $R^{(1)}=\{v\}$; at level $n=2$ regions $R^{(2)}$ contain two or more vertices; and so on, with each $R^{(n)}$ a union of regions $R^{(n-1)}\subseteq R^{(n)}$ from the previous level, terminating at $R^{(N_{\max})}=V$ for some maximal level $N_{\max}\sim\log V$. For each region $R$ at level $n$ we introduce the regional projector
$P^{(n)}_{\Gamma^{(n)}_R}$ of Eq.~\eqref{eq:P_region}.

By Appendix~\ref{app:transform}, the projectors at different levels commute, $\big[P^{(n)}_{\Gamma^{(n)}_R},P^{(n')}_{\Gamma^{(n')}_{R'}}\big]=0$ for $R^{(n)}\subseteq R^{(n')}$ or vice versa. At each fixed level $n$, projectors over disjoint regions $R\neq R'$ trivially commute since they act on disjoint sites. Hence, the joint syndrome distribution
\begin{equation}\label{eq:appL_pjoint}
    p^{(N)}\!\big(\{\Gamma^{(n)}_R\}_{n=1}^{N}\big) =\mathrm{tr}\!\qty[\rho\,\prod_{n=1}^{N}\prod_{R\,\textrm{at level }n}\!\!\!P^{(n)}_{\Gamma^{(n)}_R}]
\end{equation}
is well defined.

With the kernel
expression for each $P^{(n)}_{\Gamma^{(n)}_R}$ inserted into the trace with the density matrix \eqref{eq:rho_explicit}, the trace at a site $v$ fixes the product of all group elements associated with projectors containing $v$:
\begin{equation}
    r_v = \prod_{n = 1}^N  g^{(n)}_v l_v.
\end{equation}  
We introduce the local variable $g_v$ to be the combination that appears in the density-matrix kernel, i.e., $g_v \equiv r_v l_v^{-1}$. With this choice the link part remains the original nearest-neighbor factor $W({\bf g})=\prod_{\langle xy\rangle}w(g_xg_y^{-1})$. Using this definition, we express $g_v^{(1)} = g_v \qty(g_v^{(2)})^{-1} \qty(g_v^{(3)})^{-1} \dots \qty(g_v^{(N)})^{-1}$. Putting everything together and using the properties of the Haar measure, we obtain
\begin{equation}\label{eq:appL_general}
    {\,
    p^{(N)}\!\big(\{\Gamma^{(n)}_R\}_{n=1}^{N}\big) =\!\int\!\mathcal{D}g\,W({\bf{g}})\, \prod_{n=2}^{N}\bigg[\!\int\!\mathcal{D}g^{(n)}\!\!\prod_{R\,\textrm{at level }n}\!\!\!\!d_{\Gamma^{(n)}_R}\, \chi_{\Gamma^{(n)}_R}(g^{(n)}_R)\bigg]\, \prod_v d_{\Gamma_v}\,\chi_{\Gamma_v}\!\Big[g_v\,(g^{(2)}_{R(v)})^{-1}\!\cdots(g^{(N)}_{R(v)})^{-1}\Big]\,,\,}
\end{equation}
where $R(v)$ at level $n$ denotes the level-$n$ region containing vertex $v$. The local coupling $W({\bf{g}})$ is unmodified; the higher levels of the hierarchy enter only through the on-site syndrome insertions.

To see that marginalizing over $\Gamma^{(N)}_R$ at the highest level correctly reproduces the distribution on $N-1$ levels, we use the character completeness identity:
\begin{equation}
    \sum_{\Gamma^{(N)}_R} d_{\Gamma^{(N)}_R}\,\chi_{\Gamma^{(N)}_R}\!\big(g^{(N)}_R\big) \;=\;|G|\,\delta_{g^{(N)}_R,\,e}\,,
\end{equation}
which sets $g^{(N)}_R=e$ for every level-$N$ region. Substituting this back into \eqref{eq:appL_general} eliminates the level-$N$ variables and recovers $p^{(N-1)}$:
\begin{equation}
    \sum_{\{\Gamma^{(N)}_R\}}p^{(N)}\!\big(\{\Gamma^{(n)}_R\}_{n=1}^{N}\big) \;=\;p^{(N-1)}\!\big(\{\Gamma^{(n)}_R\}_{n=1}^{N-1}\big)\,.
\end{equation}

\section{Local ${\rm SU}(2)$ tensor construction}\label{app:tensorsSU2}

In this Appendix, we provide the expressions for the local ${\rm SU}(2)$ tensors, used in the numerical simulations. For $G = S_3$, we directly implement Eq.~\eqref{eq:tensor_I} (with $\int dg \to \frac{1}{|G|}\sum_{g \in G}$), and no additional derivations are required.

The derivation uses the following properties of the representation matrices $D^J$:
\begin{equation}
\begin{split}
  \int d\mu(g)\, D^{\Gamma}_{mn}(g)\, D^{\Gamma'}_{m'n'}(g) &= \frac{1}{d_\Gamma}\,\delta_{\Gamma',\bar\Gamma}\,\delta_{m',m}\,\delta_{n',n}, \\
  \int d\mu(g)\, D^{j_1}_{m_1 n_1}(g)\, D^{j_2}_{m_2 n_2}(g)\, D^{j_3}_{m_3 n_3}(g) &= \tj{j_1}{j_2}{j_3}{m_1}{m_2}{m_3}\tj{j_1}{j_2}{j_3}{n_1}{n_2}{n_3},
\end{split}
\end{equation}
where $\tj{j_1}{j_2}{j_3}{m_1}{m_2}{m_3}$ is the Wigner 3-j symbol~\cite{edmondsbook}. We label the bonds by $l$, $r$, $d$, $u$, and define the bond weights $w_a$:
\begin{equation}
    w_a=\begin{cases}\sqrt{1-p}, & j_a=0,\\ \sqrt{p/2}, & j_a=\tfrac12. \end{cases}
\end{equation}

\begin{itemize}
    \item With $m_{--}=-(m_l+m_d)$, $n_{--}=-(n_l+n_d)$, $m_{++}=-(m_r+m_u)$, $n_{++}=-(n_r+n_u)$, $w_{lrdu}=w_lw_rw_dw_u$ and $L_{--},L_{++}\in\{0,\tfrac12,1\}$, the explicit expression for the level-1 tensor Eq.~\eqref{eq:tensor_T} reads:
\begin{equation}
\begin{split}
  {\bf T}^{A_v B_v}_{lrdu}(J_v)
  = (2J_v+1)\, w_{lrdu} &\sum_{L_{--},\,L_{++}} (2L_{--}+1)(2L_{++}+1)
    \tj{j_l}{j_d}{L_{--}}{m_l}{m_d}{m_{--}}\tj{j_l}{j_d}{L_{--}}{n_l}{n_d}{n_{--}}
    \\
  &\times \tj{j_r}{j_u}{L_{++}}{m_r}{m_u}{m_{++}}\tj{j_r}{j_u}{L_{++}}{n_r}{n_u}{n_{++}}
    \\
  &\times (-1)^{m_{--}-n_{--}}
    \tj{L_{--}}{L_{++}}{J_v}{-m_{--}}{m_{++}}{A_v}\tj{L_{--}}{L_{++}}{J_v}{-n_{--}}{n_{++}}{B_v}.
\end{split}
\end{equation}

\item The level-1 tensor with irrep insertion Eq.~\eqref{eq:tensor_I_mod} reads:
\begin{equation}
\begin{split}
  \mathbb{I}(\Gamma_v)^{A_vB_v;\,ab}_{\{\alpha_{v,\hat\mu}\}}
  = \frac{1}{\sqrt{d_\Upsilon}} &\sum_{L_{--},\,L_{++},\,K} (2L_{--}+1)(2L_{++}+1)(2K+1)\,
     (-1)^{a_{--}-b_{--}+\kappa-\kappa'}
  \\
  &\times \tj{\Gamma_{v,l}}{\Gamma_{v,d}}{L_{--}}{a_{v,l}}{a_{v,d}}{a_{--}}
          \tj{\Gamma_{v,l}}{\Gamma_{v,d}}{L_{--}}{b_{v,l}}{b_{v,d}}{b_{--}}
  \\
  &\times \tj{\Gamma_{v,r}}{\Gamma_{v,u}}{L_{++}}{a_{v,r}}{a_{v,u}}{a_{++}}
          \tj{\Gamma_{v,r}}{\Gamma_{v,u}}{L_{++}}{b_{v,r}}{b_{v,u}}{b_{++}}
  \\
  &\times \tj{\Gamma_v}{\Upsilon}{K}{A_v}{a}{\kappa}\tj{\Gamma_v}{\Upsilon}{K}{B_v}{b}{\kappa'} \tj{L_{--}}{L_{++}}{K}{-a_{--}}{a_{++}}{-\kappa}\tj{L_{--}}{L_{++}}{K}{-b_{--}}{b_{++}}{-\kappa'},
\end{split}
\end{equation}
with $a_{--}=-(a_{v,l}+a_{v,d})$, $b_{--}=-(b_{v,l}+b_{v,d})$, $a_{++}=-(a_{v,r}+a_{v,u})$,
$b_{++}=-(b_{v,r}+b_{v,u})$, $\kappa=-(A_v+a)$ and $\kappa'=-(B_v+b)$. The ranges are
$L_{\pm\pm}\in\{0,\tfrac12,1\}$ and $K\in\Gamma_v\otimes\Upsilon$.

\item For a pair $R=\{v_1,v_2\}$ with level-1 irreps
$\Gamma_{v_1},\Gamma_{v_2}$, the tensor Eq.~\eqref{eq:tensor_I2} reads
\begin{equation}
      \mathbf{I}^{(2)}\big(\Gamma^{(2)}_R\big)^{A^{(2)}_R B^{(2)}_R}_{\{\Gamma_v,A_v,B_v\}_{v\in R}} = d_{\Gamma^{(2)}_R}\,(-1)^{A_{v_1}-B_{v_1}+A_{v_2}-B_{v_2}}
     \tj{\Gamma_{v_1}}{\Gamma_{v_2}}{\Gamma^{(2)}_R}{-A_{v_1}}{-A_{v_2}}{A^{(2)}_R}
     \tj{\Gamma_{v_1}}{\Gamma_{v_2}}{\Gamma^{(2)}_R}{-B_{v_1}}{-B_{v_2}}{B^{(2)}_R}.
\end{equation}

\end{itemize}

\section{Additional numerical details}
\label{app:addNumerics}

In this Appendix, we describe the contraction of the tensor network and discuss in more detail the two quantities used in the Main Text to characterize the transitions: the ratio $\bm{D}$ of typical correlators, Eq.~(\ref{eq:Dtypical}), including its finite-size scaling, and the mid-cut entanglement entropy of the boundary MPS.

The two-dimensional tensor network is contracted using the boundary MPS method, in which the 2D lattice is represented as the time evolution of a 1D quantum state, approximated by an MPS~\cite{vidal2004efficient}. The state is initialized in the product state $v(p)=(1/\sqrt{1-p},0,0,\dots)$ on every site, and the other boundaries are fixed by contracting the tensors with the same vector. At every time step, a row of tensors $T(\Gamma_v)$ from Eq.~\eqref{eq:tensor_T} is applied to the state, and the resulting MPS is truncated to a bond dimension $\chi$ using standard techniques~\cite{verstraete2008mps}. We also use the tensors constructed by Eq.~\eqref{eq:tensor_I2} for the level-2 tensor network contraction. We use $\chi=24$ and $32$ for $L\le 64$, and $\chi=32$ for $L=128$. The two-point correlation functions are computed by contracting the same tensor network with the tensors $\mathbb{T}(\Gamma_v)$ at $x$ and $y$ [see Eq.~\eqref{eq:tensor_I_mod} and the discussion below it].

\begin{figure}[t]
     \centering
     \includegraphics[width=\textwidth]{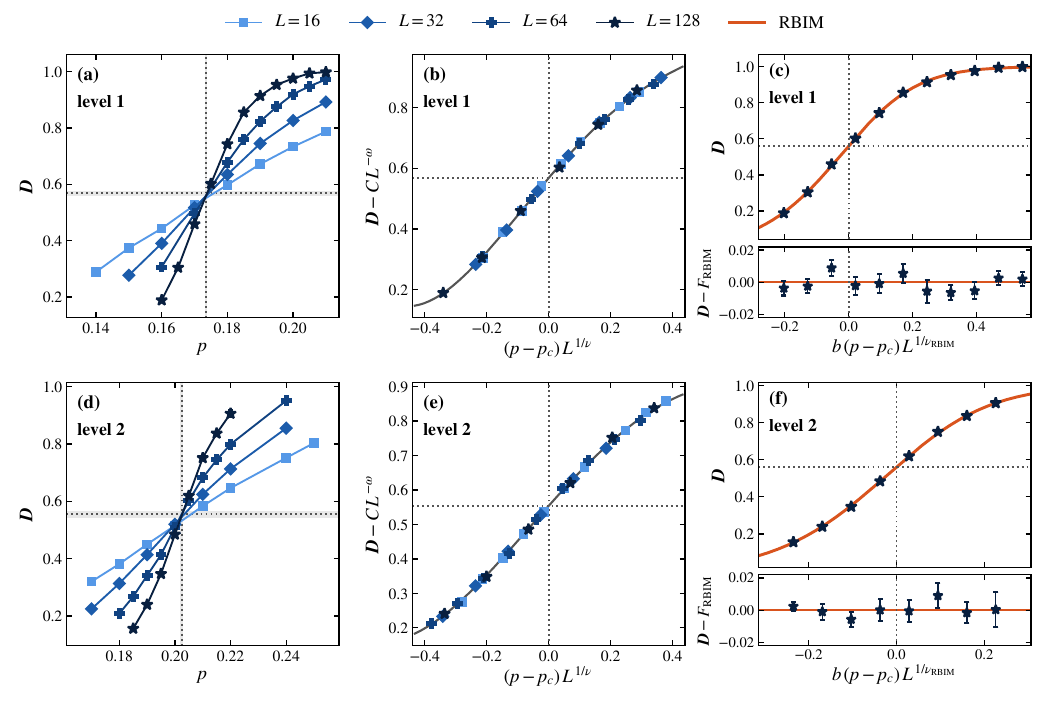}
    \caption{Finite-size scaling of the ratio $\bm{D}$ of typical correlators, Eq.~(\ref{eq:Dtypical}), for the SU(2) model with a spin-$\frac{1}{2}$ probe, on the first [(a)-(c)] and second [(d)-(f)] levels of the hierarchy.
    {\bf (a)}, {\bf (d)} $\bm{D}$ as a function of $p$. The dotted lines mark $p_c$ and $\bm{D}^*\equiv F(0)$ obtained from the fits in (b) and (e), and the shaded bands show their uncertainties.
    {\bf (b)}, {\bf (e)} Data collapse according to Eq.~(\ref{eq:FSS}), with $F$ approximated by a quartic polynomial (solid black curve), $\omega=1.35$ fixed and $C$ a constant, using all points with $|p-p_c|L^{1/\nu}\le 0.4$. The fits give $p_c^{(1)}=0.1736(2)(2)$, $\nu=1.51(1)(1)$ and $\bm{D}^*=0.568(5)(6)$, with $\chi^2/\textrm{dof}=19.7/20$, and $p_c^{(2)}=0.2024(2)(4)$, $\nu=1.47(1)(4)$ and $\bm{D}^*=0.554(5)(10)$, with $\chi^2/\textrm{dof}=25.5/23$. The first parentheses give the statistical error (bootstrap over syndrome configurations), and the second the systematic error, from varying the fitting window, the degree of the polynomial, $\omega$ and the form of the correction, and from excluding $L=16$ or $L=128$.
    {\bf (c)}, {\bf (f)} The $L=128$ data compared with the scaling function $F_{\rm RBIM}$ of the same protocol with $G=\mathbb{Z}_2$, i.e., the RBIM on the Nishimori line (orange line). $F_{\rm RBIM}$ was computed exactly at $L=128$ for $29$ values of $p$, with $15000$ syndrome configurations each, and fitted to $\frac{1}{2}[1+\tanh P]$, with $P$ a quartic polynomial. Only $p_c$ and the horizontal scale factor $b$ are fitted, without corrections to scaling, using all points shown. This gives $p_c=0.1736(1)$ and $b=0.630(7)$ on the first level, and $p_c=0.2028(1)$ and $b=0.556(6)$ on the second. The dotted lines mark $p=p_c$ and $\bm{D}^*_{\rm RBIM}=F_{\rm RBIM}(0)=0.5613(9)$.
    The lower strips show the residuals $\bm{D}-F_{\rm RBIM}$, which scatter around zero without a systematic trend and are consistent with the statistical errors, so that the SU(2) and RBIM scaling functions agree to within about $0.01$.}
    \label{fig:SU2fss}
\end{figure}

The critical points are located using the ratio $\bm{D}$ of typical correlators, Eq.~\eqref{eq:Dtypical}. Like the fidelity, $\bm{D}$ can be interpreted in terms of the distinguishability between the syndrome distributions with and without the probe pair. Consider the Kullback–Leibler (KL) divergence~\cite{gibbs2002probMetrics} between these distributions on the $n$th level of the hierarchy,
\begin{equation}
    D^{\Upsilon}_{\rm KL}(r)\equiv\sum_{\{\Gamma_v\}}p^{(n)}(\{\Gamma_v\})\log\frac{p^{(n)}(\{\Gamma_v\})}{p_{xy}^{(n)}(\{\Gamma_v\})}=-\overline{\log\expval{\frac{1}{d_\Upsilon}\chi_{\Upsilon}(g_xg_y^{-1})}_{\{\Gamma_v\}}},
\end{equation}
where $r=|x-y|$ is the separation of the probe pair, the overline denotes the average over syndrome configurations drawn from $p^{(n)}$, and the second equality follows from the same steps as in Appendix~\ref{app:rho_derivation}. Thus, $D^{\Upsilon}_{\rm KL}(r)$ is the disorder-averaged free-energy cost of inserting the probe pair. Comparing with Eq.~\eqref{eq:Dtypical},
\begin{equation}
    \bm{D}(p)=\exp\Big\{\tfrac12\Big[D^{\Upsilon}_{\rm KL}(L/4)-D^{\Upsilon}_{\rm KL}(L/2)\Big]\Big\},
\end{equation}
so that $\bm{D}$ measures how much the KL divergence grows when the separation of the probe pair is doubled.

In practice, the logarithm in Eq.~\eqref{eq:Dtypical} is averaged over several placements of the probe pair as well as over syndromes, to reduce statistical errors. We label the sites of the open $L\times L$ lattice by row $i$ and column $j$, $0 \le i,j \le L-1$. For every syndrome we insert the probe pair at $x=(i,j)$ and $y=(i,j+r)$ with rows $i \in \{L/4,\,3L/8,\,L/2,\,5L/8,\,3L/4\}$, left ends $j \in \{L/8,\,L/4,\,3L/8\}$ and separations $r \in \{L/4,\,L/2\}$. This yields 15 placements per separation, all at fixed fractions of $L$, the closest one $L/8-1$ sites from an edge. At the second level, the same sites are used, and both $x$ and $y$ are then left sites of their pairs. We use $L=16$, $32$, $64$ and $128$, with between $600$ and $26000$ syndrome configurations per point, and estimate the statistical errors by a bootstrap over syndrome configurations, resampling all placements of a given syndrome together.

\begin{figure}[t]
     \centering
        \includegraphics[width=\textwidth]{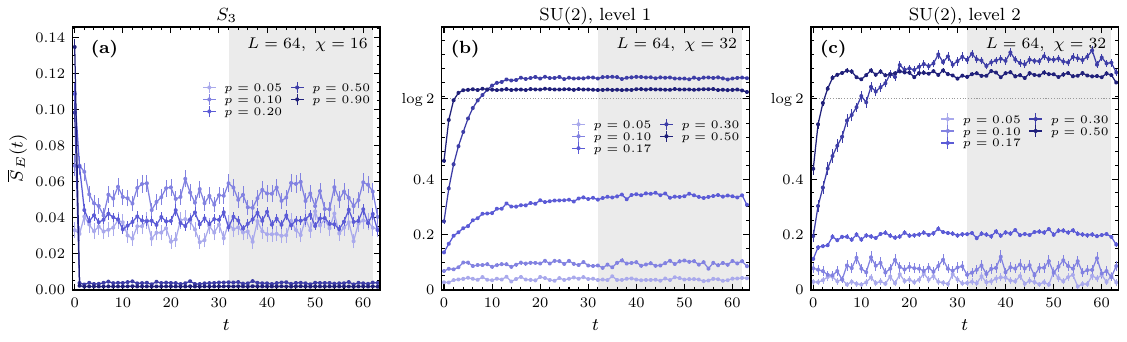}
        \caption{Disorder-average mid-cut entanglement entropy of the boundary MPS as a function of time $t$ for various $p$. The saturated plateau value of the entanglement entropy is taken as the average over the grey region from $t = L/2 + 1$ up to $t=L-1$. 
        \textbf{(a)}: entanglement entropy in the $S_3$-symmetric model from the Main Text. \textbf{(b)}: entanglement entropy in the level-1 ${\rm SU}(2)$-symmetric model from the Main Text. \textbf{(c)}: entanglement entropy in the level-2 ${\rm SU}(2)$-symmetric model from the Main Text.}
        \label{fig:entropyTime}
\end{figure}

Figure~\ref{fig:SU2fss} shows $\bm{D}(p)$ for the SU(2) model with a spin-$\frac{1}{2}$ probe, on the first [(a)-(c)] and second [(d)-(f)] levels of the hierarchy. On the first level, only the on-site irreps $\Gamma_v$ are measured. On the second level, the lattice is tiled by non-overlapping pairs of nearest-neighbor sites ($1\times 2$ blocks), and the total irrep $\Gamma^{(2)}_R$ of each pair $R$ is measured in addition, as in Fig.~\ref{fig:HierarMeas}.  On both levels, we find a $\mathbb{Z}_2$ transition in the Nishimori Ising universality class, with a threshold that increases from the first level to the second.

Since $\bm{D}$ is a ratio of correlators at two separations proportional to $L$, it is independent of $L$ at a critical point, while $\bm{D}\to 0$ below the threshold, where the typical correlator decays exponentially, and $\bm{D}\to 1$ above it. The curves for different $L$ indeed cross [Fig.~\ref{fig:SU2fss}(a,d)], with a slow drift of the crossing point with $L$. We therefore fit the data to the scaling form with the leading correction to scaling,
\begin{equation}\label{eq:FSS}
    \bm{D}(p,L)=F\big((p-p_c)L^{1/\nu}\big)+CL^{-\omega},
\end{equation}
where $F$ is the scaling function and $C$ is a constant,
with the correction exponent fixed at the Nishimori value $\omega=1.35$~\cite{wan2026Nishimori} (the results below are found to be only weakly dependent on the value of $\omega$ over a broad range). The resulting collapses [Fig.~\ref{fig:SU2fss}(b,e)] fit the data well and give $p_c^{(1)}=0.1736(2)(2)$ and $p_c^{(2)}=0.2024(2)(4)$, where the first and second parentheses give the statistical and systematic errors, respectively. Measuring the irreps of pairs therefore raises the threshold, as expected from Sec.~\ref{sec:hierarchy}. The exponents, $\nu=1.51(1)(1)$ and $1.47(1)(4)$, are consistent with the Nishimori value $\nu_{\rm RBIM}=1.534(5)$~\cite{Putz2025Nishimori,wan2026Nishimori}.

The agreement with the Nishimori point goes beyond the exponent: for the same geometry and boundary conditions, the scaling function $F$ in Eq.~(\ref{eq:FSS}) is itself universal, up to a rescaling of its argument. To test this, we computed $\bm{D}$ for the same protocol with $G=\mathbb{Z}_2$, whose syndrome distributions $p^{(1)}$ and $p^{(1)}_{xy}$ map onto partition functions of the RBIM on the Nishimori line and can be computed exactly using a free-fermion representation. This yields the scaling function $F_{\rm RBIM}$, which we compute with $L=128$. The $L=128$ data of the SU(2) model fall on $F_{\rm RBIM}$ on both levels of the hierarchy [Fig.~\ref{fig:SU2fss}(c,f)], with $\chi^2/\textrm{dof}=9.0/9$ and $3.6/6$, respectively, although only $p_c$ and the horizontal scale are fitted. In particular, there is no vertical freedom: the critical value $\bm{D}^*_{\rm RBIM}=F_{\rm RBIM}(0)=0.5613(9)$ is fixed by the RBIM, and agrees with $\bm{D}^*=F(0)=0.568(5)(6)$ and $0.554(5)(10)$ obtained from the FSS of the SU(2) data.

At every time step $t$, we compute the entanglement entropy $S^t_E(\{\Gamma_v\})$ across the middle cut of the boundary MPS from its Schmidt decomposition, and average it over syndromes,
\begin{equation}\label{eq:SEavg}
    \overline{S}^{t}_E=\sum_{\{\Gamma_v\}}p^{(n)}(\{\Gamma_v\})S_E^{t}(\{\Gamma_v\}),
\end{equation}
where $p^{(n)}$ is the syndrome distribution on the $n$th level of the hierarchy. Figure~\ref{fig:entropyTime} shows $\overline{S}^t_E$ as a function of $t$ for several values of $p$, in the $S_3$ model and on the first two levels of the SU(2) model. The entropy stops growing before $t=L/2$, and we define its plateau value as the average over $t=L/2+1,\dots,L-1$ (grey region in Fig.~\ref{fig:entropyTime}). The last time step is excluded because it is affected by the boundary.

\begin{figure}[t]
     \centering
        \includegraphics[width=\textwidth]{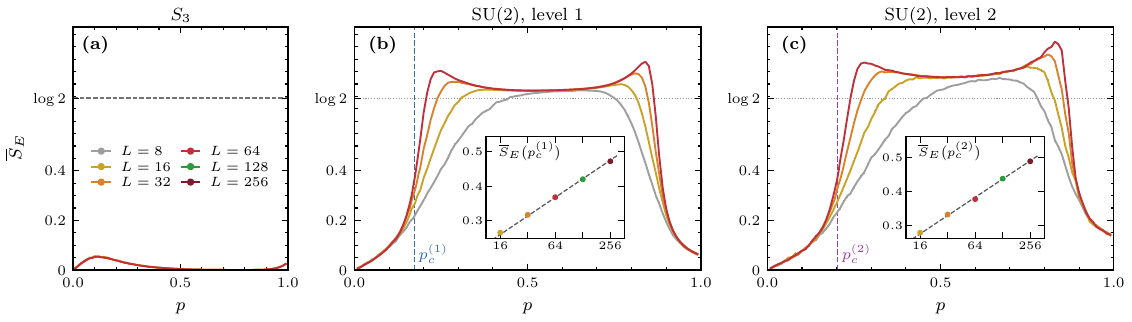}
        
        \caption{Disorder-average mid-cut entanglement entropy of the boundary MPS as a function of $p$. \textbf{(a)}: Entanglement entropy in the $S_3$-symmetric model from the Main Text. \textbf{(b)}: Entanglement entropy in the level-1 ${\rm SU}(2)$ model. Inset: critical scaling of the entanglement entropy. The slope of the dashed line is given by $c_{\rm RBIM} / 6$, where $c_{\rm RBIM} = 0.464(4)$~\cite{honecker2001universalityNishimori} is the Nishimori-Ising central charge. \textbf{(c)}: Entanglement entropy in the level-2 ${\rm SU}(2)$ model. Inset: critical scaling of the entanglement entropy. The slope is taken the same as in the inset of the panel (b).}
        \label{fig:entropyP}
\end{figure}

Figure~\ref{fig:entropyP} shows the plateau value $\overline{S}_E$ as a function of $p$. In the $S_3$ model, the entropy remains close to zero for all $p$ and does not grow with $L$, consistent with the absence of a symmetry-breaking transition at any $p>0$. In the SU(2) model, on both levels of the hierarchy, $\overline{S}_E$ is small below the threshold and rises steeply near $p_c$, approaching a value slightly above $\log 2$ in the broken phase, as expected for a cat state of the broken $\mathbb{Z}_2$ center, with an overshoot near the transitions. It drops back to zero at the second transition near $p\approx 0.88$. At the critical points, $\overline{S}_E$ grows logarithmically with $L$~\cite{holzhey1994geometric, calabrese2004entanglementQFT, calabrese2009entanglementCFT}, with a slope consistent with $c_{\rm RBIM}/6$, where $c_{\rm RBIM}=0.464(4)$ is the Nishimori Ising central charge~\cite{honecker2001universalityNishimori} [insets of Fig.~\ref{fig:entropyP}(b,c)].

\end{document}